\documentclass[twocolumn,
nofootinbib,superscriptaddress,aps,10pt,longbibliography]{revtex4-1}

\usepackage[usenames,dvipsnames]{color}
\usepackage[colorlinks=true,citecolor=blue,linkcolor=magenta]{hyperref}
\usepackage{tikz-cd}
\usepackage{xcolor}
\usepackage{amsfonts,amssymb}
\usepackage[sumlimits,intlimits]{amsmath}
\usepackage{graphics}
\usepackage{graphicx}
\usepackage{mathrsfs}
\usepackage{textcomp}
\usepackage{verbatim}
\usepackage{bm}          

\newcommand{\be}{\begin{equation}}
\newcommand{\ee}{\end{equation}}

\newcommand{\bit}{\begin{itemize}}
\newcommand{\eit}{\end{itemize}}

\newcommand{\f}{\frac}
\renewcommand{\>}{\right\rangle}
\newcommand{\<}{\left\langle}
\newcommand{\ba}{\begin{align}}
\newcommand{\ea}{\end{align}}

\newcommand{\bi}{\begin{itemize}}
\newcommand{\ei}{\end{itemize}}
\newcommand{\lf}{\left(}
\newcommand{\ri}{\right)}
\newcommand{\dd}{\mathrm{d}}

\newcommand{\Tr}{\operatorname{Tr}}

\newcommand{\bra}[1]{\< #1 \right|}
\newcommand{\ket}[1]{\left| #1 \>}

\renewcommand{\vec}[1]{\boldsymbol{\mathbf{#1}}}

\newcommand{\cp}{\mathrm{CP}}
\newcommand{\rp}{\mathrm{RP}}

\graphicspath{{}{./images/}}

\newcommand{\fourdots}[4]{
  \begin{tikzpicture}
  \fill[black] (0,0) circle (0.1cm);
  \fill[black] (0.5,0) circle (0.1cm);
    \fill[black] (1,0) circle (0.1cm);
  \fill[black] (1.5,0) circle (0.1cm);
  \node () at (0,0.4) {#1};
    \node () at (0.5,0.4) {#2};
      \node () at (1,0.4) {#3};
    \node () at (1.5,0.4) {#4};
 \end{tikzpicture} 
}

\newcommand{\fourdotssmall}[4]{
  \begin{tikzpicture}
  \fill[black] (0,0) circle (0.08cm);
  \fill[black] (0.3,0) circle (0.08cm);
    \fill[black] (0.6,0) circle (0.08cm);
  \fill[black] (0.9,0) circle (0.08cm);
  \node () at (0,0.4) {#1};
    \node () at (0.3,0.4) {#2};
      \node () at (0.6,0.4) {#3};
    \node () at (0.9,0.4) {#4};
 \end{tikzpicture} 
}

\newcommand{\fourdotssmallgap}[2]{
  \begin{tikzpicture}
  \fill[black] (0,0) circle (0.08cm);
  \fill[black] (0.9,0) circle (0.08cm);
  \node () at (0,0.4) {#1};
    \node () at (0.9,0.4) {#2};
 \end{tikzpicture} 
}

\begin{document}

\title{Entanglement and dynamics of diffusion-annihilation processes with Majorana defects}

\author{Adam Nahum}
\affiliation{
Theoretical Physics, University of Oxford, Parks Road, Oxford OX1 3PU, United Kingdom}
\author{Brian Skinner}
\affiliation{Department of Physics, Massachusetts Institute of Technology, Cambridge, MA  02139 USA}
\affiliation{Department of Physics, Ohio State University, Columbus, OH 43210, USA}

\date{\today}

\begin{abstract}
Coupling a many-body system to a thermal environment typically
destroys the quantum coherence of its state, leading to an effective classical dynamics at the longest time scales.
We show that systems with anyon-like defects can exhibit universal late-time dynamics that is stochastic, but fundamentally non-classical, because some of the quantum information about the state is topologically protected from the environment.  

Our coarse-grained model describes one-dimensional systems with domain-wall defects carrying Majorana modes. These defects undergo Brownian motion due to coupling with a bath. 
Since the fermion parity of a given pair of defects is nonlocal, it cannot be measured by the bath until the two defects happen to come into contact.
We examine how such a system anneals to zero temperature via the diffusion and pairwise annihilation of Majorana defects, and we characterize  the nontrivial  entanglement structure that arises in such stochastic processes. 

Separately, we also investigate simplified  ``quantum measurement circuits'' in one or more dimensions, involving  repeated pairwise measurement of fermion parities for a lattice of Majoranas.  The dynamics of these circuits can be solved by exact mappings to classical loop models. They yield analytically tractable examples of measurement-induced phase transitions, with critical entanglement structures that are governed by nonunitary conformal fixed points.

In the system of diffusing and annihilating Majorana defects, the relaxation to the ground state is analogous to coarsening in a classical 1D Ising model via domain wall annihilation (the classical  ``$A+A \rightarrow \emptyset$'' reaction-diffusion process).
Here, however, configurations are labeled not only by the defect positions but by a nonlocal entanglement structure.
This ``$\gamma+ \gamma \rightarrow \emptyset$'' process is a new  universality class for the coarsening of topological domain walls, whose universal properties can be obtained from an exact mapping.  
\end{abstract}

\maketitle

\section{Introduction}

Coupling a quantum many-body system to a thermal environment destroys the quantum coherence of its dynamics. 
Usually, a generic bath coupling leads to dynamics that is essentially classical at the longest timescales, such that the slow degrees of freedom are 
governed by a classical ``hydrodynamics'', either stochastic or deterministic.
In this paper we exhibit a model that avoids this classical fate,
because some of the information in the quantum state is \textit{topologically protected} \cite{kitaev2001unpaired, nayak2008non} from the environment. 
The late-time description is instead a hybrid of quantum and classical. 
Local degrees of freedom that can be ``measured'' by the bath become effectively classical, since they cannot remain long in a superposition.
But the topologically protected Hilbert space --- associated with anyon-like defects --- allows long-range entanglement to persist.
This entangled state in turn affects the slow ``classical'' degrees of freedom, which here are the positions of the defects.

The model we introduce is for a one-dimensional chain with  defect excitations that carry Majorana zero modes.
Topologically, these defects are like domain walls between the topological and trivial phases of Kitaev's $p$-wave chain \cite{kitaev2001unpaired}.
We model this system during a process of relaxation
from a high-temperature state with many defects to 
a low temperature state where the defects are absent,
via the diffusion and pairwise annihilation of defects.
During this process, the system loses energy to the surrounding bath (which is taken to be bosonic --- i.e.~fermions cannot hop into or out of the system).

The diffusive motion of a single defect is entirely classical thanks to the bath coupling.
But when the defects are widely separated --- which is mostly the case at late times --- the Hilbert space associated with the Majorana modes is protected from decoherence:  all bosonic operators on this space are nonlocal, and so are invisible to the bath.
Thus describing the state requires one to know both the positions of the defects and their entanglement structure. 

One consequence of this hybrid classical/quantum dynamics is that the density of defects decays to zero with a universal amplitude that is different from what it would be if the defects were,
say, featureless domain walls in the classical Ising model. The defect positions also exhibit different universal correlation functions.
The relaxation of the one-dimensional Ising model to zero temperature under Glauber dynamics \cite{glauber1963time}
occurs by diffusion and annihilation of domain walls --- an instance of the ``$A+A\rightarrow \emptyset$'' reaction-diffusion process \cite{Tauber_2005}. This is a nontrivial stochastic process for which a wide range of universal quantities can be obtained exactly 
\cite{bramson1980clustering,
torney1983diffusion,
lushnikov1987binary,
Bray_1990,
Tauber_2005,
derrida1994non,
derrida1996exact}.
We find that the density of defects in the Majorana model, decaying via the ``${\gamma + \gamma\rightarrow \emptyset}$ process'',
is exactly twice that of the purely classical case for a given (late) time.

The  crucial feature of the model is a stochastic evolution of the quantum state of the Majorana zero modes $\gamma_i$.
This state is able to remain a \textit{pure} state, despite the coupling to the environment, but its evolution is entirely nonunitary.
In the long time limit of the diffusion--annihilation process, the details of the environment coupling become unimportant and a simple universal description emerges.
In this limit the  environment simply effects projective measurements of the fermion parity ($i \gamma_i \gamma_{i+1}$) of pairs of adjacent Majorana modes.
Specifically, measurement takes place when two  Majorana-carrying defects approach each other in space. At such moments, when random, classical diffusion happens to bring two Majoranas into contact, their mutual fermion parity becomes a local operator that can be seen by the bath. 
The result of this projective measurement determines whether or not the Majorana domain walls are allowed to annihilate back into the ground state.

Importantly, if the measurement outcome is such as to prevent annihilation, the quantum state stores this information until the next encounter. This additional ``memory'' effect means that the universality class of the dynamics is different from the classical ${A+A\rightarrow \emptyset}$ universality class. Instead, we show that the ${\gamma + \gamma \rightarrow \emptyset}$ process can be related to two copies of the ${A+A\rightarrow \emptyset}$ process by an exact mapping.

Separately from this diffusion-annihilation dynamics, we study the process of repeated  pairwise measurement of a fixed number of Majoranas, i.e.~the evolution of the system in the absence of annihilation.
This dynamics, which involves repeated measurement of the parity of adjacent Majorana pairs, is interesting in its own right. In particular, we show that it leads to a nontrivial critically entangled state.
The simplest version of this model is a ``quantum measurement circuit'', with measurements applied randomly to a fixed lattice of Majoranas.  We show that this measurement circuit gives a tractable example of criticality induced by measurement.

In general, combining local unitary dynamics of a pure quantum state (say, for a lattice of spins) with repeated local measurements yields a phase transition in the dynamics of entanglement growth as a function of the measurement rate \cite{skinner2018measurement,li2018quantum}. Various aspects of this transition are now understood \cite{chan2018weak,choi2019quantum,szyniszewski2019entanglement,li2019measurement,cao2018collective,gullans2019dynamical,gullans2019scalable,tang2019measurement}. In the generic case this phase transition is continuous, and is characterized by a nontrivial renormalization group (RG) fixed point with connections to unconventional statistical mechanics models \cite{skinner2018measurement,ZhouNahum,li2019measurement,vasseur2018entanglement,bao2019theory,jian2019measurement}.

The projectively measured Majorana chain that we study here also flows to a nontrivial RG fixed point. 
In fact, so long as only pairs of Majoranas are measured, and translation symmetry is preserved, the dynamics is critical without the need to tune any parameter. 
However, the properties of this critical state are very different from (and simpler than) the generic case described above.
The dynamics in our Majorana models should also be contrasted with those in models of free fermions subjected to single-site measurements, where  a disentangled phase was found \cite{cao2018collective,chan2018weak}.

\begin{figure}[t]
\centering
\includegraphics[width=0.7 \linewidth]{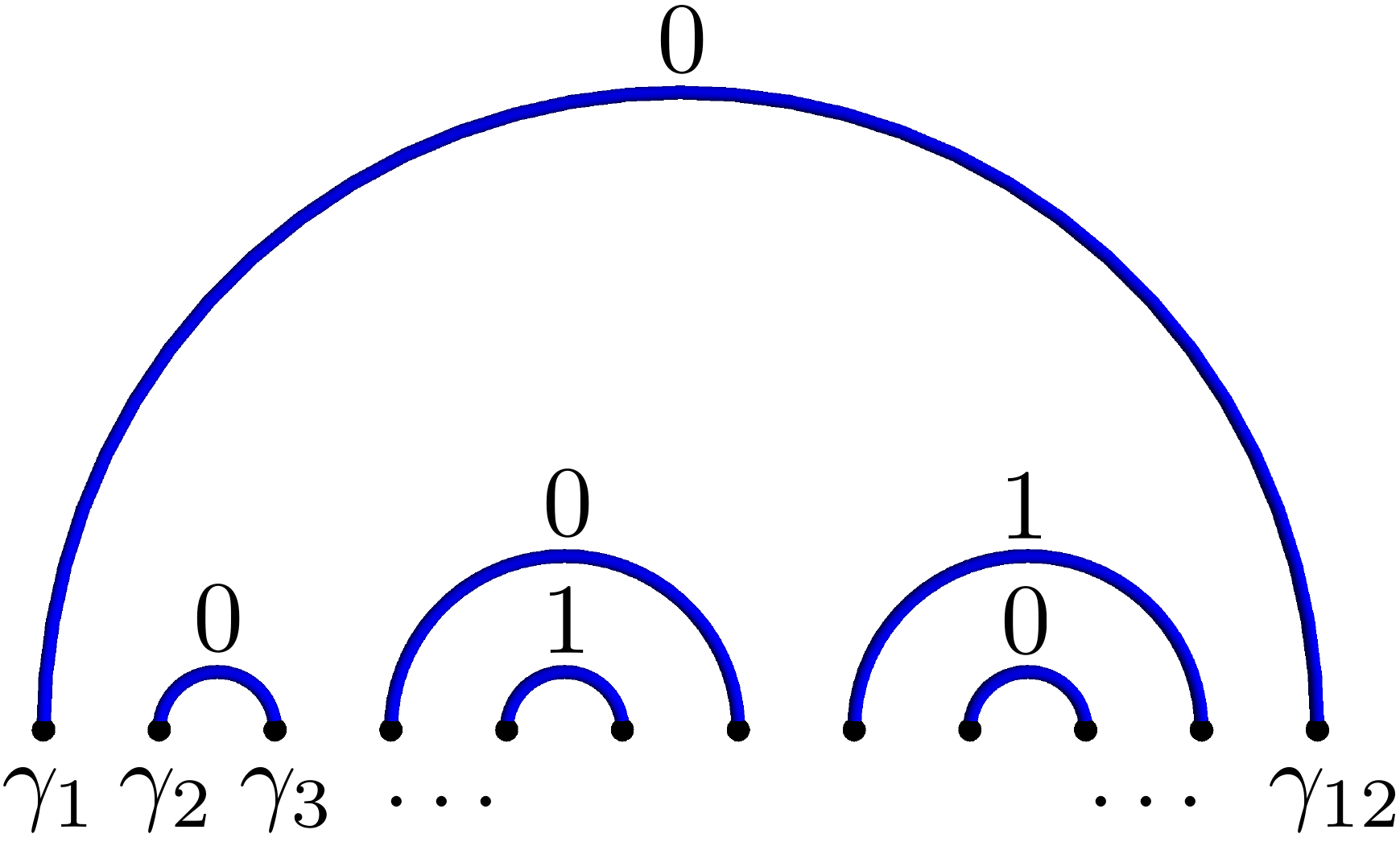}
\caption{
An example of a state for a set of Majorana modes $\gamma_1, \ldots, \gamma_{12}$, defined by a choice of pairings (an ``arc diagram'') and an assignment ${n_{ij}=(i \gamma_i \gamma_j+1)/2}=0,1$ to each arc (connecting $i$ and $j$, with $j>i$).}
\label{fig:examplearcs}
\end{figure}

Strikingly, the repeated projective measurements organize the system of Majoranas into a random state for which the entanglement of a subregion scales logarithmically with the subregion size. 
In terms of its entanglement, this state is somewhat like a random singlet state \cite{Lee1981,fisher1999phase}: 
it is described by an ``arc diagram'' like that in Fig.~\ref{fig:examplearcs}, for which only two-body entanglement is ever generated and the number of arcs crossing the midpoint of a system of length $L$ is of order $\log L$. 
But unlike the random singlet state, which is a ground state of a random Hamiltonian,
the present state is prepared by a stochastic measurement process, without any need to minimize a Hamiltonian, and we show that its universal properties are different from those of the random singlet phase.
In fact, these universal properties can be described exactly via an equivalence between the random ``updates'' to the pairing diagram produced by measurements, 
and the Temperley--Lieb operations \cite{TemperleyLieb}
that appear in the context of 
a 2D lattice model for nonintersecting loops \cite{blote1989critical,cardy2005sle, jacobsen2009conformal,saleur1987exact, cardy2000linking} and which are equivalent to a stochastic process on pairing diagrams \cite{pearce2002temperley,deGierRaiseandPeel1,deGierRaiseandPeel2}. 
We also relate the criticality of the stochastic measurement process to a Lieb-Schultz-Mattis-like theorem for disordered spin chains \cite{kimchi2018valence}.

The quantum measurement circuit dynamics can be generalized to higher dimensions, 
and remains exactly solvable by a similar mapping to a  classical loop model on a 3D lattice \cite{nahum20113d}.  The diffusion-annihilation dynamics in higher dimensions, however, are complicated by the possibility of braiding Majoranas around each other, which can alter the parity of Majorana pairs even when no measurements are performed.
The models we study could also be generalized to other types of anyon, but we leave this generalization for the future.

\tableofcontents

\section{Models and Simulation}
\label{sec:model}

This paper addresses two different types of model.

\begin{figure}[t]
\centering
\includegraphics[width=0.48 \textwidth]{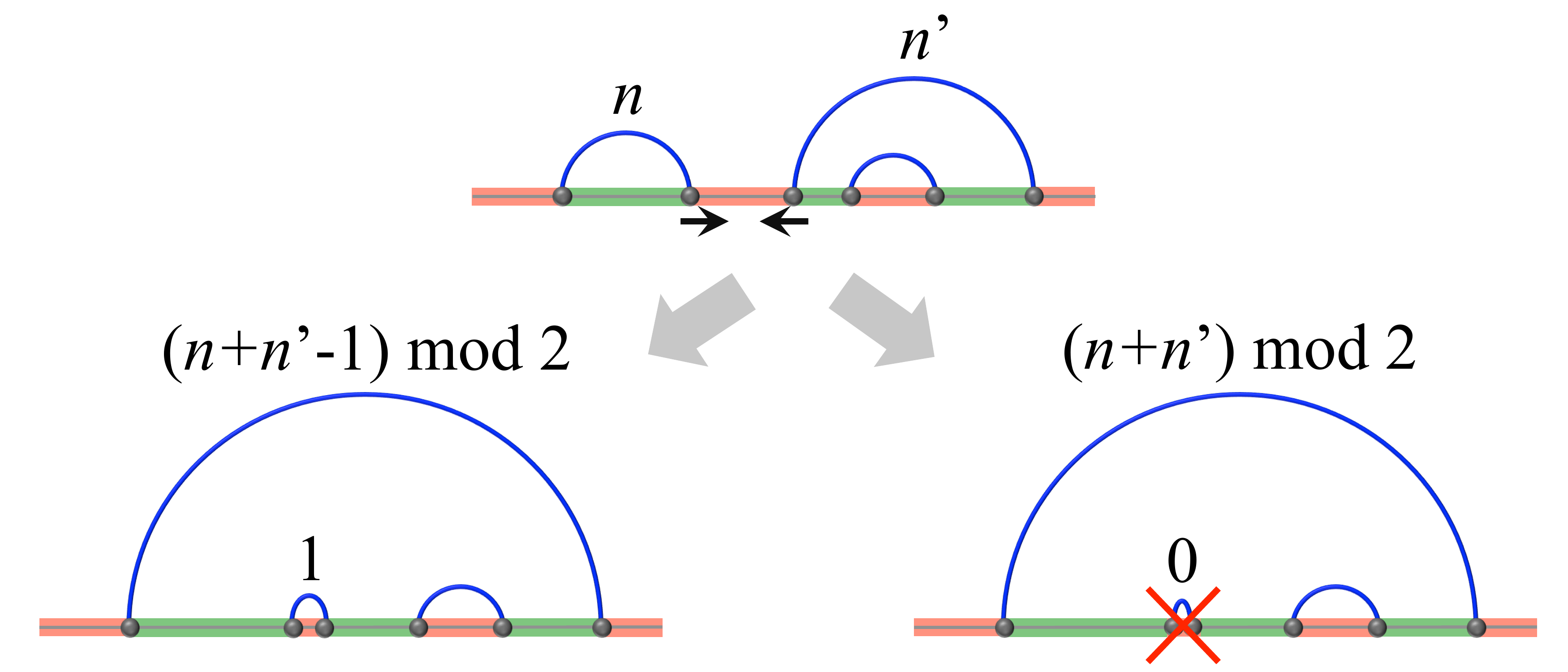}
\caption{Evolution of the number and pairing of Majoranas in the model with annihilation.  When two  Majoranas (black points) that are not paired with each other come into contact (horizontal arrows, above), the resulting local pair has fermion number either $1$ (left lower configuration) or $0$ (lower right), with each outcome having equal probability. In the latter case the pair is removed from the system.  In either case the nonlocal pair has a fermion number such that the total fermion parity of the system is conserved modulo $2$. }
\label{fig:coarsening}
\end{figure}

The first involves mobile Majorana defects which can diffuse and annihilate: we use this model to discuss a type of relaxational dynamics (Fig.~\ref{fig:coarsening}).
The relaxation we consider is to zero temperature, so that energy always goes down (or stays the same).  
This process leads to a length scale, the mean interparticle spacing, that diverges like $\sqrt t$ at late times (as in the relaxation of the classical 
1D Ising model via the annihilation of domain walls). This diverging length scale ensures that the results are universal.
The diffusion-annihilation model is introduced in Secs.~\ref{sec:diffusingdefects},~\ref{sec:microscopic} below.

The second model is simpler: we assume that there is no annihilation between Majoranas. For example, one can imagine an energy barrier arising from a repulsive short-ranged interaction, which prevents Majoranas from annihilating, but still allows them to approach each other close enough that their parity can be measured.
Or, more simply, we can place the Majoranas on a fixed lattice and apply measurements at random to pairs. 
In this case the evolution of the system is determined only by a random sequence of local projective measurements. 

\begin{figure}[t]
\centering
\includegraphics[width=0.35 \textwidth]{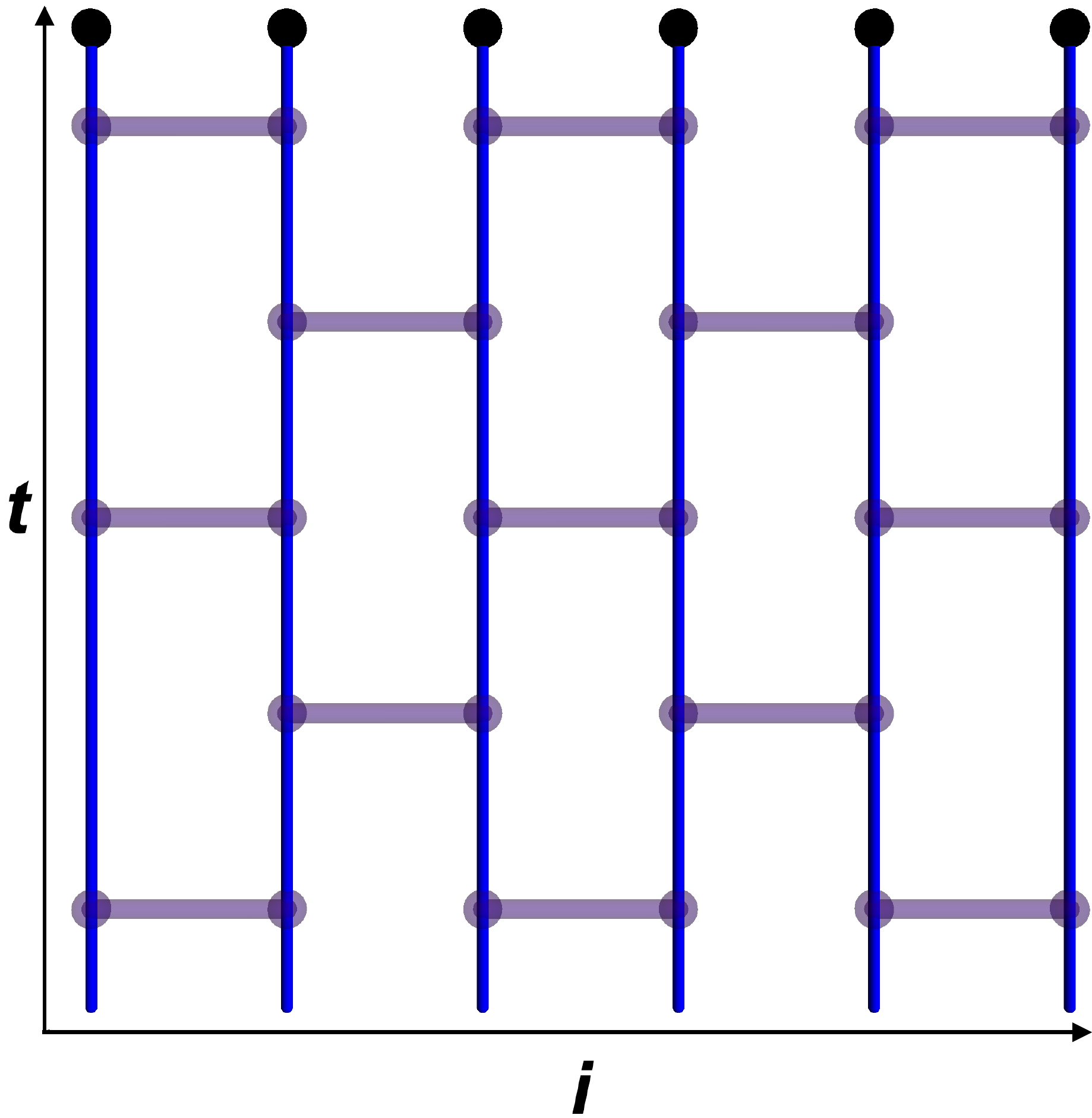}
\caption{
{Quantum measurement circuit model. The horizontal axis is the coordinate $i$ of the Majoranas and the vertical axis is the time $t$. Each horizontal bar marks a space-time location where a projective measurement of fermion parity \textit{may} be performed, with probability 1/2.}
}
\label{fig:circuit}
\end{figure}

This is a form of ``measurement-only'' random quantum circuit: see Sec.~\ref{sec:circuitmodelintro} and Fig.~\ref{fig:circuit}.
This second model is no longer a model for relaxational dynamics: we study it instead as an example of how a nontrivial entanglement structure can be produced solely by random measurement.

\subsection{Diffusing Majorana defects}
\label{sec:diffusingdefects}

We study a model of Majorana defects diffusing in 1D. 
Let us first describe a coarse-grained model, and then discuss how it can emerge from a microscopic model.
One possible physical interpretation of the defects (see Sec.~\ref{sec:microscopic}) is as \textit{domain walls} between 
the two topologically distinct phases of a Kitaev chain \cite{kitaev2001unpaired}, with additional interactions to place it at a first order transition \cite{rahmani2015emergent,rahmani2015phase,obrien2018lattice} between the trivial phase and the topological phase. (In some settings, these two states can be related by symmetry, so that no tuning of parameters is required to access the transition.)

The number of defects $N_t$ must be even but is otherwise arbitrary, and can decrease with time $t$. 
Each defect $i=1,\ldots, N_t$ has a position $x_i$ (all distinct). 
For numerical convenience we put these positions on a 1D lattice of size $L$, 
with periodic boundary conditions, though one could also imagine a continuum model.
To fully describe the state of the system we must specify the number of defects, their positions $\{ x_i\}$, and their quantum state.

Each defect contains a Majorana mode $\gamma_i$ (with anticommutators $\{\gamma_i, \gamma_j\} = 2 \delta_{ij}$). 
Any choice of pairings of Majoranas gives a basis for the total Hilbert space:
a given pair $i,j$, with $i<j$, has a two-dimensional Hilbert space spanned by states with fermion parity $i\gamma_i\gamma_j = \pm 1$. 
We will label the parity by the fermion number operator $n_{ij} = (i \gamma_i \gamma_j + 1)/2 = 0, 1$.

Thus, a labelled arc diagram such as Fig.~\ref{fig:examplearcs} completely specifies the quantum state of the system, 
so long as the instantaneous state happens to be a basis state for some choice of pairing. 
It turns out, as we explain below, that we only need to consider states of this form (rather than superpositions of such states).

The rules for annihilation and measurement of Majoranas are simple because of universality in the long-time limit; see the remainder of this section for explanations, and Sec.~\ref{sec:simulation} for the precise rules of the numerical simulations on the lattice.  The dynamics of the coarse-grained model are as follows.

Each isolated Majorana defect performs classical Brownian motion with diffusion constant $D$.
When two Majoranas $i$ and $i+1$ come into contact, their fermion parity $i\gamma_i \gamma_{i+1}$ is a local operator and is visible to the bath, which makes a projective measurement that assigns a definite value to $i\gamma_i \gamma_{i+1}$.
If $i$ and $i+1$ are already paired, this measurement produces no update to the pairing. 
If instead $i$ is paired with some other defect $j$, and $i+1$ is paired with $k$, 
then the pairings are updated so that $i$ is now paired with $i+1$, and $j$ with $k$.
This re-pairing process is illustrated in Fig.~\ref{fig:coarsening}.
The fermion number $n_{i, i+1}$ of the newly formed pair $i$ and $i+1$ is given a definite value, $n_{i,i+1} = 0 \textrm{ or } 1$, with probability $1/2$ for each option.  The fermion parity of the pair $(j,k)$ is fixed by $\mathbb{Z}_2$ conservation of fermion parity: 
the sum of the values for the initial arcs, $n_{i,j}+n_{i+1,k}$, must be equal to the sum of the new values, $n_{i,i+1}+ n_{j,k}$, modulo 2.
If we are allowing annihilation in the model, then the adjacent defects $i$ and $i+1$ annihilate \textit{if and only if} their fermion number $n_{i,i+1}$ is zero
(see Fig.~\ref{fig:coarsening}.)

The two outcomes of the measurement of $n_{i,i+1}$ are equally likely because the reduced density matrix for $n_{i,i+1}$ is maximally mixed whenever $n_{i,j}$ and $n_{i+1, k}$ have definite values (as they do before the measurement).
If the fermion number $n_{i,i+1}$ is zero the defects can annihilate into the ground state. But if it is unity, conservation of fermion number modulo 2 prevents them from annihilating. In the latter case the defects continue to diffuse.  Implicit in these dynamical rules is the hierarchy of energy scales $E_{\emptyset} < 2 E_\gamma < E_f$, where $E_{\emptyset}$ is the energy of the vacuum state, $E_\gamma$ is the energy of a single Majorana defect, and $E_f$ is the energy of a single fermion (the local ``bound state'' of two Majoranas).  The first inequality in this chain implies that Majoranas with fermion number $0$ annihilate, and the second implies that Majoranas with fermion number $1$ remain separate.

The important feature of the diffusion-annihilation dynamics that differentiates it from the purely classical case is that when $i$ and $i+1$ project into a state with $n_{i,i+1} = 1$ and diffuse apart, they \textit{remember} that they cannot annihilate. Such a pair remains in a state of definite mutual parity until one of its two constituents encounters a third defect. Correspondingly, if $i$ and $i+1$ encounter each other again, without meeting anyone else in the meantime, they will again fail to annihilate.

At late times $t$ the typical separation $r(t)$ between defects is large, and the properties of 1D random walks imply that when two defects meet once, they in fact meet a parametrically large number of times ($\mathcal{O}(r)$ times) before either of them meets a third party. 
This large number of meetings justifies having no dependence of the protocol on a microscopic annihilation rate or a microscopic bath coupling. 
So long as annihilation is allowed by the fermion parity, it will definitely happen in a time short compared to the diffusive timescale (when $r$ is large), irrespective of the microscopic rate.
(See Sec.~\ref{sec:with} for a more detailed discussion.)
The irrelevance of the microscopic annihilation rate is well-known in the relaxation of the 1D Ising model and other 1D examples; these relaxation processes are ``diffusion--limited'' rather than ``rate--limited''~\cite{Tauber_2005}.

For this same reason, we can treat the effect of the environment as a simple projective measurement. Even if the microscopic bath coupling is weak, it is amplified by the parametrically large number of repeated encounters.

Thus our model has only a single parameter, which is the diffusion constant $D$.
We will be interested in the combined evolution of the positions $\{x_i\}$ and the entanglement structure. The latter is represented simply by the geometry of the arc diagram (Fig.~\ref{fig:examplearcs}).
In particular, we will pin down the universal late-time behaviour of the density, and we will compare this with a reference classical model in which the defects are purely classical variables, representing, for example, classical Ising domain walls.

\subsection{Relation to microscopic models}
\label{sec:microscopic}

\begin{figure}[t]
\centering
\includegraphics[width=0.375 \textwidth]{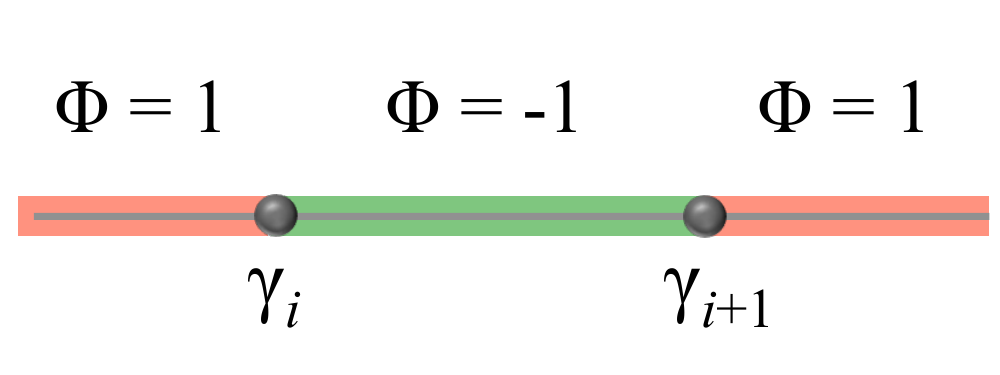}
\caption{Definition of the order parameter $\Phi(x)$. Majorana defects (denoted by $\gamma$) separate adjacent ground state domains (red/green regions), which have $\Phi = \pm 1$.}
\label{fig:correlation_domains}
\end{figure}

The Majorana defects in our coarse-grained model live at domain walls separating two types of ground state, which we can label by a local ``order parameter'' ${\Phi = \pm 1}$ (see Fig.~\ref{fig:correlation_domains}). The two ground states are assumed to have equal energy density, so that the diffusion of a domain wall is unbiased. Depending on the microscopic model, this degeneracy of the two ground states could either be due to the system being tuned to a first-order transition, or it could be enforced by a symmetry relating the two ground states (so that  $\Phi$ is the order parameter for the breaking of this symmetry).

For an example of how Majorana defects arise, consider the  Kitaev chain, which describes a chain of fermions coupled to a superconducting order parameter. Its Hamiltonian may be written \cite{kitaev2001unpaired}:
\be\label{eq:kitaevchain}
H= i \sum_{x\in \frac{1}{2} \mathbb{Z}}  \left( J +  (-1)^{2x} \delta J \right) \eta_x \eta_{x+1/2}.
\ee
Here ${J\pm \delta J}$ is the Majorana hopping amplitude, and we have labelled the sites $x$ by half-integers for convenience. The ``microscopic'' Majorana operators $\eta_i$ should be distinguished from the variables $\gamma$ used above in the coarse-grained model.
The former are related to physical fermion operators $c_x$ by combining them in pairs, such that $c_x = \frac{1}{2} \lf \eta_x + i \eta_{x+1/2}\ri$ for $x\in \mathbb{Z}$.

The two phases of the chain can be visualized simply in the limits $\delta J = J$ and $\delta J = -J$. 
In each of these limits the chain is dimerised in one of the two possible ways \cite{kitaev2001unpaired}, 
and the fermion parity of each dimer is zero. 
A single domain wall between the two phases carries an unpaired Majorana mode.

In the above model the quantum phase transition between topological and non-topological phases, which occurs at ${\delta J = 0}$, is continuous and described by free fermions. However, adding interactions to the Kitaev chain can make this transition first-order.
The details of these interactions are not important for the universal coarse-grained model, but a first-order transition is important in order to have well-defined domains. 
One method of making the transition first-order is by using a four--$\eta$ coupling (the ``Majorana Hubbard model''), as demonstrated in Refs.~\cite{rahmani2015emergent,rahmani2015phase}.
In this model the first-order transition occurs at a large interaction strength;
a different choice of four--$\eta$ interaction yields the first order transition for much smaller values of the couplings \cite{obrien2018lattice}.
The same transition can also be obtained in continuum theories, as proposed in the context of the 1D boundary of a 2D topological superconductor, with spontaneously broken time-reversal symmetry on the boundary
\cite{grover2014emergent}.

Let us briefly comment on symmetry. 
In the Kitaev chain, and the interacting versions discussed above, there is a translation symmetry ${x\rightarrow x+1/2}$ that acts as an Ising symmetry on the ``order parameter'' $\Phi$  distinguishing the topological and trivial phases. 
In the usual interpretation of the Kitaev chain, where adjacent $\eta$s are grouped to make local electron orbitals, this is not a physical symmetry \cite{jones20191d}, as is clear from the fact that it relates topologically distinct phases.
In this interpretation, a parameter must be tuned to reach the  transition point where the topological and trivial phases are degenerate in energy.

However, in other realizations of Majorana defects the two phases can be related by a physical symmetry. In the setup of Ref.~\cite{grover2014emergent}, for example, where the 1D system comprises the boundary of a 2D bulk, this symmetry is time reversal. If the Majorana modes are realized by a 1D array of defects in a 2D topological system, the translational symmetry of the Kitaev chain can be a physical translation symmetry.

Many realizations of Majorana modes have been proposed (see for example Refs.~\cite{alicea2012new,elliott2015colloquium} for an introduction), including at the vortex cores of 2D $p+ip$ superconductors \cite{read2000paired},  as excitations of the Moore Read quantum Hall state \cite{read1992fractional},  in proximity-coupled surfaces of 3D topological insulators \cite{fu2008superconducting}, and at endpoints of proximatized 1D wires (see for example Ref.\ \cite{Lutchyn2018} for a review).
For discussions of physical realizations of strongly-interacting lattice Hamiltonians for Majoranas, see for example Refs.~\cite{hassler2012strongly,chiu2015strongly, vijay2015majorana}.

We consider a model with a first-order transition, tuned to the first-order transition point. (In the presence of a symmetry relating the two phases, this simply means we are in the ordered phase for $\Phi$.)
Now we imagine that the chain is prepared at high temperature, and allowed to relax to low temperature via a coupling to a low-temperature bath.
At late times the state will consist of large domains of one phase or the other which in their interior resemble one of the two degenerate ground states. 
The typical size of these domains grows with time, by the diffusion and annihilation of domain walls (Fig.~\ref{fig:coarsening}). 
Since the system is poised right at the phase transition, neither phase is preferred, which means that the diffusion of domain walls is unbiased. (If we tuned slightly away from the transition, then one domain type would be preferred, and on average would grow at the expense of the other.)

The model described in the previous section provides a universal description of this class of models in an appropriate limit of temperature and time scale.
The temperature of the bath should not be strictly zero: in that limit the defects do not diffuse, and quantum effects other than the ones we consider may become important \cite{hakim1985quantum,sinha1992brownian,maghrebi2016flight}. However, the temperature of the bath is assumed to be parametrically small, so that at equilibrium the concentration of thermally excited defects is also parametrically small. We can therefore consider an effective model on length scales below the typical spacing between such thermally-excited defects, at which the density is treated as relaxing to zero: this is the model described in the previous subsection.
An important additional requirement is that the bath should not have low-energy (gapless) fermionic modes that can couple directly to the $\gamma$s.

\subsection{Quantum measurement circuit (no annihilation)}
\label{sec:circuitmodelintro}

We may also consider the dynamics in Sec.~\ref{sec:diffusingdefects} with diffusion but no annihilation. We will find that such dynamics lead to a critically entangled state. This annihilation-free dynamics is interesting to study even though it is not directly related to the relaxation process above.
In this limit the defects diffuse around, being projectively measured whenever they bounce into each other. We present numerical simulations of this process in Sec.~\ref{sec:entanglementoverview}.

But such a model invites an even further simplification, in which the Majoranas $\gamma_i$ are taken to be at fixed positions $i \in \mathbb{Z}$ and are subjected to repeated measurements of randomly chosen nearest-neighbor pairs. This procedure gives a quantum-circuit-like model for random projective measurement.
The simplest way to set up this ``circuit'' is as in Fig.~\ref{fig:circuit}, so that at even time steps even-numbered links have the opportunity to be measured, with a probability $p$ for a measurement to occur, and at odd time steps the odd links are measured with the same probability. 

Note that this model is for the dynamics of a pure state: i.e. the measurement outcome is chosen randomly, but we do not average over it (which would give instead a mixed state density matrix). Conceptually, we can imagine that all the measurement outcomes are recorded by an observer, who therefore has access to the pure state. See Refs.~\cite{skinner2018measurement,li2018quantum,chan2018weak,choi2019quantum,gullans2019dynamical} for discussions of these issues.

By representing the worldlines of the Majoranas in spacetime, the dynamics in this circuit model maps to a well-studied classical 2D loop model \cite{cardy2005sle, jacobsen2009conformal}, as we explain below.
Equivalently, if we neglect the fermion parity labels, the dynamics on pairing diagrams in this model map to a known stochastic process derived from the loop model \cite{pearce2002temperley,deGierRaiseandPeel1,deGierRaiseandPeel2}. 
Exact results are available for these problems that determine the universal properties of the entanglement structure generated by the circuit (Sec.~\ref{sec:loopmodel2d}).
When even and odd bonds are measured at the same rate, the model is critical, with logarithmic scaling of entanglement, while dimerizing the measurement rates leads to area-law states. This model may also be generalized to higher dimensions (Sec.~\ref{sec:higherd}).

\subsection{Simulation method}
\label{sec:simulation}

We implement a computer simulation of random-walking Majorana defects on a discrete 1D lattice with periodic boundary conditions.  The system is initialized with a large number $N_0$ of defects uniformly distributed across a system of size $L$, and the initial pairings are such that each Majorana is paired with a nearest neighbor (a dimerized configuration) and all pairs have initial parity $n = 1$. 

The system evolves by random unit hops (left or right) of randomly-selected Majoranas, such that during each time step there are $N_t$ hops, where $N_t$ is the number of Majoranas in the system at the beginning of time step $t$.  If a hop brings two Majoranas $i$ and $i+1$ onto the same site, then the hopping Majorana is reflected back to its initial site and the parity of the two Majoranas is updated according to the rules in Sec.~\ref{sec:diffusingdefects}:

If $i$ and $i+1$ are already paired, then no updates are made to the pairings in the system;

If $i$ and $i+1$ are not already paired, they become paired upon contact.  Their fermion number is set randomly to either $n_{i,i+1} = 0$ or $n_{i,i+1} = 1$ with equal probability. The previous partners of $i$ and $i+1$ are joined into a (nonlocal) pair, whose parity is fixed by fermion parity conservation (conservation of $\sum n$ modulo 2).
This is shown in Fig.~\ref{fig:coarsening}.

Below we study two cases for the evolution of the number of Majoranas.  In the first case (Sec.\ \ref{sec:without}), the number of Majoranas in the system is fixed.  In this case the arc labels  $n$ are irrelevant to the dynamics of the pairing diagram, since the rules for pairing Majoranas do not depend on the parities of the arcs involved.  
In the second case (Sec.\ \ref{sec:with}), two Majoranas are annihilated immediately if they come into contact and are measured in the $n_{i,i+1} = 0$ state. Details about the simulation parameters, including the system size and run time, are listed in the corresponding figures.

\section{Entanglement generation with a fixed number of Majoranas}
\label{sec:without}

\subsection{Overview and numerical results}
\label{sec:entanglementoverview}

\begin{figure}[t]
\centering
\includegraphics[width=0.5 \textwidth]{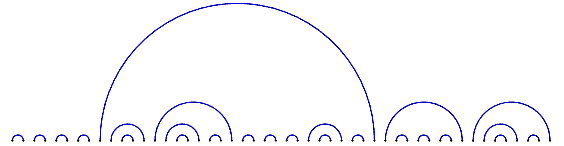}
\caption{Typical pairing configuration in the diffusing Majorana model dynamics, with no annihilation, after a long time.  This dynamics generates an arc length distribution $P(\ell) \propto 1/\ell^2$, see Sec.~\ref{sec:without}.
The abscissa is the Majorana's index $i$.
}
\label{fig:typicalarc}
\end{figure}

We begin by considering the simpler model in which annihilation of Majoranas is not allowed.
When the number of Majoranas in the system is fixed, 
there is still a nontrivial evolution of the arc diagram, i.e.\ of the wavefunction's entanglement structure.
Starting from the initial state with only nearest-neighbor pairs, longer-ranged pairs form over time via the process illustrated in Fig.~\ref{fig:coarsening}, as well as by the Brownian motion of endpoints.  One can then ask: what distribution $P(\ell)$ of arc lengths $\ell$ arises from this dynamics, and over what characteristic time scale does it develop?

The arc length distribution is directly related to the amount of entanglement in the state.
The von Neumman entanglement entropy $S_\mathcal{A}$ of a subregion $\mathcal{A}$ is
\be\label{eq:vonneumman}
S_\mathcal{A} = - \Tr_\mathcal{A}
\rho_\mathcal{A}
\log \rho_\mathcal{A},
\ee
where $\rho_\mathcal{A}$ is the reduced density matrix for the fermion orbitals $c_x$ in the subregion (see Sec.~\ref{sec:microscopic}). 
Each arc that  connects $\mathcal{A}$ to its exterior contributes half a bit of entanglement entropy. (In general there is also an order 1 contribution from the boundary of $\mathcal{A}$, which we neglect.) 
If we choose half a bit as our unit of entropy, i.e.\ if we take the logarithm in Eq.~(\ref{eq:vonneumman}) base $\sqrt 2$, then $S_\mathcal{A}$ is simply the number of arcs that leave region $\mathcal{A}$.

Let $\mathcal{A}$ be a section of $R$ contiguous lattice sites in an infinite chain, and let $\rho$ be the concentration of Majoranas in the chain (in units of the inverse lattice spacing), so that $R \rho$ is the mean number of Majoranas in $\mathcal{A}$. The mean entanglement of $\mathcal{A}$ with its surroundings is equal to the mean number of arcs exiting the region, which is
\be\label{eq:S}
S_\mathcal{A} = \, \sum_{x=1}^R \, \sum_{\ell=R-x+1}^\infty \, \rho P(\ell).
\ee
Here, $\ell$ is the integer length of the arc in units of the lattice spacing.
Notice that, if $P(\ell)$ is a power law ${P(\ell) \propto 1/\ell^\alpha}$ for arc lengths $\ell \gg \rho^{-1}$, then the value $\alpha=2$ is a critical case.
When $\alpha>2$ the above formula gives only  order 1 entanglement for $R \rho \gg 1$, i.e. a ``boundary law''.  On the other hand, a normalizable distribution with ${1<\alpha <2}$ gives $S \propto R^{2-\alpha}$.
Precisely at $\alpha=2$, the entanglement entropy scales logarithmically with the region size.

\begin{figure}[t]
\centering
\includegraphics[width=0.48 \textwidth]{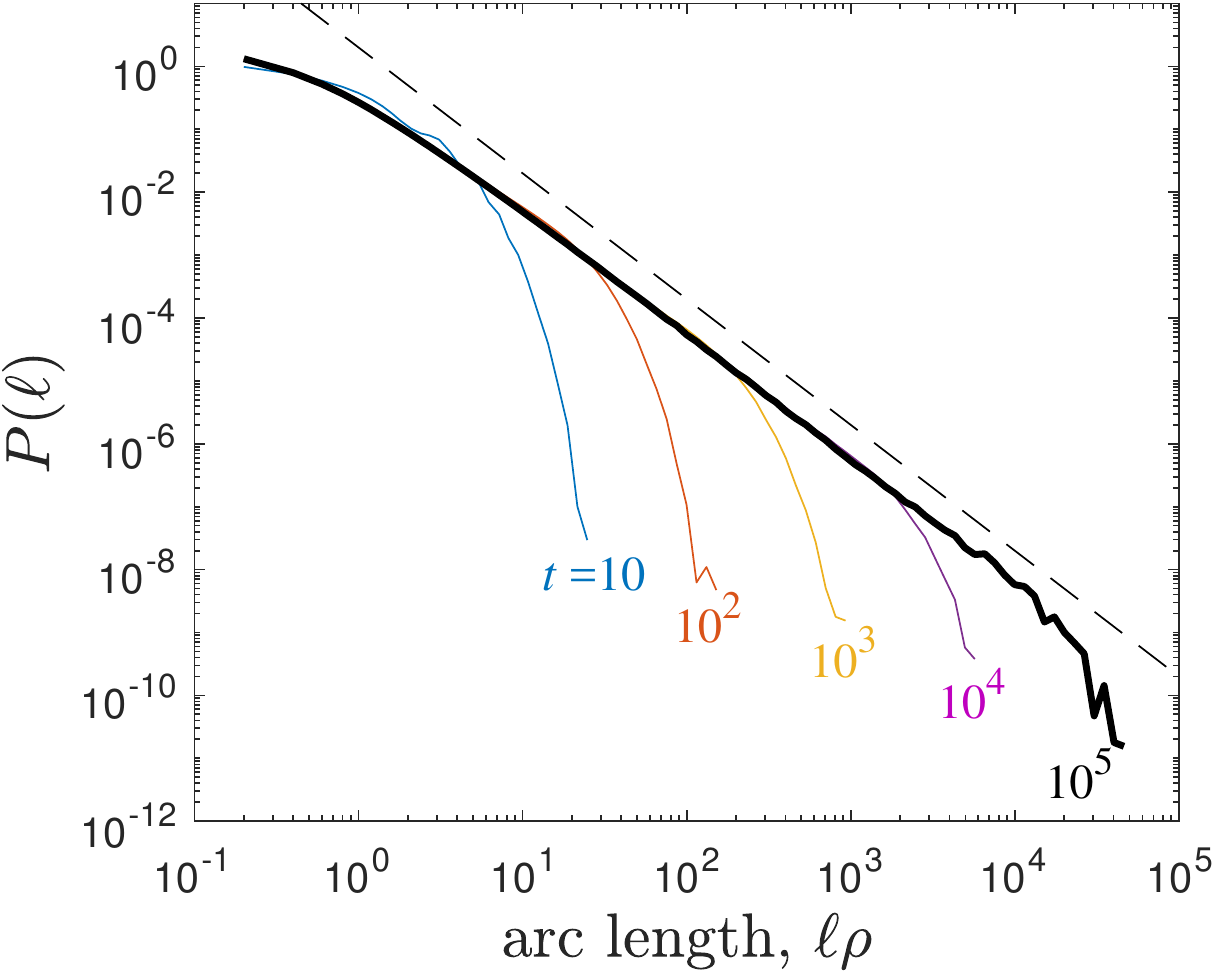}
\caption{Plot of the arc length distribution, $P(\ell)$, as measured by a simulation of $10^6$ Majoranas diffusing on a lattice of $5\times 10^6$ lattice sites with no annihilation. The arc length is plotted in units of the mean interparticle separation $\rho^{-1}$. Different curves are labeled by the value of the simulation time, and the dashed line shows $P \propto 1/\ell^2$.}
\label{fig:Pell}
\end{figure}

\begin{figure}[t]
\centering
\includegraphics[width=0.45 \textwidth]{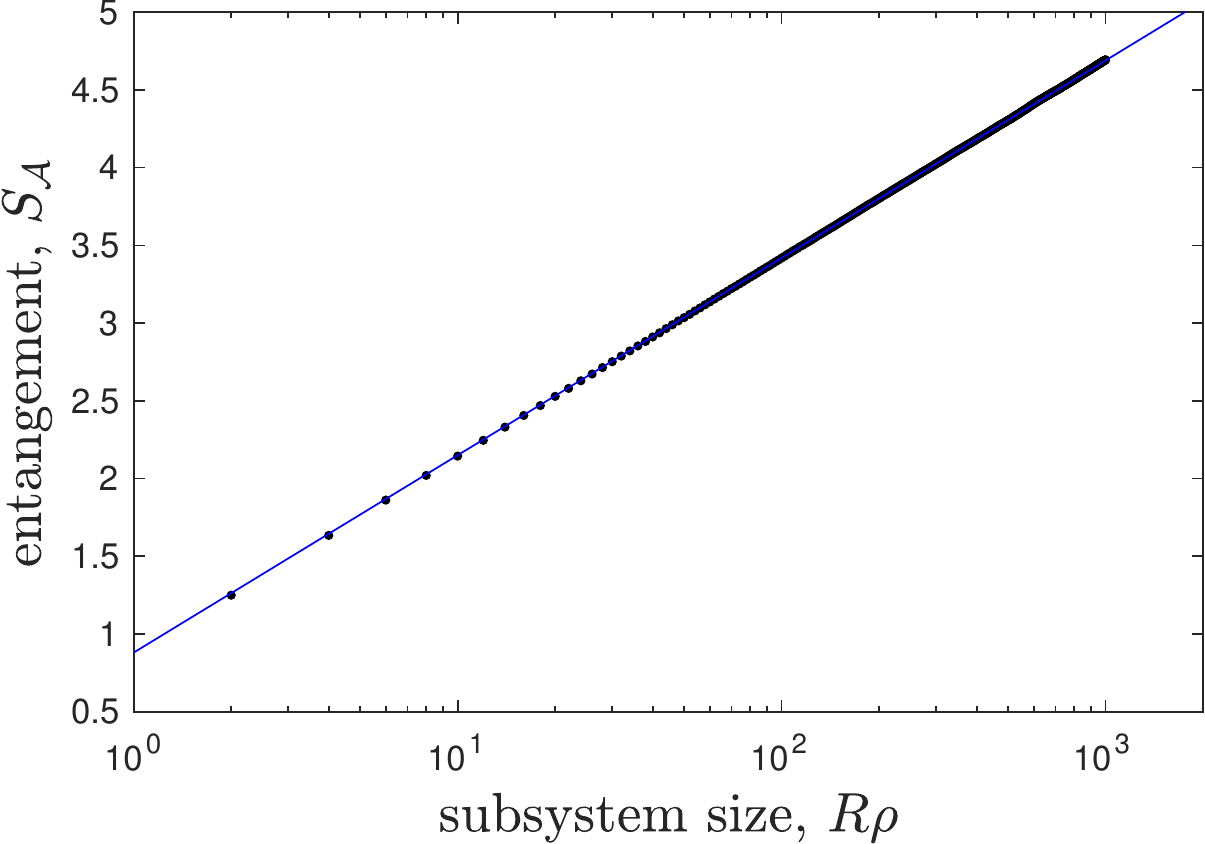}
\caption{The entanglement $S_\mathcal{A}$ of a region $\mathcal{A}$ of size $R$, plotted as a function of $R \rho$, the average number of Majoranas inside the region. The black points show the results from a simulation with $N = 3\times10^4$ Majoranas. The region $R$ is taken from the center of the system, and the entanglement is averaged over $2000 N$ time steps after equilibrating for an initial time $10 N$. The blue line shows the theory result of Eq.\ (\ref{eq:Slogwithcoeff}) with a fitted constant offset.}
\label{fig:entanglementmeasurement}
\end{figure}

We find numerically that the pairing dynamics is critical: $\alpha = 2$.
(Simulations in this subsection correspond to random-walking Majoranas; we will show below that the universal constants characterizing the entanglement are the same in the circuit model.)
Figure \ref{fig:Pell} suggests that
\ba
P(\ell) & = \f{K}{(\ell \rho)^2},
& 
K  & = 0.54 \pm 0.01.
\label{eq:KPell}
\end{align}
Therefore the entanglement scales logarithmically with a coefficient $K$.
Figure \ref{fig:entanglementmeasurement} shows the results of a direct measurement of this entanglement, yielding the consistent estimate
\ba 
S & \simeq K \ln (R \rho),
& 
K  & = 0.553 \pm 0.003.
\label{eq:Slog}
\end{align}
This logarithmic scaling of entanglement is due to each ``scale'' contributing order 1 bits of entanglement (as at a conformally-invariant 1D quantum critical point with central charge $c$, where the entanglement of a finite region is $\f{c}{3} \ln R$ \cite{calabrese2009entanglement}).
That is, fixing some constant $b>1$, there are $O(1)$  arcs crossing a given point that have $\ell \in (b, b^2)$, and similarly with $\ell \in (b^2, b^3)$, and so on.
 
The coefficient $K$ in Eqs.~(\ref{eq:KPell}) and (\ref{eq:Slog}) is dimensionless and universal.
This universality can be seen by mapping the quantum circuit model of Sec.~\ref{sec:circuitmodelintro}, in a worldline representation, to a well-known classical loop model, or equivalently to a stochastic process based on the Temperley Lieb algebra \cite{pearce2002temperley, deGierRaiseandPeel1,deGierRaiseandPeel2}. We explain this mapping in detail in Sec.~\ref{sec:loopmodel2d}. One can then argue that the universal properties survive when the effects of diffusion are included (Appendix~\ref{app:randomlattice}).
An exact result for the loop model in Ref.~\cite{cardy2000linking} (see also \cite{jacobsen2008exact,alcaraz2010shared}) equates to
\be
S_\mathcal{A} = \f{\sqrt{3}}{\pi} \ln (R \rho),
\label{eq:Slogwithcoeff}
\ee 
with $\sqrt{3}/\pi = 0.551...$, in close agreement with our numerical result. Recall that we are measuring entanglement in units of half-bits, so that $S_\mathcal{A}$ is precisely the mean number of arcs leaving $S_\mathcal{A}$. Equation (\ref{eq:Slogwithcoeff}) is for a region with two endpoints; for a finite chain of size $L$, split into two equal halves that meet at a single point, the entanglement is 
$1/2$ times the value in Eq.\ (\ref{eq:Slogwithcoeff}), with $R = L$.

The correspondences mentioned above also determine the dynamical exponent $z$ for the evolution of the entanglement:
\be
z=1.
\ee
(This dynamical exponent is also explained via a simple heuristic picture for the dynamics in the beginning of the following subsection.)
In Fig.~\ref{fig:Pell} we see the gradual approach of $P(\ell)$ to the stationary ensemble with time, with longer arcs taking longer to be generated. At a given time $t$ there is a length scale $\xi(t)$ beyond which $P(\ell)$ decays exponentially.
This length scale grows roughly linearly with time, consistent with the $z=1$ scaling $\xi(t) \sim t^{1/z} = t$. 

\begin{figure}[t]
\centering
\includegraphics[width=0.45 \textwidth]{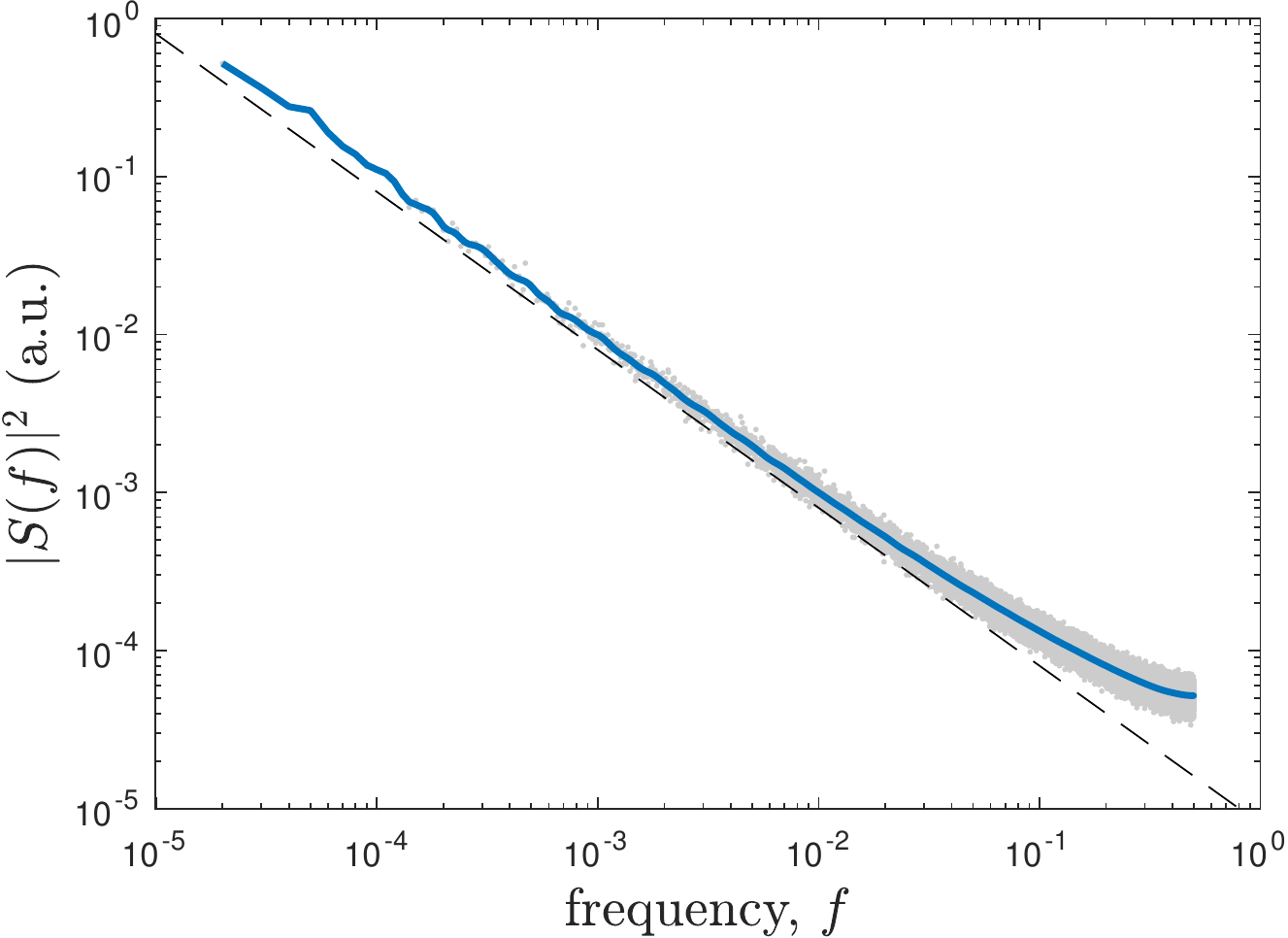}
\caption{$1/f$ noise in the temporal evolution of the entanglement.  Here the half-system entanglement $S(t)$ is measured as a function of time for a system of $10^5$ Majoranas. The system is first equilibrated for $10^5$ time steps, and then $S(t)$ is measured over another $10^5$ time steps. $S(t)$ is then Fourier-transformed and the resulting power spectrum $|S(f)|^2$ is averaged over $100$ realizations for each discrete frequency $f$ (measured in inverse time steps).  The gray points represent this averaged data, and the thick blue line is the same data smoothed by convolving with a Gaussian having a standard deviation one tenth of a decade. The dashed line shows a $1/f$ dependence.}
\label{fig:1fnoise}
\end{figure}

The structure of one arc per length scale, along with the dynamical exponent $z = 1$, also implies that the temporal fluctuations of the entanglement, $S(t)$,  produce ``$1/f$ noise'', such that $\< |S(f)|^2 \> \propto 1/f$,
where $f$ is the frequency. Equivalently, in the time  domain (and in the limit of infinite region size)
\be 
\langle [ S(t) - S(t') ]^2 \rangle \propto \log |t - t'|.
\label{eq:logtcorrelations}
\ee
This $1/f$-noise is demonstrated using simulation data in Fig.~\ref{fig:1fnoise}. 
It can be understood by noting that the entanglement of some region with large size $R$ has statistically equal contributions from each (logarithmically spaced) length scale up to  $R$. Since the number of arcs of size $\ell$ changes on a timescale of order $\ell$, this implies an equal contribution to the fluctuations from each time scale, which is equivalent to $1/f$ noise.\footnote{The equal contribution from each temporal scale, i.e. from each frequency interval ${2^k\leq f \leq 2^{k+1}}$, means ${\int_{2^k}^{2^{k+1}} \dd f \< |S(f)|^2 \>}$ is independent of $k$. This gives the $1/f$ scaling.}

In its logarithmic entanglement and two-body pairing structure, the  critically entangled state found above resembles random-singlet wavefunctions. These arise as ground states of 1D chains with quenched disorder \cite{Lee1981,Lee1982,Fisher1994,fisher1999phase,Igloi2003,refael2004entanglement,bonesteel2007infinite,shu2016properties}. In the context of disordered spin-1/2 antiferromagnets, arc diagrams like Fig.~\ref{fig:typicalarc} are used to represent  frozen long-range spin singlets in the ground state. 
Similar states arise as ground states \cite{bonesteel2007infinite} (and even highly excited states \cite{pekker2014hilbert, vasseur2015quantum}) of the Kitaev chain [Eq.~(\ref{eq:kitaevchain})] with quenched disorder, with arcs representing pairs of Majoranas of definite fermion parity.

Interestingly, however, the universal properties of these random singlet states are \textit{distinct} from those of the state obtained dynamically here. 
This difference can be seen from the universal coefficient $K$ in the entanglement entropy. In the random singlet states (both for spin-1/2 fermions and for Majorana fermions) the number of arcs leaving a region of size $R$ in an infinite chain is \cite{refael2004entanglement}
$\f{1}{3} \ln R$, in contrast to Eq.~(\ref{eq:Slogwithcoeff}).
Crudely speaking, this difference in coefficients means that the random singlet ground state of a Majorana chain is less entangled than the state obtained from the repeated pairwise measurement process. The scaling properties of this  dynamical state are related to a two-dimensional nonunitary conformal field theory: this structure is presumably quite different from that governing the random singlet phase.

\subsection{Note on translation symmetry}
\label{sec:notetranslation}

Let us comment on the role of translation symmetry in these ensembles of wavefunctions.
It is  interesting to note that the value of the exponent $\alpha=2$ exhibited by both of them --- both the random states obtained by measurement, and those obtained as ground states of random Majorana Hamiltonians ---
is the largest that is possible under the assumption that the ensemble does not spontaneously break ``translation symmetry'' in the  index $i$.\footnote{An example of an ensemble of pairing diagrams for which this symmetry \textit{is} spontaneously broken (in the sense defined below) is the ensemble supported on the two nearest-neighbor dimerized configurations, with equal probability.} 
This symmetry is a property of our dynamics with diffusion and measurement,\footnote{In our diffusive model the symmetry is a relabelling of the Majoranas, ${i\rightarrow i+1}$, rather than a physical translation. Understanding the constraints imposed by symmetries on the circuit model would require a separate analysis, since, as we have set it up, ${i\rightarrow i+1}$ is only a symmetry when combined with translation by half a time period. (This complication is avoided in a version of the circuit where measurements are applied at a fixed rate in continuous time.)}
and it is also a statistical symmetry for the random Majorana Hamiltonians (i.e., a symmetry of the ensemble of Hamiltonians) and similar random spin-1/2 models.

The basic point is that in order to avoid breaking translational symmetry, the pairing diagram must have many large pairs and an average entanglement that diverges with the system size. This statement follows from a much more general mathematical theorem in Ref.~\cite{aizenman2001bounded}, 
and for random singlet states it follows from a more general result for  random spin-1/2 chains \cite{kimchi2018valence}. But the result for pairing diagrams is a simple special case and follows directly from a basic property of such diagrams \cite{bonesteel1989valence,thouless1987fluxoid}, as we now show.

If all pairs are short-ranged, the parity $\mathcal{P}_i$ of the number of arcs ``passing over'' a given link $(i, i+1)$ between Majoranas $i$ and $i+1$ \textit{alternates} as a function of $i$ for any state in the ensemble \cite{bonesteel1989valence}. (This alternation can easily be seen by drawing a diagram.) Therefore --- again assuming all pairs are short-ranged --- this parity acts as a local ``dimerization'' order parameter, implying a spontaneous breaking of translation symmetry. 

The key question is then about the locality of this order parameter $\mathcal{P}_i$. 
We define spontaneous dimerization to be present in the pairing diagram if there exists a quantity $O_i$ (the translated version of $O_0$) which depends only on pairing information within a \textit{finite} widow around $i$, 
and whose two-point function $\overline{O_i O_j}$ has nondecaying period-2 oscillations at arbitrarily large $|j-i|$. (Note that we do not demand that $\overline{O_i O_j}$ be expressible as a correlation function of quantum operators of the form $\overline{\bra{\psi}\mathcal{O}^\text{qm}_i\mathcal{O}^\text{qm}_j\ket{\psi}}$ in the Majorana chain, and in general it will not be.)

$\mathcal{P}_i$ does not automatically satisfy the above definition, since it may include contributions from arbitrarily large arcs. But it is straightforward to argue that, if the mean number of arcs crossing a given link in the infinite chain is \textit{finite} --- i.e. if the mean entanglement is finite --- then contributions from large arcs are rare, and $\mathcal{P}_i$ may be approximated arbitrarily well by a quantity $O_i$ that is local in the above sense.

In other words, statistical translational symmetry guarantees long-range entanglement in both the random singlet phase and in the stochastic measurement process studied here. Translation symmetry can also be used to constrain the entanglement for various other examples of measurement dynamics (Sec.~\ref{sec:discussion}).

\begin{figure}[t]
\centering
\includegraphics[width=0.5 \textwidth]{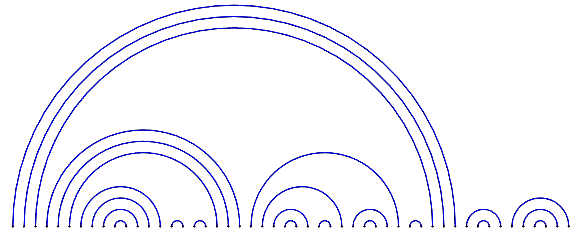}
\caption{For contrast with the pairing configurations in the Majorana model (Fig.~\ref{fig:typicalarc}), we show a configuration from the \textit{equally-weighted} ensemble of all noncrossing configurations. This ensemble has arc length distribution $P(\ell) \propto 1/\ell^{3/2}$, and so is statistically very different from the configurations generated in the Majorana model.}
\label{fig:equallyweightedarcs}
\end{figure}

For completeness, let us note that the critically-entangled ensembles mentioned above are qualitatively different from a simpler ensemble in which we choose a pairing diagram at random, with uniform probability, from all possible non-crossing pairing diagrams.\footnote{
The reason that we do not obtain all pairings with equal probability in our dynamics is that the pairing dynamics does not obey detailed balance for this distribution: the rules for merging arcs are irreversible.}
This ensemble has the arc length distribution $P(\ell) \propto 1/\ell^{3/2}$ (reviewed in Appendix \ref{sec:rwapp}).  
A typical arc diagram from this uniform distribution is presented in Fig.~\ref{fig:equallyweightedarcs}. Note that this state looks qualitatively different from the result of our stochastic dynamics (Fig.~\ref{fig:typicalarc}), and has many tightly-nested large arcs.

\subsection{Analytical treatment}
\label{sec:loopmodel2d}

The entanglement structure, encoded in the pairing diagram, undergoes a stochastic evolution that is nontrivial because of its nonlocality: when defects $i$ and $i+1$ come into contact and become paired, their  previous partners $j$ and $k$ also simultaneously become paired, irrespective of their spatial separation. 

This nonlocality means that large arcs can grow much faster than they would if their endpoints simply diffused.
For example, if an $L$-sized system of $O(L)$ defects is initialized with only short-range pairs, then arcs of order $L$ size are generated in a time of order $L^z=L$. This value $z=1$ for the dynamical exponent can be understood using a known mapping from pairing diagrams \cite{TemperleyLieb,blote1989critical} and their stochastic updates \cite{pearce2002temperley,deGierRaiseandPeel1,deGierRaiseandPeel2} to \textit{isotropic} 2D statistical mechanics models \cite{cardy2005sle,jacobsen2009conformal}. We will review this mapping below. First, however, we give an intuitive argument for the exponents governing the entanglement dynamics.

Assume to begin with that the stationary arc length distribution $P(\ell) \propto 1/\ell^\alpha$ at $\ell \rho \gg 1$, where $\alpha$ is some exponent to be determined.  Now imagine labeling a large arc in the system, and following the evolution of its length. We adopt the rule that after an update to the pairing, the label attaches to the larger of the two arcs produced. 
The endpoints of the labelled arc move both by diffusion and by coalescence of the labelled arc with other arcs, whose lengths are distributed according to $P(\ell)$. Through this latter process the length of the labeled arc can take ``steps"  much longer than the mean interparticle spacing. So long as $\alpha < 3$, the dynamics of the arc growth are dominated by rare long steps, and are therefore more like a Levy flight than a random walk.  Considering the longest expected step in a time interval $t$ gives a typical displacement $x
\sim t^{1/(\alpha-1)}$.
This sets the relationship between length and time scales: $x \sim t^{1/z}$ with dynamical exponent
\be
z = \alpha - 1.
\ee

One can now fix $\alpha = 2$ by arguing that either $\alpha < 2$ or $\alpha > 2$ leads to a contradiction. Briefly, either inequality leads to a dynamic instability that can be seen by considering the number of  arcs with length $\ell \in (L/2, L)$ in a system of large size $L$.  When $\alpha < 2$, there are many such ``$L$-sized arcs'', which results in a close nesting of $L$-sized arcs (like that in Fig.~\ref{fig:equallyweightedarcs}, for example).  Such arcs quickly annihilate each other by encounters of their endpoints, on a time scale much faster than they can be generated.\footnote{
When the endpoint of a large arc encounters the endpoint of another similar-sized arc enclosing it, both arcs are annihilated and replaced with short arcs that connect their previous endpoints.
}
On the other hand, when $\alpha > 2$, the expected number of $L$-sized arcs is much smaller than unity.  In this case there is nothing preventing the longest arc in the system from growing to become comparable to the system size, so that again the system is unstable dynamically.

In short, the distribution $P(\ell) \propto 1/\ell^2$ is the critical case with an order-unity number of arcs of length of order $\ell$ in any subsystem of length $\ell \gg \rho^{-1}$, as mentioned above.  Such a distribution provides enough long arcs to keep the upward growth of short arcs in check (via  the nested annihilation process), but not so many that long arcs are forced to nest tightly inside each other, which would lead to them rapidly annihilating.  This structure with ``one arc for every scale of length'' implies the logarithmic entanglement scaling in Eq.~(\ref{eq:Slog}), and the resulting Levy flight of arc endpoints guarantees $z = 1$.

A more precise picture can be obtained by applying results from 2D ``loop models'' and related stochastic processes. To make this mapping, we simplify the diffusion-plus-measurement process to the quantum measurement circuit model shown in Fig.~\ref{fig:circuit}, where Majorana modes $\gamma_i$ live at fixed positions $i\in \mathbb{Z}$, and adjacent pairs $(\gamma_i, \gamma_{i+1})$ are measured with probability $p$ in either even or odd time steps, depending on whether $i$ is even or odd.
At the end of this section we argue  that the universal results also apply in the model with diffusion.

If we neglect the fermion parity numbers, the ``updates'' to the pairing diagram are precisely those in the Temperley-Lieb transfer matrix representation of a  two-dimensional lattice model for nonintersecting loops  related to percolation \cite{blote1989critical,TemperleyLieb}. 
In the loop model, pairing diagrams, representing loop connectivities, can be used as a basis for the transfer matrix.
It was pointed out in Refs.~\cite{pearce2002temperley,deGierRaiseandPeel1,deGierRaiseandPeel2} that this transfer matrix can be interpreted as the transition matrix of the stochastic process on pairing diagrams.\footnote{By a correspondence between pairing diagrams and height configurations (App.~\ref{sec:rwapp}), this stochastic process can be viewed as a model for surface growth with a nonlocal update \cite{deGierRaiseandPeel1,deGierRaiseandPeel2}.}
(The transfer matrix of a lattice model cannot, in general, be re-interpreted as a stochastic process, but in geometrical models like percolation, with \textit{uncorrelated} local degrees of freedom, it can be.\footnote{This fact makes possible very efficient numerical algorithms, which are based on updating connectivity information as the configuration is grown stochastically ``slice by slice'' \cite{hoshen1976percolation,deng2005monte,nahum2013loop}.})

\begin{figure}[b]
\centering
\includegraphics[width=0.25 \textwidth]{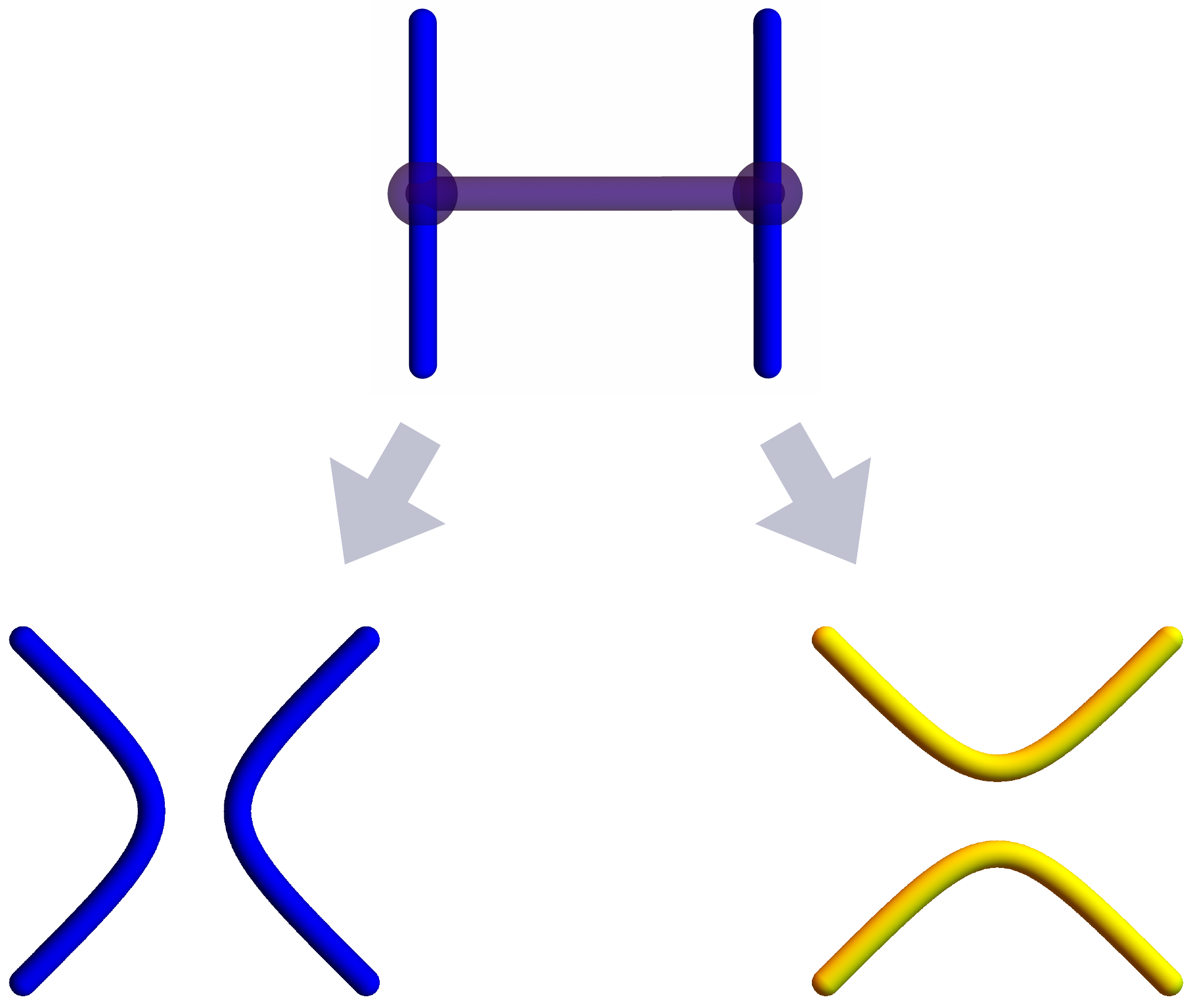}
\caption{
Mapping the measurement circuit in Fig.~\ref{fig:circuit} to a loop model (Fig.~\ref{fig:loopsandarcs}). The two possibilities, absence/presence of a measurement, map to two ways of connecting up the worldlines.
}
\label{fig:circuitelement}
\end{figure}

\begin{figure}[t]
\centering
\includegraphics[width=0.4 \textwidth]{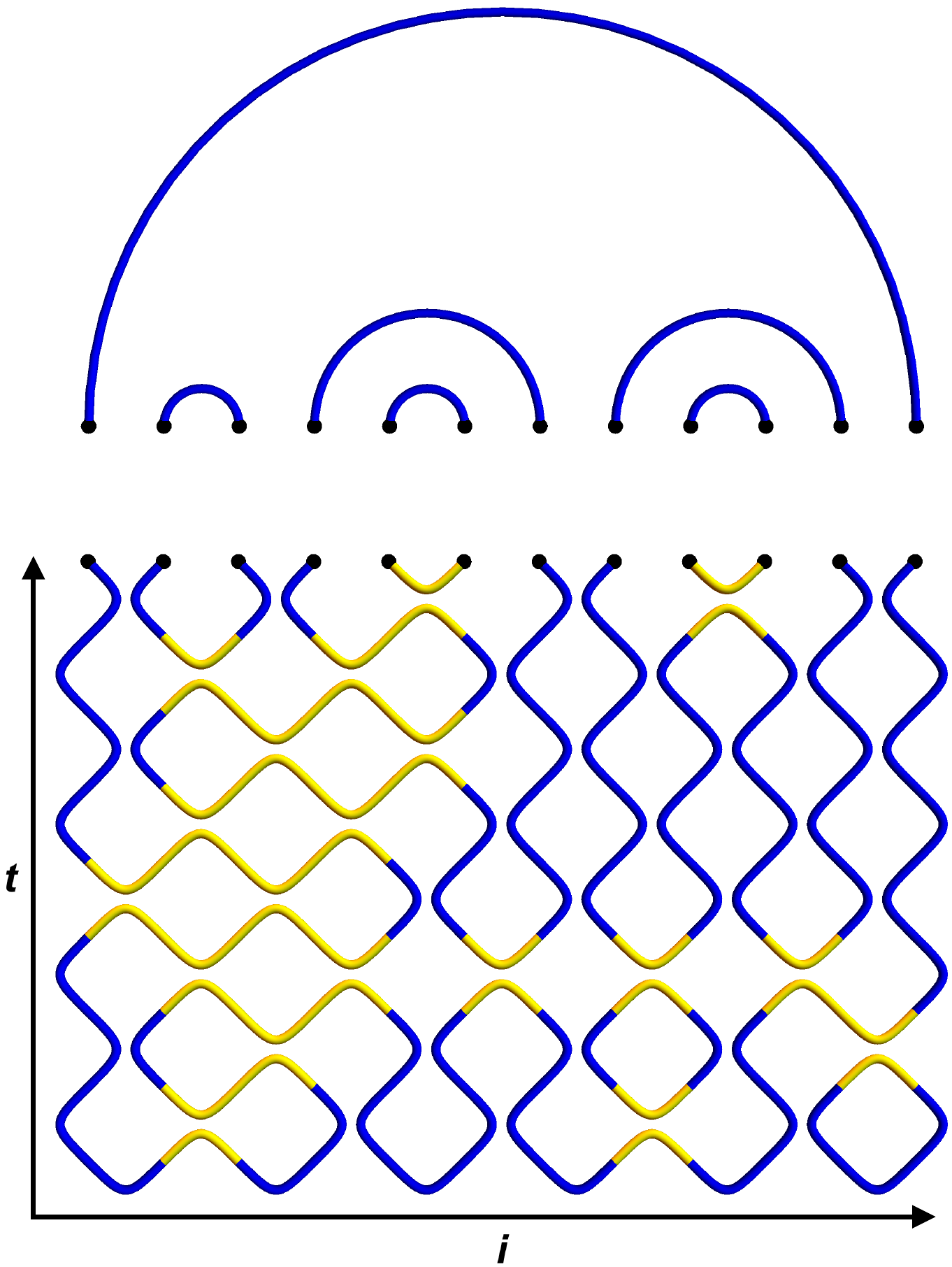}
\caption{
A history of the measurement process in the circuit maps to a loop configuration in spacetime via the protocol summarized in Figs.\ \ref{fig:circuit} and \ref{fig:circuitelement}. 
The lower part of the figure shows an example for a particular sequence of measurement locations. 
This sequence generates the pairing configuration shown in the top part of the figure.
Sites are paired if the corresponding nodes at the top boundary of the loop configuration are connected.
}
\label{fig:loopsandarcs}
\end{figure}

The relation to 2D loop configurations can be seen directly as follows.
In the circuit diagram of Fig.~\ref{fig:circuit}, each block
is either a measurement or a non-event, with probability $p$ for the former.
We represent each of these outcomes by replacing each horizontal bar with one of the two diagrams in Fig.~\ref{fig:circuitelement}:
measurements become the  reconnections shown in yellow in the loop diagram, and nonevents correspond to continuation of the vertical lines (blue).
Making this replacement for every horizontal bar, we produce a configuration of loops, plus open strings, like that in Fig.~\ref{fig:loopsandarcs}. At the bottom boundary of this diagram, representing the initial time, the strands are connected in a  pattern determined by the initial pairing of the Majoranas, which in the figure we have taken to be a dimerized configuration. The open strings in the configuration terminate on the top boundary, where they  connect the sites $i$ pairwise.
Therefore, the loop configuration specifies a pairing configuration.
It is easy to check that this pairing configuration is precisely the one resulting from the measurement dynamics. 
As noted above, the relationship between the 2D loop configurations and pairing diagrams is well-known  \cite{blote1989critical,TemperleyLieb}, including the formal relationship with a stochastic process \cite{pearce2002temperley, deGierRaiseandPeel1, deGierRaiseandPeel2}: here we provide a physical realization of this stochastic process in terms of measurement in a Majorana circuit.

When the measurement probability is $p=1/2$, the 2D loop configurations generated above are drawn from a simple ensemble
where every allowed configuration of loops is equally likely. This ensemble is scale-invariant. The loops have the same fractal structure as cluster boundaries in critical percolation \cite{saleur1987exact}, as can be seen from an explicit lattice construction (see  Refs.~\cite{cardy2005sle,jacobsen2009conformal} for reviews).
The ensemble is isotropic in 2D, and this isotropy is preserved in the scaling limit even if the measurement probability is different from 1/2, so long as we define the unit of time appropriately. Therefore the dynamical exponent for the pairing dynamics is $z = 1$ \cite{pearce2002temperley}. 
The entanglement entropy maps to the number of strands that connect section $\mathcal{A}$ of the top boundary to other parts of the top boundary.
This number has been computed exactly in Ref.~\cite{cardy2000linking}, 
giving the universal coefficient quoted in Eq.~(\ref{eq:Slogwithcoeff}).
(This quantity has also been considered in other settings in Ref.~\cite{jacobsen2008exact} and Ref.~\cite{alcaraz2010shared}, where a heuristic similarity with entanglement was noted.)
The length distribution $P(\ell)$ is (twice) the probability that two boundary sites at a distance $\ell$ are connected, which is a boundary two-point function in the conformal field theory language that decays as  $1/\ell^2$, consistent with the discussion around Eqs.~(\ref{eq:S}) and (\ref{eq:KPell}).

Above, the probability of measurement was the same on the even and odd bonds of the Majorana chain. As an aside, we note that if the probability is ``dimerized'' so that measurements are more frequent on (say) odd bonds, this drives the loop model into a non-critical phase with short loops. The corresponding correlation length exponent is that of percolation in two dimensions, ${\nu = 4/3}$. The fact that the model with equal measurement rates is critical is related to translation symmetry in the Majorana index, as discussed in the previous section. 
By contrast, the 2+1D models we introduce in Sec.~\ref{sec:loopmodel2d} --- and a variant 1+1D circuit discussed at the end of that section, which involves ``swap'' operations on Majoranas and maps to loops with crossings --- can be long-range entangled even in the absence of such a symmetry.

Finally, to complete the analysis, we must confirm that reintroducing the diffusive motion of the Majoranas (but \textit{not} the annihilation process) does not change the universal properties obtained for the circuit.
This follows from a simple renormalization group argument and is confirmed by our numerics.

Specifically, in the case with diffusion we have a loop ensemble similar to the one shown above, but rather than being defined on a regular lattice, its structure is determined by the encounters of the diffusing particles.
In Appendix~\ref{app:randomlattice} we argue that this difference introduces only renormalization-group-irrelevant perturbations of the critical theory described above for the circuit.
Therefore we do not expect the coupling to the diffusive density fluctuations to affect the leading scaling, though it will certainly contribute to subleading corrections.

\section{Diffusion--annihilation of Majorana defects: the ${\gamma + \gamma \rightarrow \emptyset}$ process}
\label{sec:with}

\subsection{Decay of the particle density}

We now address the dynamics of the ${\gamma + \gamma \rightarrow \emptyset}$ diffusion-annihilation process.
In this model, Majorana defects that come into contact and have zero parity are removed from the system (Sec.~\ref{sec:diffusingdefects}).  In the language of the microscopic quantum Hamiltonian discussed in Sec.\ \ref{sec:model}, this annihilation locally heals the ground state by removing two opposite domain walls. Thus, the model with annihilation describes a type of dissipative dynamics, in which the system is initiated in a high-energy state with many domains, and gradually relaxes to a ground state with no domain walls.

As mentioned above, this dynamics differs from a standard classical model for relaxation of the Ising model 
 \cite{glauber1963time,bramson1980clustering,
torney1983diffusion,
lushnikov1987binary,
Bray_1990,
Tauber_2005,
derrida1994non,
derrida1996exact}
 by the fact that Majoranas with fermion number $n = 1$ are prevented from annihilating: in such cases the domain walls cannot heal upon contact (see Fig.~\ref{fig:coarsening}).

Let us first review the standard classical Ising problem, which is equivalent to the ``$A+A\rightarrow \emptyset$'' reaction--diffusion process. In this problem, A single species of particles undergoes diffusion on the real line, and when two particles come into contact they annihilate with some probability $\theta$.
For this process one can show that, regardless of the value of $\theta$, and of the initial value of the particle density, the density $\rho = N_t/L$ at sufficiently long times $t$ follows
\be 
\rho_\text{cl} = \frac{1}{\sqrt{8 \pi}} \frac{1}{\sqrt{D t}},
\label{eq:rhocl}
\ee 
where $D$ is the diffusion constant. 
This exact result, in which $1/\sqrt{8\pi}$ is a \textit{universal} amplitude, has been obtained in various ways \cite{bramson1980clustering,torney1983diffusion,lushnikov1987binary}: one approach is to map the Markov transition matrix to an integrable non-Hermitian quantum spin chain \cite{lushnikov1987binary}.
However, while the value of the universal amplitude requires a more sophisticated analysis, the scaling $\rho \propto 1/\sqrt{Dt}$ can be understood simply as follows.

\begin{figure}[t]
\centering
\includegraphics[width=0.5 \textwidth]{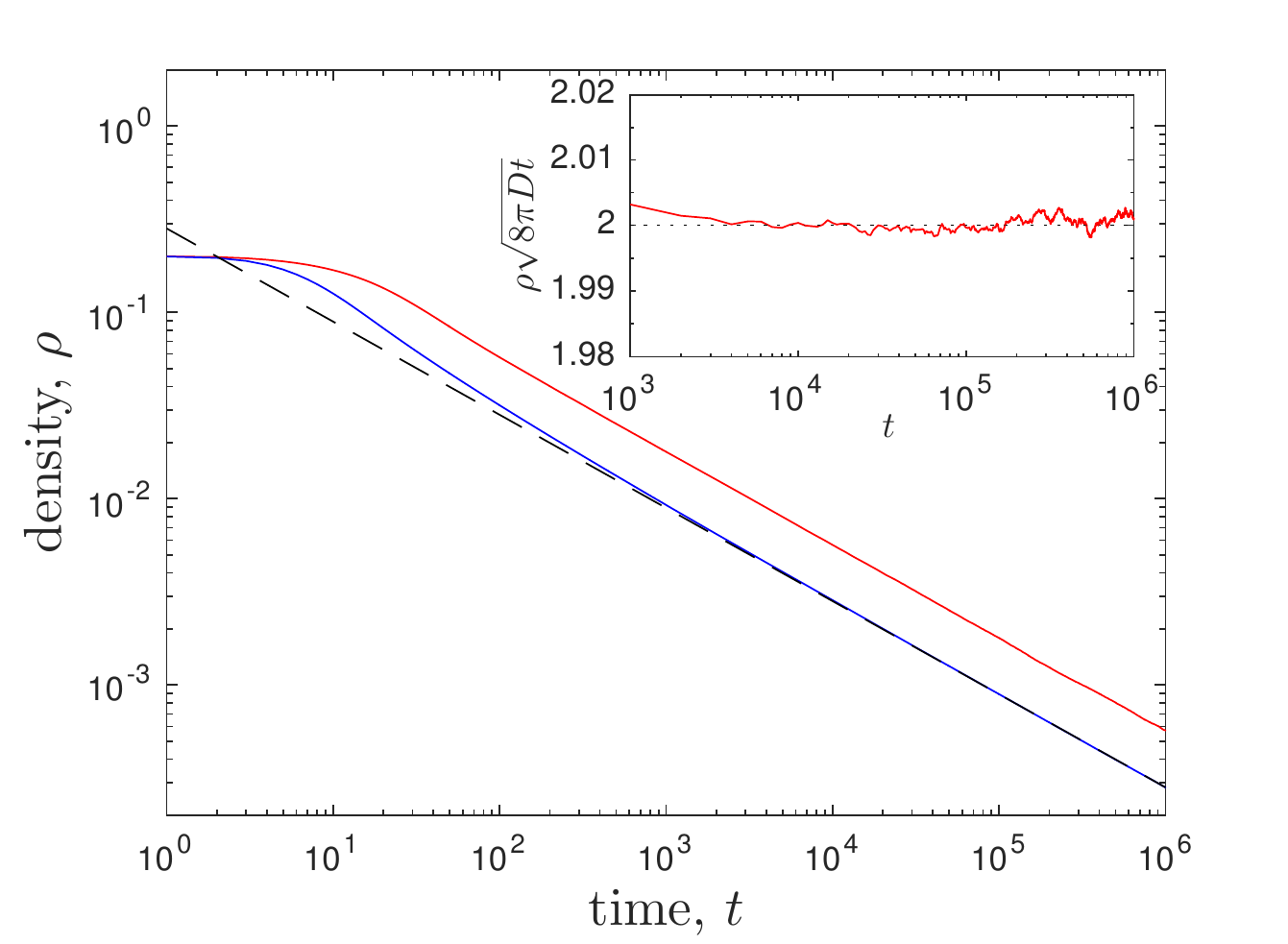}
\caption{Particle density as a function of time for the diffusion-annihilation process.  The upper (red) curve shows the result for the Majorana model, and the lower (blue) curve shows the result for the classical ``$A + A \rightarrow \emptyset$'' process.  The dashed black line shows $\rho = 1/\sqrt{8 \pi D t}$.  The inset shows the quantity $\rho \sqrt{8 \pi D t}$, which converges to $2$ at long times. Data corresponds to a simulation of $10^6$ lattice sites initialized at time $t = 0$ with a uniform, dimerized configuration of Majoranas with density $\rho(0) = 1/5$, and is averaged over $1000$ random realizations.}
\label{fig:rhot}
\end{figure}

When the density of particles $\rho \ll 1$, the typical spacing between nearest neighbors is $1/\rho \gg 1$.  Thus, the typical time required for one particle to diffuse into its nearest neighbor is $\Delta t \sim 1/(\rho^2 D)$.  In the limit where particles are very sparse, two particles that come into contact have many opportunities to collide and annihilate before one of them can diffuse away and encounter a third party. (In the absence of annihilation, two diffusing particles which collide go on to  collide $\sim 1/\rho$ times before either of them can diffuse a distance $\sim 1/\rho$.)  Thus, at small $\rho$, two particles that collide will almost certainly annihilate before either can escape the other, regardless of the value of $\theta$. 
 This means that the value of $\theta$ is irrelevant at late times (one can say that it is renormalized to unity).
 
Therefore an order 1 fraction of the particles are annihilated in the characteristic diffusive time ${\Delta t\sim 1/(\rho^2 D)\gg 1}$.  Equivalently, the change $\Delta \rho$ in this time is of the order of $\rho$ itself.
% $\Delta \rho \sim - \rho$. 
Relating $(\Delta \rho)/(\Delta t)$ to the derivative $d\rho/dt$ gives 
$d \rho/dt \sim -D \rho^3$, whose solution is $\rho \sim 1/\sqrt{D t}$.
(Note that this relation for $d\rho/dt$ should be distinguished from the mean field rate equation, which would give a different, incorrect, scaling; see e.g. Ref.~\onlinecite{Tauber_2005}.)

In our problem of diffusing Majoranas,  the power law $\rho \propto 1/\sqrt{D t}$ still holds, since the same time scale $\Delta t$ is associated with diffusion over the nearest-neighbor distance, and again an order one fraction of particles annihilate in this time window.
However, Eq.~(\ref{eq:rhocl}) no longer holds: the fraction of particles that annihilates in the time window is smaller, and the universal coefficient in Eq.~(\ref{eq:rhocl}) increases.

Unlike in the classical case, where two colliding particles will almost certainly annihilate if $t$ is large, two colliding Majoranas that happen to be projected into a state of parity $n=1$ remember this parity when the two particles collide again: no subsequent collision between them can produce annihilation, without first involving a third party.
One should therefore expect the particle density at long times to be larger than Eq.\ (\ref{eq:rhocl}).

Figure \ref{fig:rhot} shows our numerical result for $\rho(t)$ in the Majorana model with annihilation. For comparison we also plot the density from a numerical simulation of the classical model with $\theta = 1/2$.
The latter model would apply if, instead of remembering the fermion parities, we were to re-choose them at random for a pair each time they collided. The classical model has a density $\rho_\text{cl}(t)$ given at long times by Eq.~(\ref{eq:rhocl}).

The inset of Fig.~\ref{fig:rhot} shows that the Majorana model produces a density which at long times is precisely \textit{twice} that of the classical reaction-diffusion problem:
\be \label{eq:quantumdensity}
\rho(t) = \frac{1}{\sqrt{2 \pi}} \f{1}{\sqrt{D t}}.
% = 2 \rho_\text{cl}.
\ee
{We explain this surprising relation in the next subsection.}

It is worth emphasizing that this factor of  $2$ difference from the classical model cannot be understood simply as a probability for annihilation of two defects that is reduced by (say) a factor of 2 on average by the parity constraint.
Indeed, as noted above, the microscopic annihilation probability $\theta$ has no effect on the density at late times in the classical model.

\subsection{Mapping to two copies of $A+A\rightarrow \emptyset$}
\label{sec:doubledisingmapping}

Remarkably, the universality class of the Majorana diffusion-annihilation process can be characterized exactly by a mapping to an auxiliary classical diffusion-annihilation process that does not involve a nonlocal pairing structure.
Instead, this classical model contains additional ``fictitious'' labels on the particles. These labels have no meaning in the original quantum system, but by averaging over them, we %magically 
reproduce the exact dynamics of the defect positions in the original quantum system.
This classical model reduces to \textit{two independent copies} of the $A+A \rightarrow \emptyset$ process at late times. 

\begin{figure}[t]
\centering
\includegraphics[width=0.5 \textwidth]{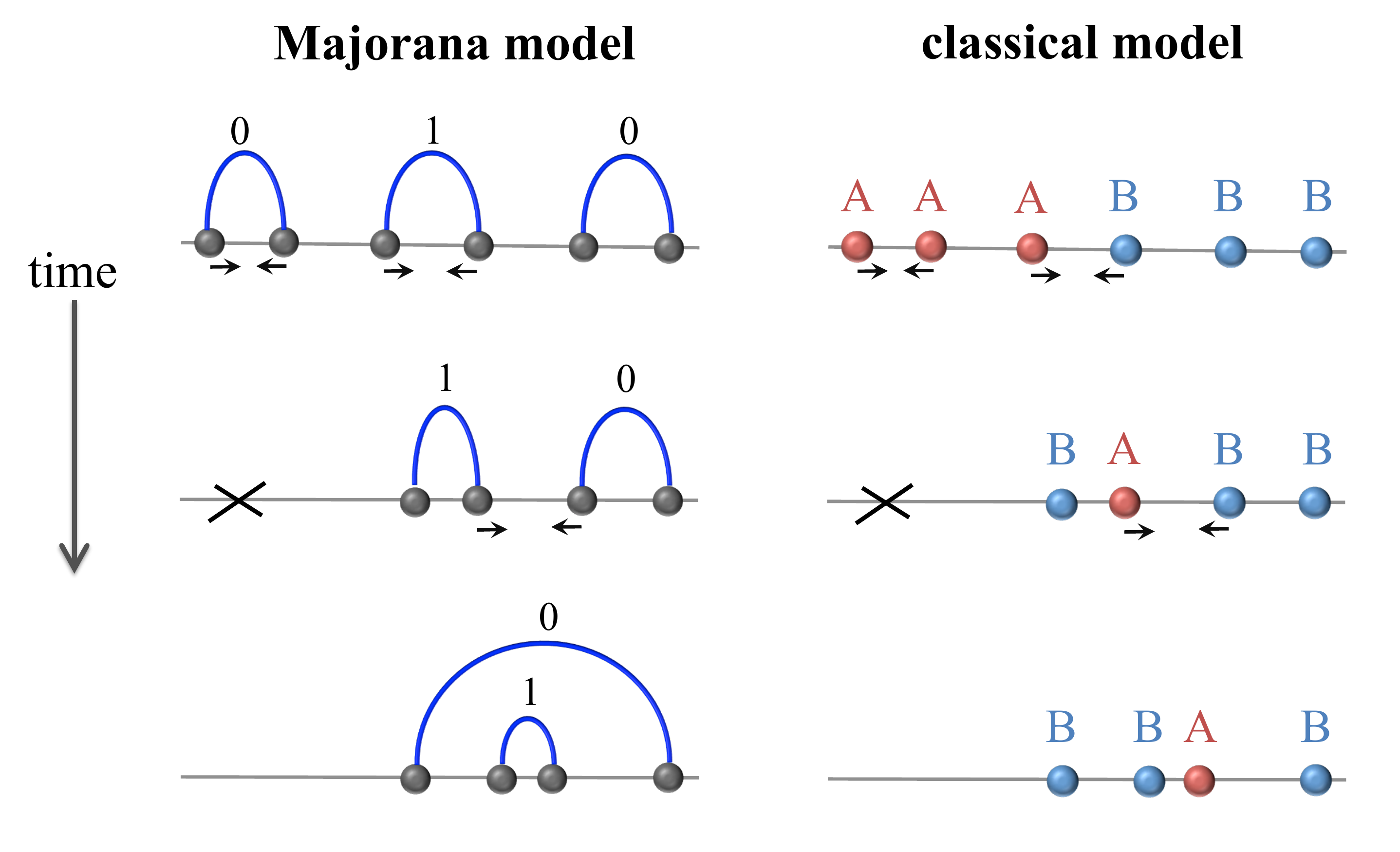}
\caption{Mapping between the Majorana and classical models. Left: a possible sequence of configurations of the former. Right: one of the corresponding configurations of the latter. A Majorana pair with $n = 0$ maps to two classical particles with the same label ($A$ or $B$), and a Majorana pair with $n = 1$ maps to two particles with opposite labels. Classical particles with the same label  annihilate on contact, while those with opposite labels do not annihilate and can pass through each other.}
\label{fig:classical_mapping}
\end{figure}

The classical model that we define involves two particle types, which we label $A$ and $B$, and the allowed annihilation processes ${A+A\rightarrow \emptyset}$ and ${B+B\rightarrow \emptyset}$. 
The classical model undergoes a stochastic process that is fully local (i.e., unlike the Majorana process, it does not require any pairing information). 
The mapping between the two models is simple: loosely speaking, each Majorana is replaced \textit{randomly} with either an $A$ or $B$ particle, but in a way that is constrained by the fermion parities in the pairing diagram. This is illustrated in  Fig.~\ref{fig:classical_mapping}.
If a pair of Majoranas has $n=0$, the two Majoranas are replaced by two classical particles with the same label, while a pair with  $n=1$  is replaced by two classical particles with opposite labels. 
We  define the mapping precisely below.

For concreteness, let us set up the Majorana dynamics on the 1D lattice as follows, in continuous time. In each infinitesimal time window $\dd t$, a given \textit{bond} of the lattice has a probability $\Gamma \dd t$ of receiving an ``update'', where $\Gamma$ is a rate.\footnote{This protocol differs in a trivial way from the numerical setup in Sec.~\ref{sec:simulation}: this is only to simplify the presentation in this section.} Updating a bond means the following takes place:

(i) If neither of the sites on the bond is occupied, nothing happens.

(ii) If one of the two sites on the bond is occupied, the occupying Majorana defect hops across the bond.

(iii) If the two sites are both occupied, and the corresponding Majoranas are in a ``1'' pair, nothing happens.

(iv) If the two sites are both occupied, and the Majoranas are in a ``0'' pair, they are annihilated.

(v) If the two sites are both occupied, but the Majoranas are not paired with each other, they are either paired into a ``1'' pair, or annihilated, with probability $1/2$ for each option; their previous partners are now paired with the appropriate fermion parity.

We now introduce a different stochastic model. We refer to it in this section as the ``classical'' model, to contrast it with the ``quantum'' dynamics, i.e. the dynamics in the Majorana model that involves the pairing information. Again we consider particles that occupy the sites of a 1D lattice, with a given site occupied by at most one particle.  In this classical model there is no pairing of particles, or any nonlocal structure. 
However, each particle now carries a new label, which is either $A$ or $B$. Again, bonds of the lattice are updated at rate $\Gamma$. When a bond is updated, the following occurs:

(i) If neither of the sites on the bond is occupied, nothing happens.

(ii) If the one of the two sites on the bond is occupied, the occupying particle hops across the bond.

(iii) If the two sites are occupied by two particles with opposite labels, then the outcomes are $AB$ or $BA$ with probability $1/2$ each: i.e.~the particles exchange places with probability $1/2$.

(iv) If the two sites are occupied by particles with the same label ($AA$ or $BB$) these are annihilated.

We now describe the mapping between the Majorana model and classical model.
As mentioned above, this mapping involves replacing each Majorana with either an $A$ or $B$ particle (Fig.~\ref{fig:classical_mapping}). The assignments are made separately for each pair in the pairing diagram (note that the two constituents of the pair may be spatially distant).
If the pair has $n=0$, its two Majoranas are replaced with classical particles with the same label: both $A$, or both $B$, with probability $1/2$ for each option.
If the pair has ${n=1}$, its two Majoranas are replaced with classical particles having opposite labels, with a probability $1/2$ for the left member of the pair to be the $A$ particle and a probability $1/2$ for the right member to be the $A$ particle.
Note that this mapping does not change the \textit{positions} of the particles: it only  replaces the {pairing structure} with a random assignment of $A$s and $B$s.

To describe things more formally, consider the evolving probability distribution in each of the processes. Let $\mathbf{P}_t^\text{qm}$ be the probability distribution for the state of the Majorana process at time $t$. This distribution tells us the probability that there are defects at given positions, with a given pairing structure, and given fermion parities. 
Let $\mathbf{P}_t^\text{cl}$ be the probability distribution for the state of the classical process at time $t$: this distribution tells us the probability that there are particles at given positions, with given labels. 

The prescription in the paragraph before last allows us to  map a  probability distribution in the Majorana model  to a probability distribution in the classical model. We call this map $\mathcal{M}$. Abusing notation slightly, we write:
\be 
\mathbf{P}_t^\text{qm} \overset{\mathcal{M}}{\longrightarrow} \mathbf{P}_t^\text{cl}.
\ee
Above, we defined a mapping for a single state of the Majorana model: this corresponds to the special case where $\mathbf{P}_t^\text{qm}$ is ``delta function'' supported on a single state. Since a general distribution is a linear superposition of such delta functions, the extension to a general $\mathbf{P}_t^\text{qm}$ is straightforward, with $\mathcal{M}$ being a linear map.

The key point is that this map is compatible with the time evolution we have  defined for each of the models. Formally, this means that the following diagram commutes:
\be\label{eq:commutativediagramdynamics}
\begin{tikzcd}[column sep=huge, row sep = large]
\mathbf{P}_t^\text{qm} \arrow[r, "\mathcal{M}"] \arrow[d, "\substack{\text{time} \\ \text{evolution}}"]
    & \mathbf{P}_t^\text{cl} \arrow[d, "\substack{\text{time} \\ \text{evolution}}"] \\
\mathbf{P}_{t+\dd t}^\text{qm} \arrow[r, "\mathcal{M}"]
&\mathbf{P}_{t+\dd t}^\text{cl} \end{tikzcd}
\ee
That is to say, it does not matter whether we make the ``quantum--classical mapping'' at the very beginning, for the initial conditions, and then run the \textit{classical} dynamics until the final time; or whether we instead run the \textit{Majorana} dynamics until the final time, and make the classical mapping at the end. In either case we get the same final probability distribution in the classical problem. 

Furthermore, the mapping preserves the positions of the defects.
Therefore, as long as we choose initial conditions correctly in the classical model, the probability distribution of the total density is \textit{identical} in the two models for all $t$.

Proving that the diagram  (\ref{eq:commutativediagramdynamics}) commutes is a matter of checking the various cases, corresponding to the various possible updates on a bond.

First note that, by the linearity of all the maps involved in (\ref{eq:commutativediagramdynamics}),  it is sufficient to check the case where the Majorana model is in a definite initial state (i.e.\ where $\mathbf{P}_t^\text{qm}$ is supported on a single state). 
Also, it is sufficient to check commutativity for the update operation on a single bond. 
We only need to go through the possible initial states for defects on the bond. 

For brevity, we discuss a couple of illustrative cases below. It is straightforward to check the remaining cases, which completes the proof. 

If the updated bond is unoccupied, or occupied by a single `particle' which then hops, the result is immediate, so we consider the cases where the updated bond is occupied by a pair of particles. 

First consider a case where the two particles occupying the bond are already paired with each other, say into the `1' state. After the mapping $\mathcal{M}$, we get two classical states with equal probability, which we denote
\be\label{eq:commuativitycheck1}
 \begin{tikzpicture}
\draw (0,0) arc (180:0:0.35);
   \fill[black] (0,0) circle (0.1cm);
  \fill[black] (0.7,0) circle (0.1cm);
  \node () at (0.35,0.5+0.2) {1};
 \end{tikzpicture}
 \quad
\overset{\mathcal{M}}{\longrightarrow}
 \quad
 \frac{1}{2} \lf 
 \begin{tikzpicture}
   \fill[black] (0,0) circle (0.1cm);
  \fill[black] (0.7,0) circle (0.1cm);
  \node () at (0,0.4) {A};
    \node () at (0.7,0.4) {B};
 \end{tikzpicture}
 +
 \begin{tikzpicture}
   \fill[black] (0,0) circle (0.1cm);
  \fill[black] (0.7,0) circle (0.1cm);
  \node () at (0,0.4) {B};
    \node () at (0.7,0.4) {A};
 \end{tikzpicture}
 \ri.
 \ee
 If we update the left-hand side, in the Majorana problem, nothing happens, since the two Majoranas are in an $n = 1$ state and cannot annihilate. Therefore, for consistency with Eq.~(\ref{eq:commutativediagramdynamics}), the right-hand side of Eq.~(\ref{eq:commuativitycheck1}) should also be invariant under the fictitious
 classical time evolution we have introduced. This is easily seen to be the case using the update rule (iii) for the classical model given above.
 
 Next, consider a case where the particles on the updated bond are \textit{not} initially paired with each other. For example, let each be in a `0' pair. We also draw their partners:
 \ba\label{eq:commuativitycheck2}
 \notag
 \begin{tikzpicture}
\draw (0,0) arc (180:0:0.25);
   \fill[black] (0,0) circle (0.1cm);
  \fill[black] (0.5,0) circle (0.1cm);
  \node () at (0.25,0.5+0) {0};
  \draw (1,0) arc (180:0:0.25);
   \fill[black] (1.5,0) circle (0.1cm);
  \fill[black] (1,0) circle (0.1cm);
  \node () at (0.25+1,0.5+0) {0};
 \end{tikzpicture}
\overset{\mathcal{M}}{\longrightarrow}
 \f{1}{4}\bigg(
 & 
  \fourdots{A}{A}{A}{A}+
   \fourdots{B}{B}{B}{B}\\
 &   \fourdots{A}{A}{B}{B}+
   \fourdots{B}{B}{A}{A}
 \bigg).
 \end{align}
After the update in the Majorana model, which acts on the two central particles, the configuration on the left becomes
(again, the coefficients are classical probabilities, not wavefunction amplitudes!):
\ba\label{eq:commuativitycheck3}
 \begin{tikzpicture}
\draw (0,0) arc (180:0:0.25);
   \fill[black] (0,0) circle (0.1cm);
  \fill[black] (0.5,0) circle (0.1cm);
  \node () at (0.25,0.5+0) {0};
  \draw (1,0) arc (180:0:0.25);
   \fill[black] (1.5,0) circle (0.1cm);
  \fill[black] (1,0) circle (0.1cm);
  \node () at (0.25+1,0.5+0) {0};
 \end{tikzpicture}
\,\, \overset{\text{update}}{\longrightarrow} \,\,
 & \f{1}{2}  \, 
  \begin{tikzpicture}
  \fill[black] (0,0) circle (0.1cm);
%  \fill[black] (0.5,0) circle (0.1cm);
   \fill[black] (1.5,0) circle (0.1cm);
%  \fill[black] (1,0) circle (0.1cm);
  \draw (0,0) arc (180:0:0.75);
%  \draw (0.5,0) arc (180:0:0.25);
  \node () at (0.75,1) {0};
%  \node () at (0.75,0.5+0) {0};
 \end{tikzpicture} 
+ \f{1}{2} \, 
  \begin{tikzpicture}
  \fill[black] (0,0) circle (0.1cm);
  \fill[black] (0.5,0) circle (0.1cm);
   \fill[black] (1.5,0) circle (0.1cm);
  \fill[black] (1,0) circle (0.1cm);
  \draw (0,0) arc (180:0:0.75);
  \draw (0.5,0) arc (180:0:0.25);
  \node () at (0.75,1) {1};
  \node () at (0.75,0.5+0) {1};
 \end{tikzpicture}.
 \end{align}
 Mapping this final state [the right-hand side of Eq.\ (\ref{eq:commuativitycheck3})] to the classical model using $\mathcal{M}$ gives:
\ba\label{eq:commuativitycheck4}
\overset{\mathcal{M}}{\longrightarrow}  \, \, & \f{1}{4} \bigg(  
\fourdotssmallgap{A}{A} +  
\fourdotssmallgap{B}{B} \bigg)  \\
 + & \f{1}{8} \bigg( 
\fourdotssmall{A}{A}{B}{B} +
 \fourdotssmall{A}{B}{A}{B} +
  \fourdotssmall{B}{A}{B}{A} +
   \fourdotssmall{B}{B}{A}{A}
\bigg).
\notag
 \end{align}
For consistency, we must obtain the same state, Eq.~(\ref{eq:commuativitycheck4}), by applying the update in the \textit{classical} model to the right hand side of Eq.~(\ref{eq:commuativitycheck2}). Using rules {(i, iii, iv)} for the classical process, we can easily check that this is the case.
The other initial configurations may be checked similarly.

Having established this mapping, the  problem is now reduced to understanding the dynamics of the classical model. 
This classical problem almost reduces to two separate diffusion--annihilation processes occurring in parallel, one for the $A$s and one for the $B$s. This is not quite the case, however: the update rule for a bond where the particle content is $AB$ or $BA$ shows that when $A$s and $B$s come into contact their hopping is correlated.  (Roughly speaking, $A$s and $B$s interact through the restriction that they cannot simultaneously occupy the same lattice site.) However, this effect is irrelevant at late times when the particles are dilute. This diluteness means that an $A$ particle is adjacent to a $B$ particle only a parametrically small fraction of the time, so that the coarse-grained dynamics of an $A$ particle (in the absence of other $A$s) remains simple Brownian motion,  with a diffusion constant $D = \Gamma a^2$ set by the hopping rate $\Gamma$ and the lattice spacing $a$. 

Thus, at late times we effectively have decoupled ${A+A\rightarrow \emptyset}$ and ${B+B\rightarrow \emptyset}$ diffusion-annihilation processes. Each of these contributes a density given by Eq.\ (\ref{eq:rhocl}) at late times (recall that this result is independent of the initial density). Summing these contributions gives Eq.~(\ref{eq:quantumdensity}).

So far we have assumed that the initial state is either a state with definite fermion parities for some choice of pairing, or a classical mixture of such states. 
One may ask how the density evolves if, instead, the initial state is a quantum superposition that cannot be written as a single pairing state, or an otherwise arbitrary density matrix.
In Appendix~\ref{app:generalinitstate}, we argue that the universal late-time results for the density and correlation functions  (presented in the following subsection) are valid for any initial state. For example, one could take the initial state to be the ground state of the critical noninteracting Kitaev chain, representing a quantum quench for an open system in a certain limit.

\subsection{Exact results for correlation functions}

A considerable amount is known about the classical ${A+A\rightarrow \emptyset}$ process, or equivalently about the relaxation of the 1D Ising model via domain wall annihilation, and these results may be translated to our present problem using the mapping above. We now discuss some examples.

In Ref.~\onlinecite{Bray_1990}, Bray calculated the correlation function ${C_\text{cl}(r;t) = \<S(x) S(x+r)\>}$ in the 1D Ising model at late time $t$ during the domain wall annihilation dynamics, where $S(x) = \pm 1$ is the Ising spin at position $x$.  The result is a universal scaling form 
\be
C_\text{cl}(r; t) = 1 - \operatorname{erf} \lf \frac{r}{\sqrt{8Dt}} \ri,
\label{eq:Cclr}
\ee 
where $D$ is the diffusion constant and $\operatorname{erf}$ is the error function.
In terms of the diffusing particles (domain walls),  $S(x) S(x+r)$ is equal to $(-1)^m$, where $m$ is the number of domain walls in between $x$ and $x+r$.  

Returning to the microscopic models of Sec.~\ref{sec:microscopic}, let us consider the Ising-like order parameter that distinguishes the two local ground states (for example of the interacting Kitaev chain). We have denoted this order parameter by $\Phi(x)$, with $\Phi(x)=\pm 1$ for the two ground states (see Fig.~\ref{fig:correlation_domains}). The correlation function 
\be
{C_\text{qm}(r) = \< \Phi(x) \Phi(x+r)\>}
\ee
is again the expectation value of $(-1)^m$, where $m$ is the number of Majorana domain walls in the interval $(x, x+r)$.
Making the classical mapping, this is equal to the expectation value of $(-1)^{m_A}\times (-1)^{m_B}$, where $m_A$ is the number of $A$ particles in between $(x,x+r)$ and equivalently for $m_B$. Since the $A$ and $B$ particles are independent at late times, we obtain the correlation function in the Kitaev chain as the product of two classical correlation functions: 
\be
C_\text{qm}(r; t) = \left[ C_\text{cl}(r; t) \right]^2.
\label{eq:Cqm}
\ee

One can also consider \textit{non}--equal-time correlation functions. For these, it turns out that the critical exponents differ between quantum and classical models. Let
\be
C^\text{qm}(r; t; t') = \< \Phi(x,t)\Phi(x+r,t')\>.
\ee
Previously we defined a map between the instantaneous probability distributions in the quantum and classical problems. But in fact one can go further and show that the probability measure for a complete \textit{history} of the particle positions, from time $0$ up to some final time $t$, is the same in the classical and the quantum problems.\footnote{So long as the labels $A$ and $B$ are assigned with the correct probability distribution at $t=0$.}
Then a simple extension of the above argument relating the quantum and classical correlation functions\footnote{This argument proceeds by considering the number of domain walls crossing a line in spacetime between $(x,t)$ and $(x+r, t')$.} gives again
\be
C_\text{qm}(r; t; t') = \left[ C_\text{cl}(r; t; t') \right]^2.
\label{eq:Cqmt}
\ee
The result for the classical Ising model \cite{Bray_1990}, at zero spatial separation and with $t\leq t'$, is 
\be
C_\text{cl}(0; t; t') = \f{2}{\pi} \arcsin \sqrt{\f{2 t}{t+t'} }.
\label{eq:Cclt}
\ee
Therefore, when $t'/t \gg 1$,
\ba
C_\text{qm}(0; t; t') & \simeq   \f{8}{\pi^2} \f{t}{t'},&
C_\text{cl}(0; t; t') & \simeq  \sqrt{\f{8}{\pi^2} \f{t}{t'}},
\end{align}
so that the power-law for the the long-time decay  (${t'\rightarrow \infty}$) is different in the two cases.

\begin{figure}[t]
\centering
\includegraphics[width=0.45 \textwidth]{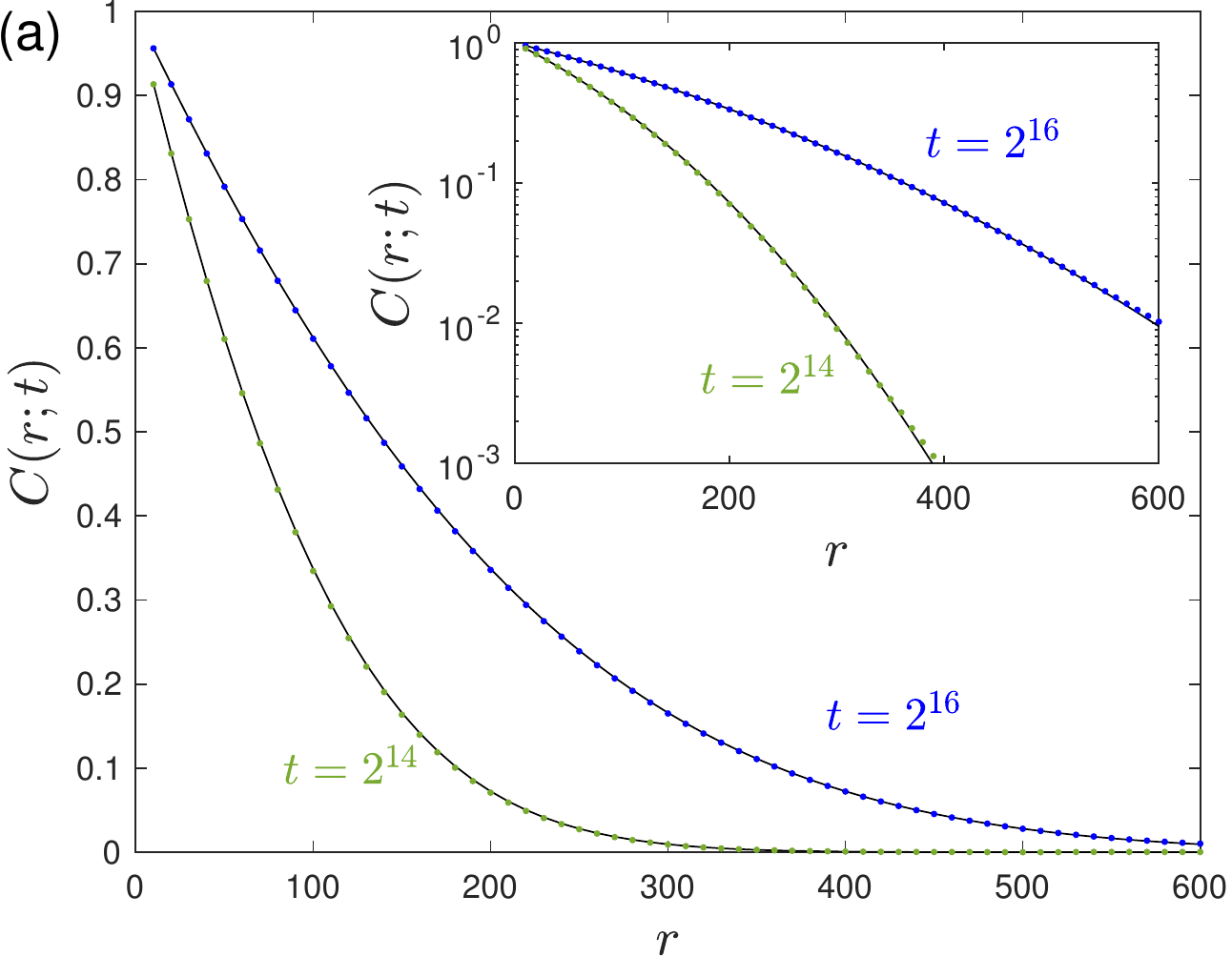}
\includegraphics[width=0.45 \textwidth]{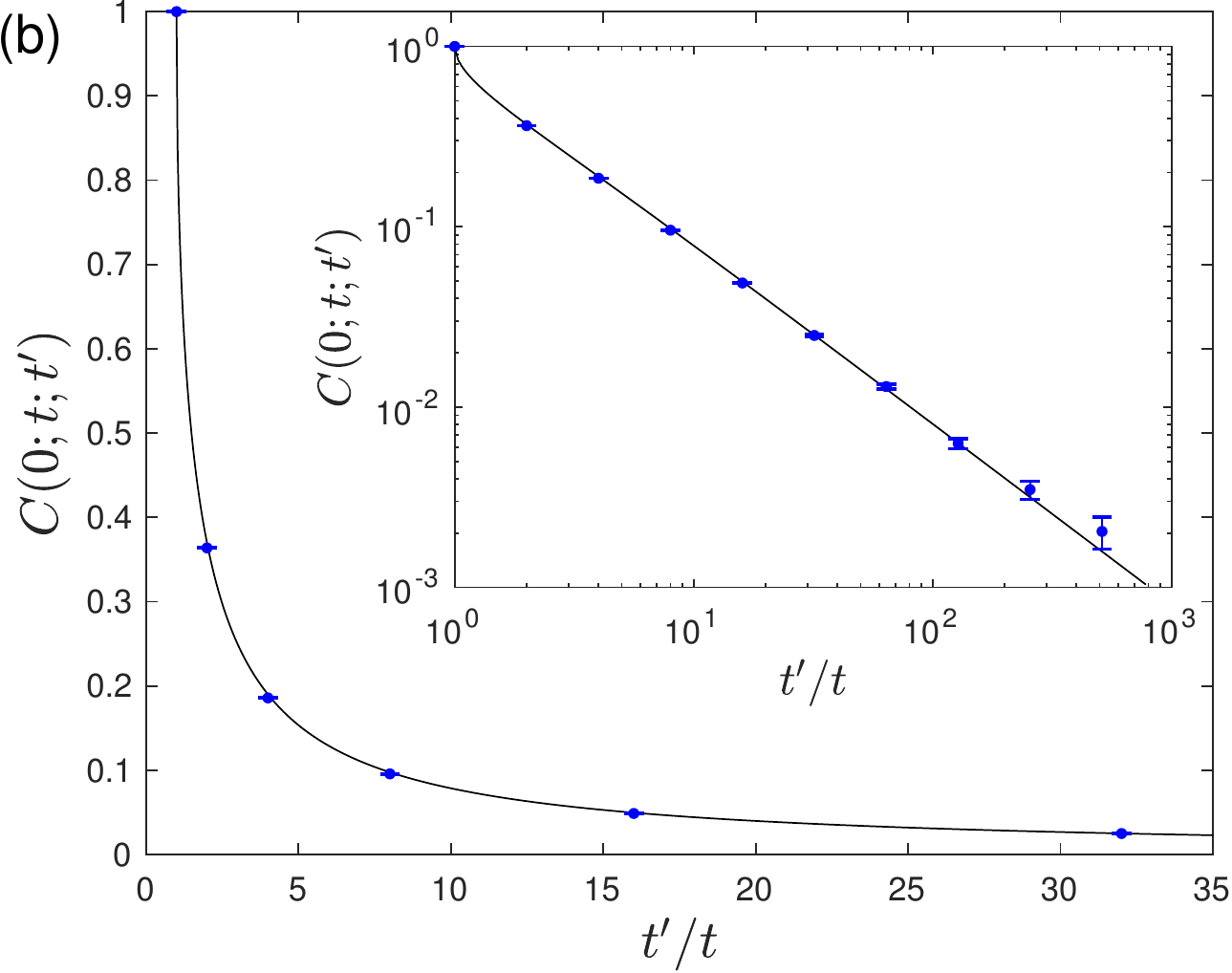}
\caption{Correlation function $C(r; t; t')$ for the ground state order parameter in the Majorana model with annihilation.  (a) The equal-time correlation function $C(r; t)$ is plotted as a function of separation $r$ at different values of the simulation time $t$.  The points correspond to simulation data (error bars are smaller than the symbol sizes) and solid black lines show the square of the classical correlation function $C_\text{cl}(r; t)$, given by Eq.\ (\ref{eq:Cclr}).  (b) The equal-position correlation function $C(0; t; t')$ is plotted as a function of $t'/t > 1$ for $t = 2^7$.  The black line shows the square of Eq.\ (\ref{eq:Cclt}). In both plots the inset shows the same data in logarithmic or semi-logarithmic scale. All data in this figure are taken from a simulation with $10^5$ lattice sites and an initial density $\rho(0) = 1/5$, and are averaged over $10^4$ random realizations.}
\label{fig:Crt}
\end{figure}

Equations (\ref{eq:Cqm}), for the equal-time correlator, and (\ref{eq:Cqmt}) for the non-equal-time correlator,  are verified directly using simulation data in Fig.~\ref{fig:Crt}.

Finally, another interesting critical index is the ``persistence exponent'',  $\Theta$ \cite{derrida1994non,
derrida1996exact}. The probability that, during the dynamics up to time $t$, the order parameter $\Phi(x)$ at a given location has \textit{never} flipped (i.e. the probability that no domain wall has ever traversed $x$) scales as $t^{-\Theta}$. The exact result in the classical Ising model is $\Theta_\text{cl}=3/8$ \cite{derrida1996exact}. Similar considerations to those above show that in the quantum model this exponent is doubled: $\Theta_\text{qm} = 3/4$.

\subsection{Entanglement in the ${\gamma+\gamma\rightarrow \emptyset}$ process}
\label{sec:entanglementwithannihilation}

In the diffusion--annihilation model, the length scale associated with entanglement is inevitably large at late times simply because the interparticle separation grows as $\sqrt t$. 
Even a ``dimerized'' pairing state would involve pairs of $O(\sqrt t)$ length. 
One can then ask: does the system contain pairs  much longer than $\sqrt{t}$, spanning many interparticle separations, at late times? It turns out that the answer to this question depends on the initial state.

In the following discussion we continue to assume that the initial state is either a state of definite pairing, or a statistical ensemble (mixture) of such states. 
This assumption excludes initial states that can only be written as quantum superpositions of different pairing states: we address these in Appendix~\ref{app:generalinitstate}.

Let $s \in \{ 1, 3, 5, 7, \ldots\}$ be an arc ``size'' measured in terms of the Majorana \textit{index}, so that an arc of size $s$ pairs Majoranas $i$ and $i+s$ for some $i$. 
This definition removes the trivial  $\sqrt t$ scaling of the physical length which comes from the large interparticle separation.
We will see in a moment that the leading effect of annihilation on the entanglement structure is to ``advect'' the arc size distribution, $P(s)$, by carrying weight from large $s$ to smaller $s$.
This advection implies that if the initial state has a finite typical arc length $\xi$, no matter how large, the typical value of $s$ at late times is only of order 1.

However if the initial state has $\xi=\infty$ --- for example if the initial state is an equilibrated state of the model without annihilation, which has $P(s)\sim s^{-2}$ --- then $\xi$ always remains infinite.

This latter result is consistent with the general fact that is impossible to have a short-range $P(s)$ unless the pairing ensemble breaks ``translational symmetry'' $i\rightarrow i+1$ in the Majorana index (see Sec.~\ref{sec:notetranslation}).
If we draw the initial state from the critical $P(s)\sim s^{-2}$ ensemble, which does \textit{not} break this symmetry, then by causality the symmetry must remain unbroken at any finite time, because symmetry breaking requires correlations over arbitrarily large distances. Therefore the tail of the size distribution must at all times decay slower than $s^{-(2+\epsilon)}$ for any $\epsilon>0$.

To understand the ``advection'' of the arc length distribution, consider the evolution of a particular large arc with $s\gg 1$. 
(As in the heuristic argument at the beginning of Sec.~\ref{sec:loopmodel2d}, in order to give the arcs a well-defined identity after a collision, we follow the larger of the two arcs produced by the collision.)
The characteristic timescale for an encounter between this arc and another is the diffusive timescale $\Delta t \sim 1/(D \rho^2)$, where $\rho$ is the instantaneous density. 

In an encounter, the chosen arc instantaneously grows or shrinks by an amount $\Delta s$. If $s\gg 1$, then typically $\Delta s \ll s$. 
On the other hand, on the same timescale $\Delta t$, an \textit{order one} fraction $1-f$ of the intervening Majoranas (lying underneath our chosen arc) is annihilated through collisions between neighbors. This annihilation changes $s$ by $s\rightarrow f s$, since we measure arc sizes by counting intervening Majoranas.
This simple rescaling dynamics can be argued to dominate over the effect of collisions involving the long arc so long as $\alpha > 2$. 
We can conclude that an arc of size $s\gg 1$ will eventually shrink to a size $s$  of order 1, and that the exponent for the power-law tail of the distribution is preserved.

If the initial distribution has a cutoff $\xi$ on the largest value of $s$, then at late times all arcs will have $s$ of order 1, and the typical entanglement length scale in the system will be set by  the nearest-neighbor spacing. 
In our simulation with a dimerized initial condition we can directly measure the evolution of the arc-size distribution $P(s)$, where $s-1$ is the number of Majoranas underneath a given arc. 
We find (see Appendix \ref{app:annihilationS}) that on a timescale of order  $1/(D \rho_0^2)$, where $\rho_0$ is the initial density, 
the distribution becomes stationary, with a tail $P(s) \propto e^{-s/s_0}$ at $s \gg1$ ($s_0 \approx 2.1$). 

In order to examine the contrasting situation where the initial state has $\xi=\infty$,
we have also run two-step simulations in which we first ``equilibrate'' the system using the dynamics without annihilation. 
This yields a power-law distribution $P(s)\sim s^{-2}$ that is cut off at large $s$ only by the total number  $N_0$ of Majoranas in the initial state.
We then switch on annihilation and observe the evolution of $P(s)$.
What we see is roughly consistent with the crude  picture  above, in which the part of the distribution with $s\gg 1$ is advected to smaller $s$ at a rate set by the decay of the particle density (see Appendix \ref{app:annihilationS} for simulation results). The distribution retains the initial power law in the range $1\ll s \ll N_t$, where $N_t$ is the total number of Majoranas at time $t$. 
Note that this means that the \textit{longest} arc in a system of  physical length $L$ always has a physical length of order $L$, so long as a nonzero number of Majoranas remain in the system, although this arc's $s$ value (which is of order $N_t$) is decreasing with time.

It is interesting to note that while the late-time density decay is insensitive to the initial state (see Sec.~\ref{sec:doubledisingmapping} and Appendix~\ref{app:generalinitstate}), the late-time entanglement structure is not. 
This dependence of the entanglement on the initial state also implies that, in the case with annihilation, the quantum state of the  Majorana defects that survive at time $t$ in general never becomes a pure state if the initial state is a mixed one \cite{gullans2019dynamical} (except in the trivial regime when $t\gtrsim L^2/D$ and no more defects remain).
By contrast the measurement-only dynamics of Sec.~\ref{sec:without} does eventually purify the initial state. In the loop picture (cf.\ Appendix~\ref{app:generalinitstate}), the full system is \textit{completely} purified when no loops connect to the initial time boundary. In the model with annihilation, this disconnection never happens while there are an extensive number of defects remaining. In the model without annihilation, it happens after a time of order $L$.

\section{Majoranas in two dimensions}
\label{sec:higherd}

There is intense interest in two-dimensional systems with Majorana-like defects, including superconducting systems with pointlike vortices that carry Majoranas \cite{read2000paired,fu2008superconducting}, and topologically ordered systems (e.g.\ fractional quantum Hall states \cite{read1992fractional} and spin liquids \cite{kitaev2006anyons})  with nonabelian ``Ising'' anyons, which are closely related to Majorana modes.
Here we do not attempt to make direct connections with these systems, since it would require many other features to be taken into account. Instead we discuss the simplest possible two-dimensional generalizations of our models.

Perhaps surprisingly, the entanglement structure of a 2+1D  quantum measurement circuit,
in which Majorana modes are subjected to a random sequence of repeated projective measurements,
remains exactly solvable in the scaling limit. 
We describe this exact solution below.

However, 2D versions of the diffusion-annihilation process are complicated by the nontrivial braiding of Majoranas. A convenient way to handle braiding is to imagine attaching branch cuts to each Majorana defect \cite{ivanov2001non} (the geometry of these branch cuts is arbitrary). One then flips the sign of the operator $\gamma_i$ whenever the $i$th Majorana defect crosses the branch cut associated with another defect.

This procedure implies that the parity of a given pair depends on the path it has taken. As a result, the classical mapping 
of Sec.~\ref{sec:doubledisingmapping} does not carry over directly to the 2D case, and 
we do not attempt to treat the 2D  $\gamma+\gamma\rightarrow\emptyset$ problem here. 

We note, however, that the simplicity of the braiding rule for Majoranas means that the 2D diffusion-annihilation process could be studied numerically with low computational cost. 
The precise formulation of the problem will depend on the system being modelled. 
For example, defects may come in two species, representing vortices and antivortices, leading to an ${A+B\rightarrow \emptyset}$ process.

Two dimensions is the upper critical dimension for such reaction diffusion problems \cite{Tauber_2005}, which exhibit logarithmic decay of the density with a  universal prefactor. For example, in the case of the $A+A\rightarrow \emptyset$ process   \cite{bramson1980clustering,lee1994renormalization}
\be
\rho(t) = \frac{1}{16\pi} \f{\ln (t/t_0)}{D t}.
\ee
It would be interesting to obtain the corresponding universal constants for the quantum models.

\begin{figure}[t]
\centering
\includegraphics[width=0.45 \textwidth]{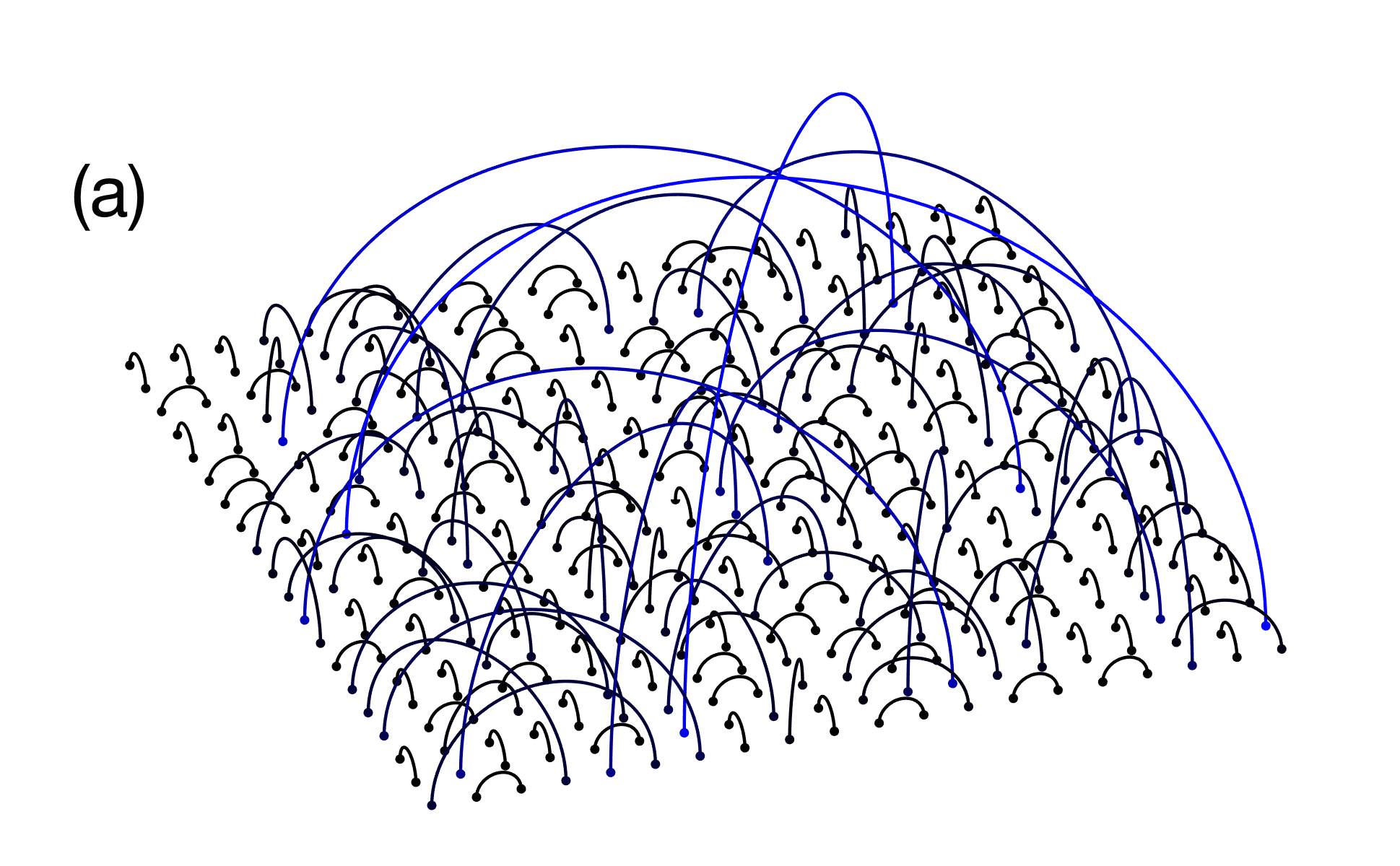}
\includegraphics[width=0.4 \textwidth]{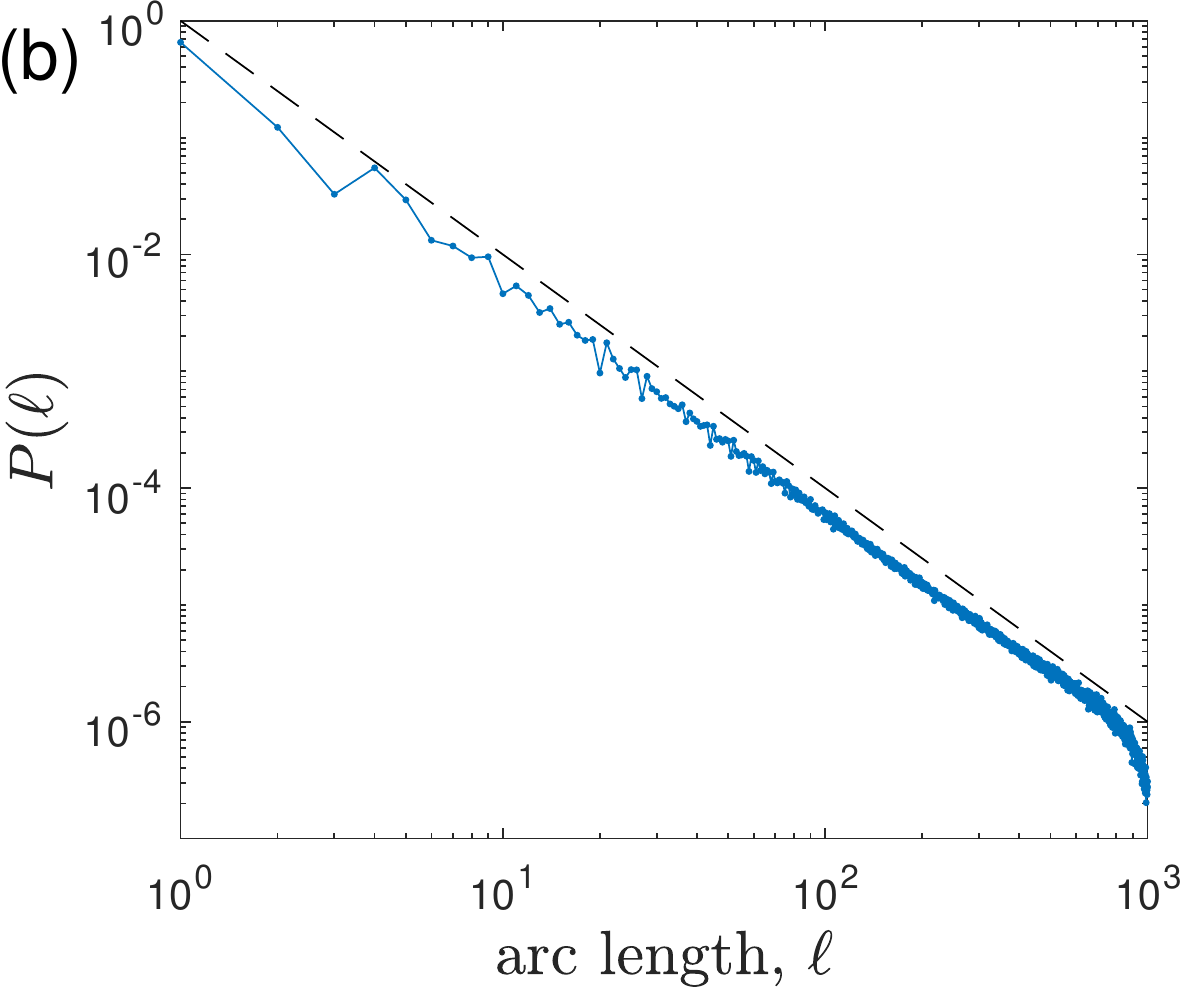}
\caption{The distribution of arcs in two dimensions for a model where Majoranas are arranged on a fixed lattice and nearest-neighbor pairs are randomly measured (no annihilation). (a) A typical configuration of arcs in a small-sized system, with lines color-coded according to the arc length. (b) The resulting distribution of arc lengths, as recorded by a simulation with a square grid of $L^2=1000 \times 1000$ Majoranas, evolved for $10^6$ time steps (each time step involves $L^2=10^6$ measurements)
with arcs reported every $1000$ time steps.  The dashed line indicates a dependence $P(\ell) \propto 1/\ell^2$.}
\label{fig:Pell2D}
\end{figure}

\begin{figure}[t]
\centering
\includegraphics[width=0.3 \textwidth]{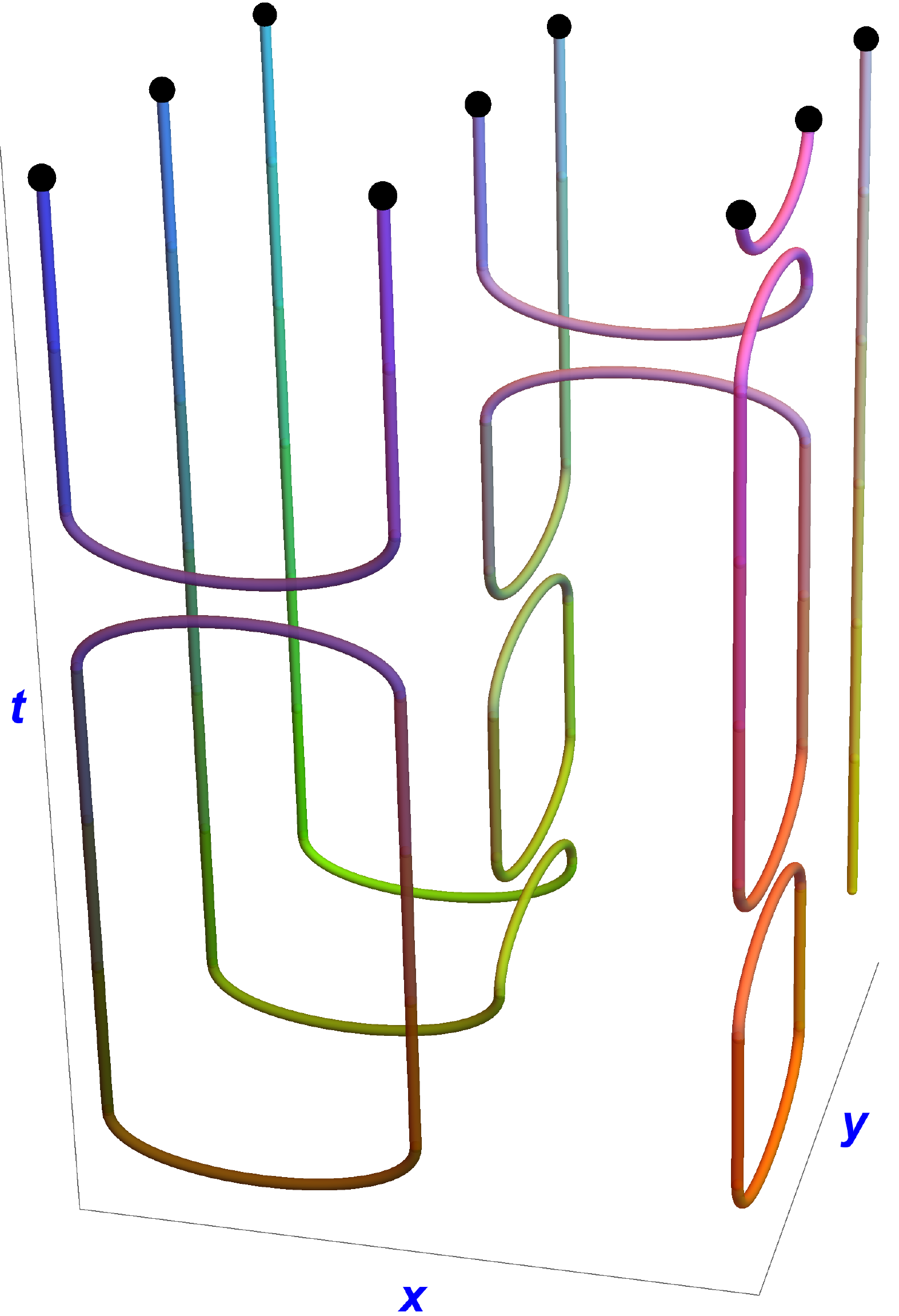}
\caption{Majorana worldlines in a 2+1D quantum measurement circuit (only a part of the system is shown).
The entanglement dynamics can be determined from the statistics of completely-packed loops in 3D.}
\label{fig:3dloops}
\end{figure}

\subsection{Higher-dimensional measurement circuits}

Let us now consider the measurement--circuit dynamics for a fixed 2D lattice of Majoranas $\gamma_{\vec r}$.
Here $\vec r$ runs over the sites of a 2D lattice that we are free to choose. 
We will find that there is a slight distinction between bipartite and non-bipartite lattices: we consider first the square lattice, as an example of the bipartite case.\footnote{If we wish to think of the circuit model as a simplification of dynamics of mobile Brownian defects undergoing random spatial encounters, then the non-bipartite case is appropriate.}

To avoid having to choose an arbitrary lattice structure in 2+1D space\textit{time}, we may imagine  applying projective measurements to arcs at random in a Poissonian fashion (so that $i\gamma_{\vec{r}}\gamma_{\vec{r}+\vec{\delta}}$ for each pair of adjacent sites
is subjected to measurements
at a rate $\Gamma$; of course only the sequence of measurements matters, not the precise times).
Using a 2+1D spacetime lattice, in analogy with Fig.~\ref{fig:circuit}, would make no difference to the long-distance properties.

First, numerical results for the square lattice are shown in Fig.~\ref{fig:Pell2D}. The top panel shows a typical configuration of the arcs at late time, and the bottom panel shows the length distribution for a randomly chosen arc. We find that it fits well to
\be\label{eq:2ddistribution}
P(\ell)\sim \ell^{-2}
\ee
at large $\ell$, where $\ell$ is the Cartesian distance between endpoints of the arc. This power law is fortuitously the same as the 1D case, but the underlying universal structure is different.

The power law can again be explained using a mapping to a classical statistical model of nonintersecting loops \cite{nahum20113d}, now in 3D spacetime. This mapping goes through in complete analogy to the 1+1D case. We obtain loop configurations like that shown in Fig.~\ref{fig:3dloops}, with a loop ``turnaround'' for every measurement event.

A field theory for 3D loop models of this kind was derived in \cite{nahum20113d,nahum2013phase}. The field theory is a ``replica limit'' of a nonlinear sigma model with continuous symmetry, in the phase\footnote{A phase transition out of this phase can also be engineered, see the comment towards the end of this subsection.} where the continuous symmetry is spontaneously broken. (The symmetry is $\mathrm{SU}(N)$ and the target space is complex projective space,  $\cp^{N -1}$. The replica limit is $N \rightarrow 1$.)
However, the details of this field theory need not concern us here, because the leading scaling turns out to be simple. The fact that the ordered phase can be described with weakly interacting Goldstone modes implies that a long loop resembles a Brownian path at large scales \cite{nahum20113d}, somewhat like a polymer in a melt \cite{de1979scaling}, with simple Brownian exponents.

This fact can be used to compute the length distribution $P(\ell)$ for the arcs in the stationary state. $P(\ell)$ is the probability that, if we follow a Brownian curve that is initiated at a point on the final time surface of the 2+1D region, its \textit{first} return to this surface is at a Euclidean distance $\ell$ from the starting point. Calculating this probability is simplified by the fact that the various components of the Brownian path, corresponding to the three spacetime axes, are statistically independent. 
A straightforward calculation gives Eq.~(\ref{eq:2ddistribution})
at large $\ell$. 

This behavior $P(\ell) \sim 1/\ell^2$ is the same power law as in the 1D case, but at least in this approach that seems to be a coincidence (since in 1+1D a loop is not Brownian).
In 2+1D it also leads to a different scaling of the entanglement. For a compact 2D region, say a disc of size $R$ in an infinite 2D system, an appropriate integral of Eq.\ (\ref{eq:2ddistribution}) shows that the entanglement scales as  ($a$ is the lattice spacing)
\be\label{eq:rlogr}
S(R)\sim \lf \frac{R}{a} \ri  \ln \lf  \frac{R}{a} \ri,
\ee
with an $O(1)$ prefactor that we do not determine here.

This $R \ln R$ entanglement scaling coincides with that of a free fermion ground state with a Fermi surface \cite{WolfViolation2006,EntanglementGioev2006,SwingleEntanglementFermi2010}, though the wavefunction (a translationally-invariant Slater determinant) is of a very different nature there.

The case with a non-bipartite spatial lattice is similar. Strictly speaking, the loop model is in a different universality class as compared to the bipartite case\footnote{The key difference is that in the bipartite case there is a consistent way to orient the Majorana worldlines in spacetime, such that on one spatial sublattice the worldlines are upgoing and on the other sublattice they are downgoing. This leads to a model of oriented loops.} \cite{nahum20113d,nahum2012universal}. 
This difference leads to a different field theory (the $\rp^{N - 1}$ model with $N \rightarrow 1$).
However, there is again a phase where the loops are Brownian at long scales, leading to the same power law for $P(\ell)$.

A difference between the two cases appears if we dimerize the measurement rates, in analogy to the 1+1D phase diagram discussed in Sec.~\ref{sec:loopmodel2d}.
This dimerization can be used to drive a phase transition into an area law state with only short-range pairs. Unlike in the 1+1D model in Sec.~\ref{sec:without}, the critical value of the dimerization is nonzero, so that even in the absence of translation symmetry there is a \textit{stable phase} with the scaling in Eq.~(\ref{eq:rlogr}).
The phase transition out of this phase is in a different universality class in bipartite and non-bipartite models.

These results in fact generalize immediately to spatial dimensions higher than two, where there is again a phase where the spacetime loops are Brownian, the arc length distribution is $P(\ell)\sim \ell^{-2}$, and the entanglement of a $d$-dimensional ball is scales as $(R/a)^{d-1}\ln (R/a)$.

To complete the discussion of universality classes, let us describe a model that yields a different 1+1D universality class from that discussed in Sec.~\ref{sec:without}.
Such a model can be obtained by making the one-dimensional system effectively non-bipartite. The simplest way to do so is to randomly replace some of the measurements (of adjacent Majoranas) with ``swap'' operations. These are unitary transformations that exchange the Majoranas $\gamma_i$ and $\gamma_{i+1}$.\footnote{Recall that the phase induced by the unitary operation is not important for the dynamics we discuss.} In this case, the loop model mapping of Sec.~\ref{sec:loopmodel2d} goes through as before, but now with an additional node configuration to be added to the two shown in Fig.~\ref{fig:circuitelement}. This configuration is one where the worldines of the two $\gamma$s \textit{cross}, which leads to a 2D loop model with crossings \cite{nahum2013loop,jacobsen2003dense,martins1998intersecting}. The crossings are a renormalization-group-relevant perturbation, leading to a new universality class of loop gas. In this model we expect the half-system entanglement to scale not as $\ln L$ but as $(\ln L)^2$.\footnote{See the discussion of the ``spanning number'' in Ref.~\cite{nahum2013loop}.} If we further perturb the model by ``dimerizing'' the measurement probabilities, so that even links are measured more frequently than odd links, then there is a continuous phase transition into a short-range entangled state at a nonzero critical value of the dimerization \cite{nahum2013loop}.

\section{Outlook}
\label{sec:discussion}

In a system of nonabelian anyons or Majorana modes, some quantum information is protected from a decohering environment. We have shown that this protection leads to new universality classes for relaxation that are not entirely classical, since the active degrees of freedom at late time include a nonlocal entanglement structure.
The ${\gamma+\gamma \rightarrow \emptyset}$ coarsening process, which is solvable by the  mapping in Sec.~\ref{sec:with}, is perhaps the simplest case, but we expect there are other interesting examples.

Our analysis of the dynamics of entanglement in this process has also led us to study related models involving measurement only. These models do not directly describe relaxation in many-body systems, but they illustrate that random measurements can lead to a nontrivial, critically-entangled pure state.  The late-time statistics of the state can be understood exactly, including a logarithmic violation of the area law for entanglement in any number of dimensions.

Let us suggest some directions for future research.

An interesting feature of the 1+1D measurement-only models studied here is the role of statistical translation symmetry in the Majorana index, $i \rightarrow i + 1$. Of course, the individual states generated by the dynamics are not translation-invariant, because a particular realization of the measurement process yields random, non-translation-symmetric outcomes. However, as noted in the text, the translation symmetry of the \textit{ensemble} of states still imposes a constraint on the entanglement structure, guaranteeing greater-than-boundary-law entanglement. Similar observations can be used to constrain the phase diagram of various dynamical protocols involving both measurements and unitary operations.

For example, consider a spin-1/2 chain subjected to unitary/measurement dynamics that is invariant under spin $\mathrm{SU}(2)$ symmetry (i.e. involving measurements only of spin-singlet operators) and under statistical translation symmetry. 
The stationary ensemble of states in the spin-zero sector must then have super-area-law entanglement, by a Lieb-Schultz-Mattis-like constraint on wavefunctions for spin-1/2s \cite{kimchi2018valence}. 

For the Majorana chain, heuristic arguments also suggest that statistical translation symmetry is enough to  guarantee super-area-law entanglement for more general unitary/measurement dynamics, whether free or interacting (we will discuss this elsewhere). 
It would be interesting to generalize such statements to other settings.

The critical 1+1D measurement circuit can be viewed as sitting at a topological phase transition driven by dimerization in the measurement rates. Making the rate stronger for either even or odd bonds leads to distinct area law phases for the state produced by the dynamics at long times. (An equivalent transition can be obtained in the model of diffusing domain walls by tuning the relative energy density of the two phases, $\Phi=\pm 1$.)
On an open chain of size $L$, these phases differ in a manner analogous to the two phases of the equilibrium Kitaev chain \cite{kitaev2001unpaired}: in one of them all Majoranas are paired in short range arcs, while in the other there are two Majoranas, one close to each boundary, that have no nearby partner. In the measurement dynamics these modes are instead paired by a system-spanning arc. This distinction between the two phases can be detected by the average mutual information between the first and last sites, or an equivalent correlation function \cite{skinner2018measurement}. (In the loop model, the phases are distinguished by the presence or absence of a large strand traveling around the lattice boundary  \cite{chalker1988percolation,gruzberg1999exact,beamond2002quantum}.) 
It would be interesting to explore connections between this transition (or relatives of it, in which time-translational invariance is artificially imposed via postselection of measurement outcomes) and recent work on non-Hermitian topological phases \cite{bergholtz2019exceptional}.

The models considered here can be extended to other types of anyon, for example Fibonacci or $\mathrm{SU}(2)_k$ anyons \cite{feiguin2007interacting,trebst2008collective,vasseur2015quantum}.
This kind of extension could give interesting stochastic models for 
entanglement structures  represented by  fusion trees. We expect that long-range-entangled states can be obtained by measurement here too, and that there are again interesting diffusion-annihilation processes. 
Two-dimensional versions of these problems are also interesting, and have the additional ingredient of braiding, with braiding rules that become more complex when one goes beyond Majoranas. 
An important distinction will be between models where braiding \cite{nayak2008non} and measurement \cite{bonderson2008measurement,cui2015universal,levaillant2015universal} together yield a computationally universal set of operations and models where they do not, since in the former case the full Hilbert space for a set of anyons can be explored, while in the latter case only a discrete subset can be accessed.

Even for  Majoranas, which have the simplest braiding rules
(such that the measurement-only entanglement dynamics  remains tractable in 2D)
braiding has a nontrivial effect on the 2D diffusion-annihilation process. The resulting dynamics may admit an interesting solution.
%{\red we have not solved here and which} it would be interesting to solve. 

\acknowledgments 

We thank Steven H. Simon for a useful discussion.
AN acknowledges support from the Gordon and Betty Moore Foundation under the EPiQS initiative (grant No.~GBMF4303), from EPSRC Grant No.~EP/N028678/1, and from a Royal Society University Research Fellowship. BS acknowledges the support of Argonne National Laboratory during a postdoctoral fellowship, and support by the NSF STC ``Center for Integrated Quantum Materials" under Cooperative Agreement No.~DMR-1231319.

\appendix

\section{Uniform ensemble of pairings}
\label{sec:rwapp}

As noted in the text the statistics of the uniform ensemble of pairings are different from the statistics of the pairings generated in our (measurement-only) dynamics. For completeness we review the standard fact that in the uniform pairing ensemble the probability for a length--$\ell$ arc  scales like $\ell^{-3/2}$. This follows from mapping the arc diagram to the trajectory of a 1D random walker whose position coordinate, or ``height'', which we denote $h$, is non-negative. The indices $i$ in the arc diagram correspond to time steps of the random walk. If  $i$ is the left-hand-side of an arc, this maps to a step with $\Delta h=1$, and if $i$ is the right-hand-side of an arc, this maps to a step with $\Delta h = -1$. The initial coordinate of the walker, prior  to the first ($i=1$) step, is taken to be zero. This gives a bijection between arc diagrams and trajectories of the walker that start and end at $h=0$, take steps $\Delta h =\pm 1$, and are never negative. Arcs map to sections of the walk that start and end at the same height $h'$ for some $h'$ and are strictly greater than $h'$ in between. Therefore (neglecting boundary effects, e.g. for arcs in the interior of a  large system) the probability of an arc having length $\ell$ scales like the probability of first return to the origin for a 1D random walker, as $\ell^{-3/2}$.

\section{Brownian wandering of loops}
\label{app:randomlattice}

Here we argue that the universal entanglement structure in the model with diffusing Majoranas, which are  measured when they come into contact, is the same as that in the circuit model, up to subleading corrections. It does not matter for the following discussion whether the diffusing Majoranas are placed on a lattice or in the continuum.

In the mapping to loops, each encounter between two random-walking Majoranas yields a node like the right-hand (yellow) configuration in Fig.~\ref{fig:circuitelement}. Repeated encounters between the same two particles (without meeting other particles in between) do not change the pairing, and so can be grouped into a single node. After this grouping, we effectively have a loop model on a random lattice whose characteristic spatial ``lattice spacing'' is $1/\rho$, where $\rho$ is the density of Majorana defects, with corresponding temporal spacing $1/(\rho^2 D)$. These are finite numbers and can be set to unity, on average, by an appropriate choice of units. 

However, there remain stochastic local fluctuations of the density, so that the ``lattice constants'' vary from place to place in the 2D spacetime plane. 
The coarse-grained statistics of these density fluctuations follow from a stochastic diffusion equation.

We treat these fluctuations as a perturbation of the conformal field theory for the loop ensemble and argue that this perturbation is irrelevant. (This conformal field theory is nonunitary, but many of its properties are known; see Refs.~\onlinecite{cardy2005sle, jacobsen2009conformal} for reviews.)
Strictly speaking this only shows that the universal properties are unchanged when the perturbation is small, but it would be  surprising if a larger  perturbation changed the universality class.

We must consider the most relevant operators in the conformal field theory that can couple to density fluctuations. There are two candidate operators.
First, fluctuations can couple to the stress tensor $T_{\mu\nu}$ of the conformal field theory. This leads to a perturbation of the schematic form ${\mathcal{S}_\text{CFT}+ \int \dd x \dd t \, m_{\mu\nu}(x,t) T_{\mu\nu}(x,t)}$, where $m_{\mu \nu}$ is a random perturbation which depends on the coarse-grained density and its derivatives. Since the scaling dimension of the stress tensor is two, we can easily check (for example by averaging over $m$ using the replica trick \cite{cardy1996scaling}) that this perturbation is renormalization-group irrelevant so long as $\<\< m_{\mu\nu}(x,t) m_{\mu'\nu'}(x',t')\>\>$ decays with $|x-x'|$ and $|t-t'|$, which it of course does.

Next, the conformal field theory contains a scalar operator $\epsilon$ with scaling dimension $x=5/4$. $\epsilon$ is the operator which drives the model away from criticality when the measurement rates are dimerized, as mentioned briefly in Sec.~\ref{sec:loopmodel2d}. 
In the off-lattice context, we
may make sense of dimerization by labelling the Majoranas with the index $i$ (starting at the left of the system) and distinguishing between encounters of adjacent Majoranas $(i,i+1)$ with even versus odd $i$.
More simply, we can take our local measure of dimerization to be  the physical order parameter $\Phi$ for which which the defects are domain walls, since this changes sign under translation of the Majorana index $i\rightarrow i+1$. 
We must then consider the perturbation ${\sim \int \dd x \dd t \, \Phi(x,t) \epsilon(x,t)}$.
The question is again how rapidly correlations in $\Phi$ decay. However, one can argue that they decay faster than any power law.\footnote{%
This argument may proceed as follows. If we neglect pairing information, the system of defect trajectories $x_i$ can be mapped to a system of noninteracting random walks, by a relabelling of walks, and  ${\Phi(x,t)\Phi(x,0)=\pm 1}$ is determined by the parity of the number of walks that cross position $x$ during the time interval $(0,t)$. This number is a product of independent contributions from $O(\sqrt{t})$ walks, giving a decay like $\exp (-c \sqrt{t})$ after averaging, where $c$ is a constant.
}
So, again, we conclude that Brownian fluctuations produce an irrelevant perturbation. Other allowed perturbations involve  operators that are irrelevant even when they are added to the action with uniform couplings \cite{jacobsen2009conformal, read2001exact, nahum2016universality}. 

Therefore we do not expect the coupling to the diffusive density fluctuations to affect the leading scaling of the entanglement, though it will will contribute to subleading corrections.

\section{Initial states in the ${\gamma+\gamma \rightarrow \emptyset}$ process} 
\label{app:generalinitstate} 

The treatment of the diffusion-annihilation process in Sec.~\ref{sec:doubledisingmapping} assumes that the initial state is one with definite fermion parities in some basis, or that it is a classical mixture of such states. Here we argue that the late-time results for the density and order-parameter correlation functions  hold for general initial states, i.e.~for arbitrary choices of the Majorana modes' initial density matrix (whether pure or mixed). Our argument is heuristic, rather than rigorous.

It is convenient to consider the system in terms of loop configurations, analogous to the one shown in Fig.~\ref{fig:loopsandarcs}. 
However, these loops are not completely packed, since at late times the Majoranas are at low density on the lattice, and the loop shapes are determined by the diffusive wandering of the particle trajectories as well as by reconnection events. Most importantly, in addition to local reconnections like those in Fig.~\ref{fig:circuitelement}, there are local annihilation events which cause the number of strands to decrease with time.
To avoid confusion, note that the loops we are talking about are not quite the same thing as the  trajectories of defects, which are directed paths in spacetime. The loops are  \textit{not} directed paths, because the reconnection and annihilation events are places where strands ``turn around''.   We will refer to these (non-directed) loops as a worldlines.

We claim that at  late times  $t \gg 1/[D \rho(t=0)^2]$, when many annihilation events have taken place, the dynamics of the density is statistically independent of the initial quantum state of the Majoranas. 

To see this independence, let us first consider the extent to which the state at time $t$ can be inferred from the sequence of previous encounters, and to what extent there is an additional dependence on the initial state. 
If we think of the loop diagram as defining a tensor network, with the initial state\footnote{The initial state is defined by a density matrix.} attaching to the initial time boundary, the question is how, given a fixed structure for the tensor network at times greater than zero, the final state depends on which initial state is attached at the lower boundary.

In discussing the dependence on the initial state, it is useful to divide the Majoranas that are present at time $t$ into two classes: ``type 1'' Majoranas,  whose worldlines connect to the initial-time boundary of the spacetime region, and ``type 2'' Majoranas, whose worldlines connect to another Majorana on the final-time boundary.  The tensor network picture shows that the density matrix of the subsystem of type 2 Majoranas is independent of the initial state (given the previous history of the measurement process), while that of the subsystem of type 1 Majoranas is not. 

When two Majoranas collide, whether they annihilate is decided by the outcome of a projective measurement.  \emph{Fixing} the loop configuration and the outcomes of previous measurements, let the probability that this measurement gives fermion parity zero be $p_0$. One may check that the only case where $p_0$ depends on the initial state is when the two colliding Majoranas are both of type 1. 
In this case, $p_0$ is fixed by the expectation value of a nonlocal fermion parity operator $i\gamma_j \gamma_k$ in the initial state, where $j$ and $k$ are the two Majoranas in the initial state that the two worldines connect to. 
These Majoranas will typically be widely separated in the initial state if $t$ is large.

When the loop configuration, i.e.~the history of encounters, is fixed, $p_0$ depends on the initial state via $\< i \gamma_i \gamma_j\>$. However, the quantities of interest involve averages over all possible histories. Let us average over all the past histories that are consistent with a given set of defect positions at time $t$, a given assignment of types to the Majoranas, and a given pairing diagram for the type 2 Majoranas. The expectation value $\< i \gamma_i \gamma_j\>$ in the initial state must then also be averaged over a random choice of the pair $(i,j)$ 
(with $i<j$). The details of the associated probability distribution  do not matter, only the fact that the indices $i$ and $j$ are randomly chosen from a parametrically large spatial region and are weakly correlated.  This implies that the average is much smaller than unity (as can be shown using the fact that $i\gamma_i \gamma_j$ and $i\gamma_i \gamma_k$ anticommute for $j\neq k$). 
Therefore, once we average over histories, $p_0$ is close to 1/2 at late times, independently of the initial state, and consequently the dependence on the initial state disappears.
In other words, the late-time dynamics of the  particle positions is independent of the choice of initial wavefunction or density matrix for the Majorana modes.

\begin{figure}
\centering
\includegraphics[width=0.45 \textwidth]{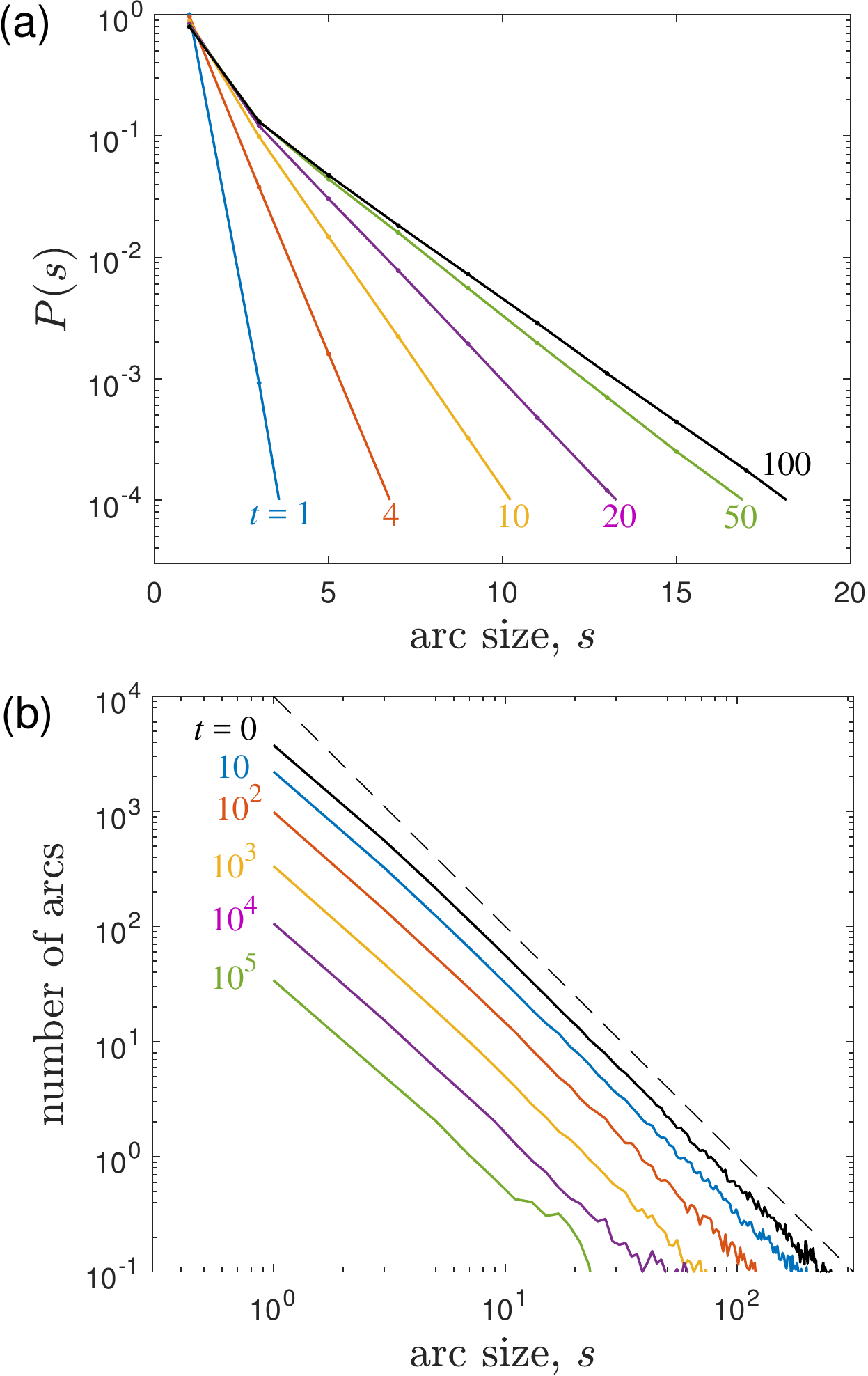}
\caption{The time evolution of the distribution of arc size $s$ for the diffusion-annihilation model. (a) $P(s)$ is plotted in semilogarithmic scale for different times, beginning from a dimerized state with all arcs having $s = 1$. Each curve is labeled by the corresponding simulation time. Data corresponding to $t > 100$ produce a curve $P(s)$ that is indistinguishable from the one labeled $t = 100$.  (b) The total number of arcs having a given size $s$ is plotted in double-logarithmic scale as a function of time, for a system that starts in the fully-equilibrated state of the model without annihilation.  At time $t = 0$ the annihilation is turned on, and the total number of arcs starts to decline.  However, the distribution of arc sizes retains the dependence $1/s^2$ (illustrated by the dashed line).  All data in this plot correspond to a simulation of 5000 Majoranas at time $t=0$, and is averaged over 1000 independent realizations.}
\label{fig:Ps-annihilation}
\end{figure}

\section{Simulation of the entanglement structure in the diffusion-annihilation model} 
\label{app:annihilationS}

As discussed in Sec.\ \ref{sec:entanglementwithannihilation}, the entanglement structure of the diffusion-annihilation process with Majoranas depends on the initial state. If the initial state is dimerized, or otherwise has a finite typical arc length $\xi$, then the distribution of arc sizes $P(s)$ quickly settles to a fixed distribution with a finite, short range.  (Here $s$ refers to the integer difference in the index $i$ of paired Majoranas, so that $s - 1$ is the number of Majoranas that lie underneath a particular arc.)  On the other hand, if the initial state is long-ranged, such as the stationary distribution $P_0(s) \propto 1/s^2$ of the model without annihilation, then the distribution $P_t(s)$ at a later time $t$ remains long-ranged for all $t$. The crude advection picture in Sec.~\ref{sec:entanglementwithannihilation} can be argued to be self-consistent for $\alpha>2$, and for large $s$ it gives:
\be
P_t(s)\sim \frac{1}{f_t^2} P_0 (s/f_t)
\ee
where $f_t = \rho(t)/\rho(0)$ is the fraction of particles that have not been annihilated. This suggests that if $P_0(s)$ is a power law $s^{-\alpha}$ at large $s$ then the exponent is preserved under the dynamics (and in the special case $\alpha=2$ the amplitude is also preserved).

In this Appendix we provide simulation results to support these claims by examining the cases where $P_0(s)$ is either short-ranged or given by the critical distribution $P_0(s)\sim s^{-2}$.

Figure \ref{fig:Ps-annihilation}(a) shows the distribution $P(s)$ as a function of time, with time $t = 0$ corresponding to a dimerized state, such that $s = 1$ for all pairs.  As time increases, $P(s)$ settles into a limiting distribution ${P(s) \propto \exp(-s/s_0)}$, with the characteristic range ${s_0 \approx 2.1}$.  

Figure \ref{fig:Ps-annihilation}(b), on the other hand, shows the time evolution of $P(s)$ starting from the fully-equilibrated state of the model without annihilation.  At time $t = 0$, the annihilation is turned on, and the number of Majoranas begins to decline. However, the distribution $P(s)$ retains its functional form $\sim 1/s^2$, and is cut off only by the total number of Majoranas remaining in the system.

\bibliography{majoranamodel}

\begin{thebibliography}{94}%
\makeatletter
\providecommand \@ifxundefined [1]{%
 \@ifx{#1\undefined}
}%
\providecommand \@ifnum [1]{%
 \ifnum #1\expandafter \@firstoftwo
 \else \expandafter \@secondoftwo
 \fi
}%
\providecommand \@ifx [1]{%
 \ifx #1\expandafter \@firstoftwo
 \else \expandafter \@secondoftwo
 \fi
}%
\providecommand \natexlab [1]{#1}%
\providecommand \enquote  [1]{``#1''}%
\providecommand \bibnamefont  [1]{#1}%
\providecommand \bibfnamefont [1]{#1}%
\providecommand \citenamefont [1]{#1}%
\providecommand \href@noop [0]{\@secondoftwo}%
\providecommand \href [0]{\begingroup \@sanitize@url \@href}%
\providecommand \@href[1]{\@@startlink{#1}\@@href}%
\providecommand \@@href[1]{\endgroup#1\@@endlink}%
\providecommand \@sanitize@url [0]{\catcode `\\12\catcode `\$12\catcode
  `\&12\catcode `\#12\catcode `\^12\catcode `\_12\catcode `\%12\relax}%
\providecommand \@@startlink[1]{}%
\providecommand \@@endlink[0]{}%
\providecommand \url  [0]{\begingroup\@sanitize@url \@url }%
\providecommand \@url [1]{\endgroup\@href {#1}{\urlprefix }}%
\providecommand \urlprefix  [0]{URL }%
\providecommand \Eprint [0]{\href }%
\providecommand \doibase [0]{http://dx.doi.org/}%
\providecommand \selectlanguage [0]{\@gobble}%
\providecommand \bibinfo  [0]{\@secondoftwo}%
\providecommand \bibfield  [0]{\@secondoftwo}%
\providecommand \translation [1]{[#1]}%
\providecommand \BibitemOpen [0]{}%
\providecommand \bibitemStop [0]{}%
\providecommand \bibitemNoStop [0]{.\EOS\space}%
\providecommand \EOS [0]{\spacefactor3000\relax}%
\providecommand \BibitemShut  [1]{\csname bibitem#1\endcsname}%
\let\auto@bib@innerbib\@empty
%</preamble>
\bibitem [{\citenamefont {Kitaev}(2001)}]{kitaev2001unpaired}%
  \BibitemOpen
  \bibfield  {author} {\bibinfo {author} {\bibfnamefont {A~Yu}\ \bibnamefont
  {Kitaev}},\ }\bibfield  {title} {\enquote {\bibinfo {title} {Unpaired
  majorana fermions in quantum wires},}\ }\href@noop {} {\bibfield  {journal}
  {\bibinfo  {journal} {Physics-Uspekhi}\ }\textbf {\bibinfo {volume} {44}},\
  \bibinfo {pages} {131} (\bibinfo {year} {2001})}\BibitemShut {NoStop}%
\bibitem [{\citenamefont {Nayak}\ \emph {et~al.}(2008)\citenamefont {Nayak},
  \citenamefont {Simon}, \citenamefont {Stern}, \citenamefont {Freedman},\ and\
  \citenamefont {Sarma}}]{nayak2008non}%
  \BibitemOpen
  \bibfield  {author} {\bibinfo {author} {\bibfnamefont {Chetan}\ \bibnamefont
  {Nayak}}, \bibinfo {author} {\bibfnamefont {Steven~H}\ \bibnamefont {Simon}},
  \bibinfo {author} {\bibfnamefont {Ady}\ \bibnamefont {Stern}}, \bibinfo
  {author} {\bibfnamefont {Michael}\ \bibnamefont {Freedman}}, \ and\ \bibinfo
  {author} {\bibfnamefont {Sankar~Das}\ \bibnamefont {Sarma}},\ }\bibfield
  {title} {\enquote {\bibinfo {title} {Non-abelian anyons and topological
  quantum computation},}\ }\href@noop {} {\bibfield  {journal} {\bibinfo
  {journal} {Reviews of Modern Physics}\ }\textbf {\bibinfo {volume} {80}},\
  \bibinfo {pages} {1083} (\bibinfo {year} {2008})}\BibitemShut {NoStop}%
\bibitem [{\citenamefont {Glauber}(1963)}]{glauber1963time}%
  \BibitemOpen
  \bibfield  {author} {\bibinfo {author} {\bibfnamefont {Roy~J}\ \bibnamefont
  {Glauber}},\ }\bibfield  {title} {\enquote {\bibinfo {title} {Time-dependent
  statistics of the ising model},}\ }\href@noop {} {\bibfield  {journal}
  {\bibinfo  {journal} {Journal of mathematical physics}\ }\textbf {\bibinfo
  {volume} {4}},\ \bibinfo {pages} {294--307} (\bibinfo {year}
  {1963})}\BibitemShut {NoStop}%
\bibitem [{\citenamefont {Tauber}\ \emph {et~al.}(2005)\citenamefont {Tauber},
  \citenamefont {Howard},\ and\ \citenamefont {Vollmayr-Lee}}]{Tauber_2005}%
  \BibitemOpen
  \bibfield  {author} {\bibinfo {author} {\bibfnamefont {Uwe~C}\ \bibnamefont
  {Tauber}}, \bibinfo {author} {\bibfnamefont {Martin}\ \bibnamefont {Howard}},
  \ and\ \bibinfo {author} {\bibfnamefont {Benjamin~P}\ \bibnamefont
  {Vollmayr-Lee}},\ }\bibfield  {title} {\enquote {\bibinfo {title}
  {Applications of field-theoretic renormalization group methods to
  reaction{\textendash}diffusion problems},}\ }\href {\doibase
  10.1088/0305-4470/38/17/r01} {\bibfield  {journal} {\bibinfo  {journal}
  {Journal of Physics A: Mathematical and General}\ }\textbf {\bibinfo {volume}
  {38}},\ \bibinfo {pages} {R79--R131} (\bibinfo {year} {2005})}\BibitemShut
  {NoStop}%
\bibitem [{\citenamefont {Bramson}\ and\ \citenamefont
  {Griffeath}(1980)}]{bramson1980clustering}%
  \BibitemOpen
  \bibfield  {author} {\bibinfo {author} {\bibfnamefont {Maury}\ \bibnamefont
  {Bramson}}\ and\ \bibinfo {author} {\bibfnamefont {David}\ \bibnamefont
  {Griffeath}},\ }\bibfield  {title} {\enquote {\bibinfo {title} {Clustering
  and dispersion rates for some interacting particle systems on z},}\
  }\href@noop {} {\bibfield  {journal} {\bibinfo  {journal} {The Annals of
  Probability}\ }\textbf {\bibinfo {volume} {8}},\ \bibinfo {pages} {183--213}
  (\bibinfo {year} {1980})}\BibitemShut {NoStop}%
\bibitem [{\citenamefont {Torney}\ and\ \citenamefont
  {McConnell}(1983)}]{torney1983diffusion}%
  \BibitemOpen
  \bibfield  {author} {\bibinfo {author} {\bibfnamefont {David~C}\ \bibnamefont
  {Torney}}\ and\ \bibinfo {author} {\bibfnamefont {Harden~M}\ \bibnamefont
  {McConnell}},\ }\bibfield  {title} {\enquote {\bibinfo {title}
  {Diffusion-limited reactions in one dimension},}\ }\href@noop {} {\bibfield
  {journal} {\bibinfo  {journal} {The Journal of Physical Chemistry}\ }\textbf
  {\bibinfo {volume} {87}},\ \bibinfo {pages} {1941--1951} (\bibinfo {year}
  {1983})}\BibitemShut {NoStop}%
\bibitem [{\citenamefont {Lushnikov}(1987)}]{lushnikov1987binary}%
  \BibitemOpen
  \bibfield  {author} {\bibinfo {author} {\bibfnamefont {AA}~\bibnamefont
  {Lushnikov}},\ }\bibfield  {title} {\enquote {\bibinfo {title} {Binary
  reaction 1+ 1→ 0 in one dimension},}\ }\href@noop {} {\bibfield  {journal}
  {\bibinfo  {journal} {Physics Letters A}\ }\textbf {\bibinfo {volume}
  {120}},\ \bibinfo {pages} {135--137} (\bibinfo {year} {1987})}\BibitemShut
  {NoStop}%
\bibitem [{\citenamefont {Bray}(1990)}]{Bray_1990}%
  \BibitemOpen
  \bibfield  {author} {\bibinfo {author} {\bibfnamefont {A~J}\ \bibnamefont
  {Bray}},\ }\bibfield  {title} {\enquote {\bibinfo {title} {Universal scaling
  function for domain growth in the glauber-ising chain},}\ }\href {\doibase
  10.1088/0305-4470/23/2/005} {\bibfield  {journal} {\bibinfo  {journal}
  {Journal of Physics A: Mathematical and General}\ }\textbf {\bibinfo {volume}
  {23}},\ \bibinfo {pages} {L67--L72} (\bibinfo {year} {1990})}\BibitemShut
  {NoStop}%
\bibitem [{\citenamefont {Derrida}\ \emph {et~al.}(1994)\citenamefont
  {Derrida}, \citenamefont {Bray},\ and\ \citenamefont
  {Godreche}}]{derrida1994non}%
  \BibitemOpen
  \bibfield  {author} {\bibinfo {author} {\bibfnamefont {B}~\bibnamefont
  {Derrida}}, \bibinfo {author} {\bibfnamefont {AJ}~\bibnamefont {Bray}}, \
  and\ \bibinfo {author} {\bibfnamefont {C}~\bibnamefont {Godreche}},\
  }\bibfield  {title} {\enquote {\bibinfo {title} {Non-trivial exponents in the
  zero temperature dynamics of the 1d ising and potts models},}\ }\href@noop {}
  {\bibfield  {journal} {\bibinfo  {journal} {Journal of Physics A:
  Mathematical and General}\ }\textbf {\bibinfo {volume} {27}},\ \bibinfo
  {pages} {L357} (\bibinfo {year} {1994})}\BibitemShut {NoStop}%
\bibitem [{\citenamefont {Derrida}\ \emph {et~al.}(1996)\citenamefont
  {Derrida}, \citenamefont {Hakim},\ and\ \citenamefont
  {Pasquier}}]{derrida1996exact}%
  \BibitemOpen
  \bibfield  {author} {\bibinfo {author} {\bibfnamefont {Bernard}\ \bibnamefont
  {Derrida}}, \bibinfo {author} {\bibfnamefont {Vincent}\ \bibnamefont
  {Hakim}}, \ and\ \bibinfo {author} {\bibfnamefont {Vincent}\ \bibnamefont
  {Pasquier}},\ }\bibfield  {title} {\enquote {\bibinfo {title} {Exact exponent
  for the number of persistent spins in the zero-temperature dynamics of the
  one-dimensional potts model},}\ }\href@noop {} {\bibfield  {journal}
  {\bibinfo  {journal} {Journal of statistical physics}\ }\textbf {\bibinfo
  {volume} {85}},\ \bibinfo {pages} {763--797} (\bibinfo {year}
  {1996})}\BibitemShut {NoStop}%
\bibitem [{\citenamefont {Skinner}\ \emph {et~al.}(2019)\citenamefont
  {Skinner}, \citenamefont {Ruhman},\ and\ \citenamefont
  {Nahum}}]{skinner2018measurement}%
  \BibitemOpen
  \bibfield  {author} {\bibinfo {author} {\bibfnamefont {Brian}\ \bibnamefont
  {Skinner}}, \bibinfo {author} {\bibfnamefont {Jonathan}\ \bibnamefont
  {Ruhman}}, \ and\ \bibinfo {author} {\bibfnamefont {Adam}\ \bibnamefont
  {Nahum}},\ }\bibfield  {title} {\enquote {\bibinfo {title}
  {Measurement-induced phase transitions in the dynamics of entanglement},}\
  }\href {\doibase 10.1103/PhysRevX.9.031009} {\bibfield  {journal} {\bibinfo
  {journal} {Phys. Rev. X}\ }\textbf {\bibinfo {volume} {9}},\ \bibinfo {pages}
  {031009} (\bibinfo {year} {2019})}\BibitemShut {NoStop}%
\bibitem [{\citenamefont {Li}\ \emph {et~al.}(2018)\citenamefont {Li},
  \citenamefont {Chen},\ and\ \citenamefont {Fisher}}]{li2018quantum}%
  \BibitemOpen
  \bibfield  {author} {\bibinfo {author} {\bibfnamefont {Yaodong}\ \bibnamefont
  {Li}}, \bibinfo {author} {\bibfnamefont {Xiao}\ \bibnamefont {Chen}}, \ and\
  \bibinfo {author} {\bibfnamefont {Matthew~PA}\ \bibnamefont {Fisher}},\
  }\bibfield  {title} {\enquote {\bibinfo {title} {Quantum zeno effect and the
  many-body entanglement transition},}\ }\href@noop {} {\bibfield  {journal}
  {\bibinfo  {journal} {Physical Review B}\ }\textbf {\bibinfo {volume} {98}},\
  \bibinfo {pages} {205136} (\bibinfo {year} {2018})}\BibitemShut {NoStop}%
\bibitem [{\citenamefont {Chan}\ \emph {et~al.}(2019)\citenamefont {Chan},
  \citenamefont {Nandkishore}, \citenamefont {Pretko},\ and\ \citenamefont
  {Smith}}]{chan2018weak}%
  \BibitemOpen
  \bibfield  {author} {\bibinfo {author} {\bibfnamefont {Amos}\ \bibnamefont
  {Chan}}, \bibinfo {author} {\bibfnamefont {Rahul~M.}\ \bibnamefont
  {Nandkishore}}, \bibinfo {author} {\bibfnamefont {Michael}\ \bibnamefont
  {Pretko}}, \ and\ \bibinfo {author} {\bibfnamefont {Graeme}\ \bibnamefont
  {Smith}},\ }\bibfield  {title} {\enquote {\bibinfo {title}
  {Unitary-projective entanglement dynamics},}\ }\href {\doibase
  10.1103/PhysRevB.99.224307} {\bibfield  {journal} {\bibinfo  {journal} {Phys.
  Rev. B}\ }\textbf {\bibinfo {volume} {99}},\ \bibinfo {pages} {224307}
  (\bibinfo {year} {2019})}\BibitemShut {NoStop}%
\bibitem [{\citenamefont {Choi}\ \emph {et~al.}(2019)\citenamefont {Choi},
  \citenamefont {Bao}, \citenamefont {Qi},\ and\ \citenamefont
  {Altman}}]{choi2019quantum}%
  \BibitemOpen
  \bibfield  {author} {\bibinfo {author} {\bibfnamefont {Soonwon}\ \bibnamefont
  {Choi}}, \bibinfo {author} {\bibfnamefont {Yimu}\ \bibnamefont {Bao}},
  \bibinfo {author} {\bibfnamefont {Xiao-Liang}\ \bibnamefont {Qi}}, \ and\
  \bibinfo {author} {\bibfnamefont {Ehud}\ \bibnamefont {Altman}},\ }\bibfield
  {title} {\enquote {\bibinfo {title} {Quantum error correction and
  entanglement phase transition in random unitary circuits with projective
  measurements},}\ }\href@noop {} {\bibfield  {journal} {\bibinfo  {journal}
  {arXiv preprint arXiv:1903.05124}\ } (\bibinfo {year} {2019})}\BibitemShut
  {NoStop}%
\bibitem [{\citenamefont {Szyniszewski}\ \emph {et~al.}(2019)\citenamefont
  {Szyniszewski}, \citenamefont {Romito},\ and\ \citenamefont
  {Schomerus}}]{szyniszewski2019entanglement}%
  \BibitemOpen
  \bibfield  {author} {\bibinfo {author} {\bibfnamefont {M.}~\bibnamefont
  {Szyniszewski}}, \bibinfo {author} {\bibfnamefont {A.}~\bibnamefont
  {Romito}}, \ and\ \bibinfo {author} {\bibfnamefont {H.}~\bibnamefont
  {Schomerus}},\ }\bibfield  {title} {\enquote {\bibinfo {title} {Entanglement
  transition from variable-strength weak measurements},}\ }\href {\doibase
  10.1103/PhysRevB.100.064204} {\bibfield  {journal} {\bibinfo  {journal}
  {Phys. Rev. B}\ }\textbf {\bibinfo {volume} {100}},\ \bibinfo {pages}
  {064204} (\bibinfo {year} {2019})}\BibitemShut {NoStop}%
\bibitem [{\citenamefont {Li}\ \emph {et~al.}(2019)\citenamefont {Li},
  \citenamefont {Chen},\ and\ \citenamefont {Fisher}}]{li2019measurement}%
  \BibitemOpen
  \bibfield  {author} {\bibinfo {author} {\bibfnamefont {Yaodong}\ \bibnamefont
  {Li}}, \bibinfo {author} {\bibfnamefont {Xiao}\ \bibnamefont {Chen}}, \ and\
  \bibinfo {author} {\bibfnamefont {Matthew}\ \bibnamefont {Fisher}},\
  }\bibfield  {title} {\enquote {\bibinfo {title} {Measurement-driven
  entanglement transition in hybrid quantum circuits},}\ }\href@noop {}
  {\bibfield  {journal} {\bibinfo  {journal} {arXiv preprint arXiv:1901.08092}\
  } (\bibinfo {year} {2019})}\BibitemShut {NoStop}%
\bibitem [{\citenamefont {Cao}\ \emph {et~al.}(2019)\citenamefont {Cao},
  \citenamefont {Tilloy},\ and\ \citenamefont {Luca}}]{cao2018collective}%
  \BibitemOpen
  \bibfield  {author} {\bibinfo {author} {\bibfnamefont {Xiangyu}\ \bibnamefont
  {Cao}}, \bibinfo {author} {\bibfnamefont {Antoine}\ \bibnamefont {Tilloy}}, \
  and\ \bibinfo {author} {\bibfnamefont {Andrea~De}\ \bibnamefont {Luca}},\
  }\bibfield  {title} {\enquote {\bibinfo {title} {{Entanglement in a fermion
  chain under continuous monitoring}},}\ }\href {\doibase
  10.21468/SciPostPhys.7.2.024} {\bibfield  {journal} {\bibinfo  {journal}
  {SciPost Phys.}\ }\textbf {\bibinfo {volume} {7}},\ \bibinfo {pages} {24}
  (\bibinfo {year} {2019})}\BibitemShut {NoStop}%
\bibitem [{\citenamefont {Gullans}\ and\ \citenamefont
  {Huse}(2019{\natexlab{a}})}]{gullans2019dynamical}%
  \BibitemOpen
  \bibfield  {author} {\bibinfo {author} {\bibfnamefont {Michael~J}\
  \bibnamefont {Gullans}}\ and\ \bibinfo {author} {\bibfnamefont {David~A}\
  \bibnamefont {Huse}},\ }\bibfield  {title} {\enquote {\bibinfo {title}
  {Dynamical purification phase transition induced by quantum measurements},}\
  }\href@noop {} {\bibfield  {journal} {\bibinfo  {journal} {arXiv preprint
  arXiv:1905.05195}\ } (\bibinfo {year} {2019}{\natexlab{a}})}\BibitemShut
  {NoStop}%
\bibitem [{\citenamefont {Gullans}\ and\ \citenamefont
  {Huse}(2019{\natexlab{b}})}]{gullans2019scalable}%
  \BibitemOpen
  \bibfield  {author} {\bibinfo {author} {\bibfnamefont {Michael~J}\
  \bibnamefont {Gullans}}\ and\ \bibinfo {author} {\bibfnamefont {David~A}\
  \bibnamefont {Huse}},\ }\bibfield  {title} {\enquote {\bibinfo {title}
  {Scalable probes of measurement-induced criticality},}\ }\href@noop {}
  {\bibfield  {journal} {\bibinfo  {journal} {arXiv preprint arXiv:1910.00020}\
  } (\bibinfo {year} {2019}{\natexlab{b}})}\BibitemShut {NoStop}%
\bibitem [{\citenamefont {Tang}\ and\ \citenamefont
  {Zhu}(2019)}]{tang2019measurement}%
  \BibitemOpen
  \bibfield  {author} {\bibinfo {author} {\bibfnamefont {Qicheng}\ \bibnamefont
  {Tang}}\ and\ \bibinfo {author} {\bibfnamefont {W}~\bibnamefont {Zhu}},\
  }\bibfield  {title} {\enquote {\bibinfo {title} {Measurement-induced phase
  transition: A case study in the non-integrable model by density-matrix
  renormalization group calculations},}\ }\href@noop {} {\bibfield  {journal}
  {\bibinfo  {journal} {arXiv preprint arXiv:1908.11253}\ } (\bibinfo {year}
  {2019})}\BibitemShut {NoStop}%
\bibitem [{\citenamefont {Zhou}\ and\ \citenamefont {Nahum}(2019)}]{ZhouNahum}%
  \BibitemOpen
  \bibfield  {author} {\bibinfo {author} {\bibfnamefont {Tianci}\ \bibnamefont
  {Zhou}}\ and\ \bibinfo {author} {\bibfnamefont {Adam}\ \bibnamefont
  {Nahum}},\ }\bibfield  {title} {\enquote {\bibinfo {title} {Emergent
  statistical mechanics of entanglement in random unitary circuits},}\ }\href
  {\doibase 10.1103/PhysRevB.99.174205} {\bibfield  {journal} {\bibinfo
  {journal} {Phys. Rev. B}\ }\textbf {\bibinfo {volume} {99}},\ \bibinfo
  {pages} {174205} (\bibinfo {year} {2019})}\BibitemShut {NoStop}%
\bibitem [{\citenamefont {Vasseur}\ \emph {et~al.}(2019)\citenamefont
  {Vasseur}, \citenamefont {Potter}, \citenamefont {You},\ and\ \citenamefont
  {Ludwig}}]{vasseur2018entanglement}%
  \BibitemOpen
  \bibfield  {author} {\bibinfo {author} {\bibfnamefont {Romain}\ \bibnamefont
  {Vasseur}}, \bibinfo {author} {\bibfnamefont {Andrew~C.}\ \bibnamefont
  {Potter}}, \bibinfo {author} {\bibfnamefont {Yi-Zhuang}\ \bibnamefont {You}},
  \ and\ \bibinfo {author} {\bibfnamefont {Andreas W.~W.}\ \bibnamefont
  {Ludwig}},\ }\bibfield  {title} {\enquote {\bibinfo {title} {Entanglement
  transitions from holographic random tensor networks},}\ }\href {\doibase
  10.1103/PhysRevB.100.134203} {\bibfield  {journal} {\bibinfo  {journal}
  {Phys. Rev. B}\ }\textbf {\bibinfo {volume} {100}},\ \bibinfo {pages}
  {134203} (\bibinfo {year} {2019})}\BibitemShut {NoStop}%
\bibitem [{\citenamefont {Bao}\ \emph {et~al.}(2019)\citenamefont {Bao},
  \citenamefont {Choi},\ and\ \citenamefont {Altman}}]{bao2019theory}%
  \BibitemOpen
  \bibfield  {author} {\bibinfo {author} {\bibfnamefont {Yimu}\ \bibnamefont
  {Bao}}, \bibinfo {author} {\bibfnamefont {Soonwon}\ \bibnamefont {Choi}}, \
  and\ \bibinfo {author} {\bibfnamefont {Ehud}\ \bibnamefont {Altman}},\
  }\bibfield  {title} {\enquote {\bibinfo {title} {Theory of the phase
  transition in random unitary circuits with measurements},}\ }\href@noop {}
  {\bibfield  {journal} {\bibinfo  {journal} {arXiv preprint arXiv:1908.04305}\
  } (\bibinfo {year} {2019})}\BibitemShut {NoStop}%
\bibitem [{\citenamefont {Jian}\ \emph {et~al.}(2019)\citenamefont {Jian},
  \citenamefont {You}, \citenamefont {Vasseur},\ and\ \citenamefont
  {Ludwig}}]{jian2019measurement}%
  \BibitemOpen
  \bibfield  {author} {\bibinfo {author} {\bibfnamefont {Chao-Ming}\
  \bibnamefont {Jian}}, \bibinfo {author} {\bibfnamefont {Yi-Zhuang}\
  \bibnamefont {You}}, \bibinfo {author} {\bibfnamefont {Romain}\ \bibnamefont
  {Vasseur}}, \ and\ \bibinfo {author} {\bibfnamefont {Andreas~WW}\
  \bibnamefont {Ludwig}},\ }\bibfield  {title} {\enquote {\bibinfo {title}
  {Measurement-induced criticality in random quantum circuits},}\ }\href@noop
  {} {\bibfield  {journal} {\bibinfo  {journal} {arXiv preprint
  arXiv:1908.08051}\ } (\bibinfo {year} {2019})}\BibitemShut {NoStop}%
\bibitem [{\citenamefont {Bhatt}\ and\ \citenamefont {Lee}(1981)}]{Lee1981}%
  \BibitemOpen
  \bibfield  {author} {\bibinfo {author} {\bibfnamefont {R.~N.}\ \bibnamefont
  {Bhatt}}\ and\ \bibinfo {author} {\bibfnamefont {P.~A.}\ \bibnamefont
  {Lee}},\ }\bibfield  {title} {\enquote {\bibinfo {title} {A scaling method
  for low temperature behavior of random antiferromagnetic systems
  (invited)},}\ }\href {\doibase 10.1063/1.329684} {\bibfield  {journal}
  {\bibinfo  {journal} {Journal of Applied Physics}\ }\textbf {\bibinfo
  {volume} {52}},\ \bibinfo {pages} {1703--1707} (\bibinfo {year} {1981})},\
  \Eprint
  {http://arxiv.org/abs/http://aip.scitation.org/doi/pdf/10.1063/1.329684}
  {http://aip.scitation.org/doi/pdf/10.1063/1.329684} \BibitemShut {NoStop}%
\bibitem [{\citenamefont {Fisher}(1999)}]{fisher1999phase}%
  \BibitemOpen
  \bibfield  {author} {\bibinfo {author} {\bibfnamefont {Daniel~S}\
  \bibnamefont {Fisher}},\ }\bibfield  {title} {\enquote {\bibinfo {title}
  {Phase transitions and singularities in random quantum systems},}\
  }\href@noop {} {\bibfield  {journal} {\bibinfo  {journal} {Physica A:
  Statistical Mechanics and its Applications}\ }\textbf {\bibinfo {volume}
  {263}},\ \bibinfo {pages} {222--233} (\bibinfo {year} {1999})}\BibitemShut
  {NoStop}%
\bibitem [{\citenamefont {Temperley}\ and\ \citenamefont
  {Lieb}(1971)}]{TemperleyLieb}%
  \BibitemOpen
  \bibfield  {author} {\bibinfo {author} {\bibfnamefont {H.~N.~V.}\
  \bibnamefont {Temperley}}\ and\ \bibinfo {author} {\bibfnamefont {H.~E.}\
  \bibnamefont {Lieb}},\ }\bibfield  {title} {\enquote {\bibinfo {title}
  {Relations between the 'percolation' and 'colouring' problem and other
  graph-theoretical problems associated with regular planar lattices: some
  exact results for the 'percolation' problem},}\ }\href {\doibase
  10.1098/rspa.1971.0067} {\bibfield  {journal} {\bibinfo  {journal}
  {Proceedings of the Royal Society of London. A. Mathematical and Physical
  Sciences}\ }\textbf {\bibinfo {volume} {322}},\ \bibinfo {pages} {251--280}
  (\bibinfo {year} {1971})}\BibitemShut {NoStop}%
\bibitem [{\citenamefont {Blote}\ and\ \citenamefont
  {Nienhuis}(1989)}]{blote1989critical}%
  \BibitemOpen
  \bibfield  {author} {\bibinfo {author} {\bibfnamefont {HWJ}\ \bibnamefont
  {Blote}}\ and\ \bibinfo {author} {\bibfnamefont {Bernard}\ \bibnamefont
  {Nienhuis}},\ }\bibfield  {title} {\enquote {\bibinfo {title} {Critical
  behaviour and conformal anomaly of the o (n) model on the square lattice},}\
  }\href@noop {} {\bibfield  {journal} {\bibinfo  {journal} {Journal of Physics
  A: Mathematical and General}\ }\textbf {\bibinfo {volume} {22}},\ \bibinfo
  {pages} {1415} (\bibinfo {year} {1989})}\BibitemShut {NoStop}%
\bibitem [{\citenamefont {Cardy}(2005)}]{cardy2005sle}%
  \BibitemOpen
  \bibfield  {author} {\bibinfo {author} {\bibfnamefont {John}\ \bibnamefont
  {Cardy}},\ }\bibfield  {title} {\enquote {\bibinfo {title} {Sle for
  theoretical physicists},}\ }\href@noop {} {\bibfield  {journal} {\bibinfo
  {journal} {Annals of Physics}\ }\textbf {\bibinfo {volume} {318}},\ \bibinfo
  {pages} {81--118} (\bibinfo {year} {2005})}\BibitemShut {NoStop}%
\bibitem [{\citenamefont {Jacobsen}(2009)}]{jacobsen2009conformal}%
  \BibitemOpen
  \bibfield  {author} {\bibinfo {author} {\bibfnamefont {Jesper~Lykke}\
  \bibnamefont {Jacobsen}},\ }\bibfield  {title} {\enquote {\bibinfo {title}
  {Conformal field theory applied to loop models},}\ }in\ \href@noop {} {\emph
  {\bibinfo {booktitle} {Polygons, polyominoes and polycubes}}}\ (\bibinfo
  {publisher} {Springer},\ \bibinfo {year} {2009})\ pp.\ \bibinfo {pages}
  {347--424}\BibitemShut {NoStop}%
\bibitem [{\citenamefont {Saleur}\ and\ \citenamefont
  {Duplantier}(1987)}]{saleur1987exact}%
  \BibitemOpen
  \bibfield  {author} {\bibinfo {author} {\bibfnamefont {Hubert}\ \bibnamefont
  {Saleur}}\ and\ \bibinfo {author} {\bibfnamefont {Bertrand}\ \bibnamefont
  {Duplantier}},\ }\bibfield  {title} {\enquote {\bibinfo {title} {Exact
  determination of the percolation hull exponent in two dimensions},}\
  }\href@noop {} {\bibfield  {journal} {\bibinfo  {journal} {Physical review
  letters}\ }\textbf {\bibinfo {volume} {58}},\ \bibinfo {pages} {2325}
  (\bibinfo {year} {1987})}\BibitemShut {NoStop}%
\bibitem [{\citenamefont {Cardy}(2000)}]{cardy2000linking}%
  \BibitemOpen
  \bibfield  {author} {\bibinfo {author} {\bibfnamefont {John}\ \bibnamefont
  {Cardy}},\ }\bibfield  {title} {\enquote {\bibinfo {title} {Linking numbers
  for self-avoiding loops and percolation: Application to the spin quantum hall
  transition},}\ }\href@noop {} {\bibfield  {journal} {\bibinfo  {journal}
  {Physical review letters}\ }\textbf {\bibinfo {volume} {84}},\ \bibinfo
  {pages} {3507} (\bibinfo {year} {2000})}\BibitemShut {NoStop}%
\bibitem [{\citenamefont {Pearce}\ \emph {et~al.}(2002)\citenamefont {Pearce},
  \citenamefont {Rittenberg}, \citenamefont {De~Gier},\ and\ \citenamefont
  {Nienhuis}}]{pearce2002temperley}%
  \BibitemOpen
  \bibfield  {author} {\bibinfo {author} {\bibfnamefont {Paul~A}\ \bibnamefont
  {Pearce}}, \bibinfo {author} {\bibfnamefont {Vladimir}\ \bibnamefont
  {Rittenberg}}, \bibinfo {author} {\bibfnamefont {Jan}\ \bibnamefont
  {De~Gier}}, \ and\ \bibinfo {author} {\bibfnamefont {Bernard}\ \bibnamefont
  {Nienhuis}},\ }\bibfield  {title} {\enquote {\bibinfo {title}
  {Temperley--lieb stochastic processes},}\ }\href@noop {} {\bibfield
  {journal} {\bibinfo  {journal} {Journal of Physics A: Mathematical and
  General}\ }\textbf {\bibinfo {volume} {35}},\ \bibinfo {pages} {L661}
  (\bibinfo {year} {2002})}\BibitemShut {NoStop}%
\bibitem [{\citenamefont {de~Gier}\ \emph {et~al.}(2003)\citenamefont
  {de~Gier}, \citenamefont {Nienhuis}, \citenamefont {Pearce},\ and\
  \citenamefont {Rittenberg}}]{deGierRaiseandPeel1}%
  \BibitemOpen
  \bibfield  {author} {\bibinfo {author} {\bibfnamefont {Jan}\ \bibnamefont
  {de~Gier}}, \bibinfo {author} {\bibfnamefont {Bernard}\ \bibnamefont
  {Nienhuis}}, \bibinfo {author} {\bibfnamefont {Paul~A.}\ \bibnamefont
  {Pearce}}, \ and\ \bibinfo {author} {\bibfnamefont {Vladimir}\ \bibnamefont
  {Rittenberg}},\ }\bibfield  {title} {\enquote {\bibinfo {title} {Stochastic
  processes and conformal invariance},}\ }\href {\doibase
  10.1103/PhysRevE.67.016101} {\bibfield  {journal} {\bibinfo  {journal} {Phys.
  Rev. E}\ }\textbf {\bibinfo {volume} {67}},\ \bibinfo {pages} {016101}
  (\bibinfo {year} {2003})}\BibitemShut {NoStop}%
\bibitem [{\citenamefont {de~Gier}\ \emph {et~al.}(2004)\citenamefont
  {de~Gier}, \citenamefont {Nienhuis}, \citenamefont {Pearce},\ and\
  \citenamefont {Rittenberg}}]{deGierRaiseandPeel2}%
  \BibitemOpen
  \bibfield  {author} {\bibinfo {author} {\bibfnamefont {Jan}\ \bibnamefont
  {de~Gier}}, \bibinfo {author} {\bibfnamefont {Bernard}\ \bibnamefont
  {Nienhuis}}, \bibinfo {author} {\bibfnamefont {Paul~A.}\ \bibnamefont
  {Pearce}}, \ and\ \bibinfo {author} {\bibfnamefont {Vladimir}\ \bibnamefont
  {Rittenberg}},\ }\bibfield  {title} {\enquote {\bibinfo {title} {The raise
  and peel model of a fluctuating interface},}\ }\href {\doibase
  10.1023/B:JOSS.0000003102.81727.fd} {\bibfield  {journal} {\bibinfo
  {journal} {Journal of Statistical Physics}\ }\textbf {\bibinfo {volume}
  {114}},\ \bibinfo {pages} {1--35} (\bibinfo {year} {2004})}\BibitemShut
  {NoStop}%
\bibitem [{\citenamefont {Kimchi}\ \emph {et~al.}(2018)\citenamefont {Kimchi},
  \citenamefont {Nahum},\ and\ \citenamefont {Senthil}}]{kimchi2018valence}%
  \BibitemOpen
  \bibfield  {author} {\bibinfo {author} {\bibfnamefont {Itamar}\ \bibnamefont
  {Kimchi}}, \bibinfo {author} {\bibfnamefont {Adam}\ \bibnamefont {Nahum}}, \
  and\ \bibinfo {author} {\bibfnamefont {T}~\bibnamefont {Senthil}},\
  }\bibfield  {title} {\enquote {\bibinfo {title} {Valence bonds in random
  quantum magnets: theory and application to ybmggao 4},}\ }\href@noop {}
  {\bibfield  {journal} {\bibinfo  {journal} {Physical Review X}\ }\textbf
  {\bibinfo {volume} {8}},\ \bibinfo {pages} {031028} (\bibinfo {year}
  {2018})}\BibitemShut {NoStop}%
\bibitem [{\citenamefont {Nahum}\ \emph {et~al.}(2011)\citenamefont {Nahum},
  \citenamefont {Chalker}, \citenamefont {Serna}, \citenamefont {Ortuno},\ and\
  \citenamefont {Somoza}}]{nahum20113d}%
  \BibitemOpen
  \bibfield  {author} {\bibinfo {author} {\bibfnamefont {Adam}\ \bibnamefont
  {Nahum}}, \bibinfo {author} {\bibfnamefont {JT}~\bibnamefont {Chalker}},
  \bibinfo {author} {\bibfnamefont {P}~\bibnamefont {Serna}}, \bibinfo {author}
  {\bibfnamefont {M}~\bibnamefont {Ortuno}}, \ and\ \bibinfo {author}
  {\bibfnamefont {AM}~\bibnamefont {Somoza}},\ }\bibfield  {title} {\enquote
  {\bibinfo {title} {3d loop models and the cp n- 1 sigma model},}\ }\href@noop
  {} {\bibfield  {journal} {\bibinfo  {journal} {Physical review letters}\
  }\textbf {\bibinfo {volume} {107}},\ \bibinfo {pages} {110601} (\bibinfo
  {year} {2011})}\BibitemShut {NoStop}%
\bibitem [{\citenamefont {Rahmani}\ \emph
  {et~al.}(2015{\natexlab{a}})\citenamefont {Rahmani}, \citenamefont {Zhu},
  \citenamefont {Franz},\ and\ \citenamefont {Affleck}}]{rahmani2015emergent}%
  \BibitemOpen
  \bibfield  {author} {\bibinfo {author} {\bibfnamefont {Armin}\ \bibnamefont
  {Rahmani}}, \bibinfo {author} {\bibfnamefont {Xiaoyu}\ \bibnamefont {Zhu}},
  \bibinfo {author} {\bibfnamefont {Marcel}\ \bibnamefont {Franz}}, \ and\
  \bibinfo {author} {\bibfnamefont {Ian}\ \bibnamefont {Affleck}},\ }\bibfield
  {title} {\enquote {\bibinfo {title} {Emergent supersymmetry from strongly
  interacting majorana zero modes},}\ }\href@noop {} {\bibfield  {journal}
  {\bibinfo  {journal} {Physical review letters}\ }\textbf {\bibinfo {volume}
  {115}},\ \bibinfo {pages} {166401} (\bibinfo {year}
  {2015}{\natexlab{a}})}\BibitemShut {NoStop}%
\bibitem [{\citenamefont {Rahmani}\ \emph
  {et~al.}(2015{\natexlab{b}})\citenamefont {Rahmani}, \citenamefont {Zhu},
  \citenamefont {Franz},\ and\ \citenamefont {Affleck}}]{rahmani2015phase}%
  \BibitemOpen
  \bibfield  {author} {\bibinfo {author} {\bibfnamefont {Armin}\ \bibnamefont
  {Rahmani}}, \bibinfo {author} {\bibfnamefont {Xiaoyu}\ \bibnamefont {Zhu}},
  \bibinfo {author} {\bibfnamefont {Marcel}\ \bibnamefont {Franz}}, \ and\
  \bibinfo {author} {\bibfnamefont {Ian}\ \bibnamefont {Affleck}},\ }\bibfield
  {title} {\enquote {\bibinfo {title} {Phase diagram of the interacting
  majorana chain model},}\ }\href@noop {} {\bibfield  {journal} {\bibinfo
  {journal} {Physical Review B}\ }\textbf {\bibinfo {volume} {92}},\ \bibinfo
  {pages} {235123} (\bibinfo {year} {2015}{\natexlab{b}})}\BibitemShut
  {NoStop}%
\bibitem [{\citenamefont {O’Brien}\ and\ \citenamefont
  {Fendley}(2018)}]{obrien2018lattice}%
  \BibitemOpen
  \bibfield  {author} {\bibinfo {author} {\bibfnamefont {Edward}\ \bibnamefont
  {O’Brien}}\ and\ \bibinfo {author} {\bibfnamefont {Paul}\ \bibnamefont
  {Fendley}},\ }\bibfield  {title} {\enquote {\bibinfo {title} {Lattice
  supersymmetry and order-disorder coexistence in the tricritical ising
  model},}\ }\href@noop {} {\bibfield  {journal} {\bibinfo  {journal} {Physical
  review letters}\ }\textbf {\bibinfo {volume} {120}},\ \bibinfo {pages}
  {206403} (\bibinfo {year} {2018})}\BibitemShut {NoStop}%
\bibitem [{\citenamefont {Grover}\ \emph {et~al.}(2014)\citenamefont {Grover},
  \citenamefont {Sheng},\ and\ \citenamefont
  {Vishwanath}}]{grover2014emergent}%
  \BibitemOpen
  \bibfield  {author} {\bibinfo {author} {\bibfnamefont {Tarun}\ \bibnamefont
  {Grover}}, \bibinfo {author} {\bibfnamefont {DN}~\bibnamefont {Sheng}}, \
  and\ \bibinfo {author} {\bibfnamefont {Ashvin}\ \bibnamefont {Vishwanath}},\
  }\bibfield  {title} {\enquote {\bibinfo {title} {Emergent space-time
  supersymmetry at the boundary of a topological phase},}\ }\href@noop {}
  {\bibfield  {journal} {\bibinfo  {journal} {Science}\ }\textbf {\bibinfo
  {volume} {344}},\ \bibinfo {pages} {280--283} (\bibinfo {year}
  {2014})}\BibitemShut {NoStop}%
\bibitem [{\citenamefont {Jones}\ and\ \citenamefont
  {Metlitski}(2019)}]{jones20191d}%
  \BibitemOpen
  \bibfield  {author} {\bibinfo {author} {\bibfnamefont {Robert~A}\
  \bibnamefont {Jones}}\ and\ \bibinfo {author} {\bibfnamefont {Max~A}\
  \bibnamefont {Metlitski}},\ }\bibfield  {title} {\enquote {\bibinfo {title}
  {1d lattice models for the boundary of 2d" majorana" fermion spts:
  Kramers-wannier duality as an exact $ z\_2 $ symmetry},}\ }\href@noop {}
  {\bibfield  {journal} {\bibinfo  {journal} {arXiv preprint arXiv:1902.05957}\
  } (\bibinfo {year} {2019})}\BibitemShut {NoStop}%
\bibitem [{\citenamefont {Alicea}(2012)}]{alicea2012new}%
  \BibitemOpen
  \bibfield  {author} {\bibinfo {author} {\bibfnamefont {Jason}\ \bibnamefont
  {Alicea}},\ }\bibfield  {title} {\enquote {\bibinfo {title} {New directions
  in the pursuit of majorana fermions in solid state systems},}\ }\href@noop {}
  {\bibfield  {journal} {\bibinfo  {journal} {Reports on progress in physics}\
  }\textbf {\bibinfo {volume} {75}},\ \bibinfo {pages} {076501} (\bibinfo
  {year} {2012})}\BibitemShut {NoStop}%
\bibitem [{\citenamefont {Elliott}\ and\ \citenamefont
  {Franz}(2015)}]{elliott2015colloquium}%
  \BibitemOpen
  \bibfield  {author} {\bibinfo {author} {\bibfnamefont {Steven~R}\
  \bibnamefont {Elliott}}\ and\ \bibinfo {author} {\bibfnamefont {Marcel}\
  \bibnamefont {Franz}},\ }\bibfield  {title} {\enquote {\bibinfo {title}
  {Colloquium: Majorana fermions in nuclear, particle, and solid-state
  physics},}\ }\href@noop {} {\bibfield  {journal} {\bibinfo  {journal}
  {Reviews of Modern Physics}\ }\textbf {\bibinfo {volume} {87}},\ \bibinfo
  {pages} {137} (\bibinfo {year} {2015})}\BibitemShut {NoStop}%
\bibitem [{\citenamefont {Read}\ and\ \citenamefont
  {Green}(2000)}]{read2000paired}%
  \BibitemOpen
  \bibfield  {author} {\bibinfo {author} {\bibfnamefont {Nicholas}\
  \bibnamefont {Read}}\ and\ \bibinfo {author} {\bibfnamefont {Dmitry}\
  \bibnamefont {Green}},\ }\bibfield  {title} {\enquote {\bibinfo {title}
  {Paired states of fermions in two dimensions with breaking of parity and
  time-reversal symmetries and the fractional quantum hall effect},}\
  }\href@noop {} {\bibfield  {journal} {\bibinfo  {journal} {Physical Review
  B}\ }\textbf {\bibinfo {volume} {61}},\ \bibinfo {pages} {10267} (\bibinfo
  {year} {2000})}\BibitemShut {NoStop}%
\bibitem [{\citenamefont {Read}\ and\ \citenamefont
  {Moore}(1992)}]{read1992fractional}%
  \BibitemOpen
  \bibfield  {author} {\bibinfo {author} {\bibfnamefont {N}~\bibnamefont
  {Read}}\ and\ \bibinfo {author} {\bibfnamefont {G}~\bibnamefont {Moore}},\
  }\bibfield  {title} {\enquote {\bibinfo {title} {Fractional quantum hall
  effect and nonabelian statistics},}\ }\href@noop {} {\bibfield  {journal}
  {\bibinfo  {journal} {Progress of Theoretical Physics Supplement}\ }\textbf
  {\bibinfo {volume} {107}},\ \bibinfo {pages} {157--166} (\bibinfo {year}
  {1992})}\BibitemShut {NoStop}%
\bibitem [{\citenamefont {Fu}\ and\ \citenamefont
  {Kane}(2008)}]{fu2008superconducting}%
  \BibitemOpen
  \bibfield  {author} {\bibinfo {author} {\bibfnamefont {Liang}\ \bibnamefont
  {Fu}}\ and\ \bibinfo {author} {\bibfnamefont {Charles~L}\ \bibnamefont
  {Kane}},\ }\bibfield  {title} {\enquote {\bibinfo {title} {Superconducting
  proximity effect and majorana fermions at the surface of a topological
  insulator},}\ }\href@noop {} {\bibfield  {journal} {\bibinfo  {journal}
  {Physical review letters}\ }\textbf {\bibinfo {volume} {100}},\ \bibinfo
  {pages} {096407} (\bibinfo {year} {2008})}\BibitemShut {NoStop}%
\bibitem [{\citenamefont {Lutchyn}\ \emph {et~al.}(2018)\citenamefont
  {Lutchyn}, \citenamefont {Bakkers}, \citenamefont {Kouwenhoven},
  \citenamefont {Krogstrup}, \citenamefont {Marcus},\ and\ \citenamefont
  {Oreg}}]{Lutchyn2018}%
  \BibitemOpen
  \bibfield  {author} {\bibinfo {author} {\bibfnamefont {R.~M.}\ \bibnamefont
  {Lutchyn}}, \bibinfo {author} {\bibfnamefont {E.~P. A.~M.}\ \bibnamefont
  {Bakkers}}, \bibinfo {author} {\bibfnamefont {L.~P.}\ \bibnamefont
  {Kouwenhoven}}, \bibinfo {author} {\bibfnamefont {P.}~\bibnamefont
  {Krogstrup}}, \bibinfo {author} {\bibfnamefont {C.~M.}\ \bibnamefont
  {Marcus}}, \ and\ \bibinfo {author} {\bibfnamefont {Y.}~\bibnamefont
  {Oreg}},\ }\bibfield  {title} {\enquote {\bibinfo {title} {Majorana zero
  modes in superconductor-semiconductor heterostructures},}\ }\href {\doibase
  10.1038/s41578-018-0003-1} {\bibfield  {journal} {\bibinfo  {journal} {Nature
  Reviews Materials}\ }\textbf {\bibinfo {volume} {3}},\ \bibinfo {pages}
  {52--68} (\bibinfo {year} {2018})}\BibitemShut {NoStop}%
\bibitem [{\citenamefont {Hassler}\ and\ \citenamefont
  {Schuricht}(2012)}]{hassler2012strongly}%
  \BibitemOpen
  \bibfield  {author} {\bibinfo {author} {\bibfnamefont {Fabian}\ \bibnamefont
  {Hassler}}\ and\ \bibinfo {author} {\bibfnamefont {Dirk}\ \bibnamefont
  {Schuricht}},\ }\bibfield  {title} {\enquote {\bibinfo {title} {Strongly
  interacting majorana modes in an array of josephson junctions},}\ }\href@noop
  {} {\bibfield  {journal} {\bibinfo  {journal} {New Journal of Physics}\
  }\textbf {\bibinfo {volume} {14}},\ \bibinfo {pages} {125018} (\bibinfo
  {year} {2012})}\BibitemShut {NoStop}%
\bibitem [{\citenamefont {Chiu}\ \emph {et~al.}(2015)\citenamefont {Chiu},
  \citenamefont {Pikulin},\ and\ \citenamefont {Franz}}]{chiu2015strongly}%
  \BibitemOpen
  \bibfield  {author} {\bibinfo {author} {\bibfnamefont {Ching-Kai}\
  \bibnamefont {Chiu}}, \bibinfo {author} {\bibfnamefont {DI}~\bibnamefont
  {Pikulin}}, \ and\ \bibinfo {author} {\bibfnamefont {M}~\bibnamefont
  {Franz}},\ }\bibfield  {title} {\enquote {\bibinfo {title} {Strongly
  interacting majorana fermions},}\ }\href@noop {} {\bibfield  {journal}
  {\bibinfo  {journal} {Physical Review B}\ }\textbf {\bibinfo {volume} {91}},\
  \bibinfo {pages} {165402} (\bibinfo {year} {2015})}\BibitemShut {NoStop}%
\bibitem [{\citenamefont {Vijay}\ \emph {et~al.}(2015)\citenamefont {Vijay},
  \citenamefont {Hsieh},\ and\ \citenamefont {Fu}}]{vijay2015majorana}%
  \BibitemOpen
  \bibfield  {author} {\bibinfo {author} {\bibfnamefont {Sagar}\ \bibnamefont
  {Vijay}}, \bibinfo {author} {\bibfnamefont {Timothy~H}\ \bibnamefont
  {Hsieh}}, \ and\ \bibinfo {author} {\bibfnamefont {Liang}\ \bibnamefont
  {Fu}},\ }\bibfield  {title} {\enquote {\bibinfo {title} {Majorana fermion
  surface code for universal quantum computation},}\ }\href@noop {} {\bibfield
  {journal} {\bibinfo  {journal} {Physical Review X}\ }\textbf {\bibinfo
  {volume} {5}},\ \bibinfo {pages} {041038} (\bibinfo {year}
  {2015})}\BibitemShut {NoStop}%
\bibitem [{\citenamefont {Hakim}\ and\ \citenamefont
  {Ambegaokar}(1985)}]{hakim1985quantum}%
  \BibitemOpen
  \bibfield  {author} {\bibinfo {author} {\bibfnamefont {Vincent}\ \bibnamefont
  {Hakim}}\ and\ \bibinfo {author} {\bibfnamefont {Vinay}\ \bibnamefont
  {Ambegaokar}},\ }\bibfield  {title} {\enquote {\bibinfo {title} {Quantum
  theory of a free particle interacting with a linearly dissipative
  environment},}\ }\href@noop {} {\bibfield  {journal} {\bibinfo  {journal}
  {Physical Review A}\ }\textbf {\bibinfo {volume} {32}},\ \bibinfo {pages}
  {423} (\bibinfo {year} {1985})}\BibitemShut {NoStop}%
\bibitem [{\citenamefont {Sinha}\ and\ \citenamefont
  {Sorkin}(1992)}]{sinha1992brownian}%
  \BibitemOpen
  \bibfield  {author} {\bibinfo {author} {\bibfnamefont {Supurna}\ \bibnamefont
  {Sinha}}\ and\ \bibinfo {author} {\bibfnamefont {Rafael~D}\ \bibnamefont
  {Sorkin}},\ }\bibfield  {title} {\enquote {\bibinfo {title} {Brownian motion
  at absolute zero},}\ }\href@noop {} {\bibfield  {journal} {\bibinfo
  {journal} {Physical Review B}\ }\textbf {\bibinfo {volume} {45}},\ \bibinfo
  {pages} {8123} (\bibinfo {year} {1992})}\BibitemShut {NoStop}%
\bibitem [{\citenamefont {Maghrebi}\ \emph {et~al.}(2016)\citenamefont
  {Maghrebi}, \citenamefont {Kr{\"u}ger},\ and\ \citenamefont
  {Kardar}}]{maghrebi2016flight}%
  \BibitemOpen
  \bibfield  {author} {\bibinfo {author} {\bibfnamefont {Mohammad~F}\
  \bibnamefont {Maghrebi}}, \bibinfo {author} {\bibfnamefont {Matthias}\
  \bibnamefont {Kr{\"u}ger}}, \ and\ \bibinfo {author} {\bibfnamefont {Mehran}\
  \bibnamefont {Kardar}},\ }\bibfield  {title} {\enquote {\bibinfo {title}
  {Flight of a heavy particle nonlinearly coupled to a quantum bath},}\
  }\href@noop {} {\bibfield  {journal} {\bibinfo  {journal} {Physical Review
  B}\ }\textbf {\bibinfo {volume} {93}},\ \bibinfo {pages} {014309} (\bibinfo
  {year} {2016})}\BibitemShut {NoStop}%
\bibitem [{\citenamefont {Calabrese}\ and\ \citenamefont
  {Cardy}(2009)}]{calabrese2009entanglement}%
  \BibitemOpen
  \bibfield  {author} {\bibinfo {author} {\bibfnamefont {Pasquale}\
  \bibnamefont {Calabrese}}\ and\ \bibinfo {author} {\bibfnamefont {John}\
  \bibnamefont {Cardy}},\ }\bibfield  {title} {\enquote {\bibinfo {title}
  {Entanglement entropy and conformal field theory},}\ }\href@noop {}
  {\bibfield  {journal} {\bibinfo  {journal} {Journal of Physics A:
  Mathematical and Theoretical}\ }\textbf {\bibinfo {volume} {42}},\ \bibinfo
  {pages} {504005} (\bibinfo {year} {2009})}\BibitemShut {NoStop}%
\bibitem [{\citenamefont {Jacobsen}\ and\ \citenamefont
  {Saleur}(2008)}]{jacobsen2008exact}%
  \BibitemOpen
  \bibfield  {author} {\bibinfo {author} {\bibfnamefont {Judith~Lone}\
  \bibnamefont {Jacobsen}}\ and\ \bibinfo {author} {\bibfnamefont {Hubert}\
  \bibnamefont {Saleur}},\ }\bibfield  {title} {\enquote {\bibinfo {title}
  {Exact valence bond entanglement entropy and probability distribution in the
  x x x spin chain and the potts model},}\ }\href@noop {} {\bibfield  {journal}
  {\bibinfo  {journal} {Physical review letters}\ }\textbf {\bibinfo {volume}
  {100}},\ \bibinfo {pages} {087205} (\bibinfo {year} {2008})}\BibitemShut
  {NoStop}%
\bibitem [{\citenamefont {Alcaraz}\ and\ \citenamefont
  {Rittenberg}(2010)}]{alcaraz2010shared}%
  \BibitemOpen
  \bibfield  {author} {\bibinfo {author} {\bibfnamefont {Francisco~C}\
  \bibnamefont {Alcaraz}}\ and\ \bibinfo {author} {\bibfnamefont {Vladimir}\
  \bibnamefont {Rittenberg}},\ }\bibfield  {title} {\enquote {\bibinfo {title}
  {Shared information in stationary states at criticality},}\ }\href@noop {}
  {\bibfield  {journal} {\bibinfo  {journal} {Journal of Statistical Mechanics:
  Theory and Experiment}\ }\textbf {\bibinfo {volume} {2010}},\ \bibinfo
  {pages} {P03024} (\bibinfo {year} {2010})}\BibitemShut {NoStop}%
\bibitem [{\citenamefont {Bhatt}\ and\ \citenamefont {Lee}(1982)}]{Lee1982}%
  \BibitemOpen
  \bibfield  {author} {\bibinfo {author} {\bibfnamefont {R.~N.}\ \bibnamefont
  {Bhatt}}\ and\ \bibinfo {author} {\bibfnamefont {P.~A.}\ \bibnamefont
  {Lee}},\ }\bibfield  {title} {\enquote {\bibinfo {title} {Scaling studies of
  highly disordered spin-\textonehalf{} antiferromagnetic systems},}\ }\href
  {\doibase 10.1103/PhysRevLett.48.344} {\bibfield  {journal} {\bibinfo
  {journal} {Phys. Rev. Lett.}\ }\textbf {\bibinfo {volume} {48}},\ \bibinfo
  {pages} {344--347} (\bibinfo {year} {1982})}\BibitemShut {NoStop}%
\bibitem [{\citenamefont {Fisher}(1994)}]{Fisher1994}%
  \BibitemOpen
  \bibfield  {author} {\bibinfo {author} {\bibfnamefont {Daniel~S.}\
  \bibnamefont {Fisher}},\ }\bibfield  {title} {\enquote {\bibinfo {title}
  {Random antiferromagnetic quantum spin chains},}\ }\href {\doibase
  10.1103/PhysRevB.50.3799} {\bibfield  {journal} {\bibinfo  {journal} {Phys.
  Rev. B}\ }\textbf {\bibinfo {volume} {50}},\ \bibinfo {pages} {3799--3821}
  (\bibinfo {year} {1994})}\BibitemShut {NoStop}%
\bibitem [{\citenamefont {Lin}\ \emph {et~al.}(2003)\citenamefont {Lin},
  \citenamefont {M\'elin}, \citenamefont {Rieger},\ and\ \citenamefont
  {Igl\'oi}}]{Igloi2003}%
  \BibitemOpen
  \bibfield  {author} {\bibinfo {author} {\bibfnamefont {Y.-C.}\ \bibnamefont
  {Lin}}, \bibinfo {author} {\bibfnamefont {R.}~\bibnamefont {M\'elin}},
  \bibinfo {author} {\bibfnamefont {H.}~\bibnamefont {Rieger}}, \ and\ \bibinfo
  {author} {\bibfnamefont {F.}~\bibnamefont {Igl\'oi}},\ }\bibfield  {title}
  {\enquote {\bibinfo {title} {Low-energy fixed points of random heisenberg
  models},}\ }\href {\doibase 10.1103/PhysRevB.68.024424} {\bibfield  {journal}
  {\bibinfo  {journal} {Phys. Rev. B}\ }\textbf {\bibinfo {volume} {68}},\
  \bibinfo {pages} {024424} (\bibinfo {year} {2003})}\BibitemShut {NoStop}%
\bibitem [{\citenamefont {Refael}\ and\ \citenamefont
  {Moore}(2004)}]{refael2004entanglement}%
  \BibitemOpen
  \bibfield  {author} {\bibinfo {author} {\bibfnamefont {Gil}\ \bibnamefont
  {Refael}}\ and\ \bibinfo {author} {\bibfnamefont {Joel~E}\ \bibnamefont
  {Moore}},\ }\bibfield  {title} {\enquote {\bibinfo {title} {Entanglement
  entropy of random quantum critical points in one dimension},}\ }\href@noop {}
  {\bibfield  {journal} {\bibinfo  {journal} {Physical review letters}\
  }\textbf {\bibinfo {volume} {93}},\ \bibinfo {pages} {260602} (\bibinfo
  {year} {2004})}\BibitemShut {NoStop}%
\bibitem [{\citenamefont {Bonesteel}\ and\ \citenamefont
  {Yang}(2007)}]{bonesteel2007infinite}%
  \BibitemOpen
  \bibfield  {author} {\bibinfo {author} {\bibfnamefont {NE}~\bibnamefont
  {Bonesteel}}\ and\ \bibinfo {author} {\bibfnamefont {Kun}\ \bibnamefont
  {Yang}},\ }\bibfield  {title} {\enquote {\bibinfo {title}
  {Infinite-randomness fixed points for chains of non-abelian
  quasiparticles},}\ }\href@noop {} {\bibfield  {journal} {\bibinfo  {journal}
  {Physical review letters}\ }\textbf {\bibinfo {volume} {99}},\ \bibinfo
  {pages} {140405} (\bibinfo {year} {2007})}\BibitemShut {NoStop}%
\bibitem [{\citenamefont {Shu}\ \emph {et~al.}(2016)\citenamefont {Shu},
  \citenamefont {Yao}, \citenamefont {Ke}, \citenamefont {Lin},\ and\
  \citenamefont {Sandvik}}]{shu2016properties}%
  \BibitemOpen
  \bibfield  {author} {\bibinfo {author} {\bibfnamefont {Yu-Rong}\ \bibnamefont
  {Shu}}, \bibinfo {author} {\bibfnamefont {Dao-Xin}\ \bibnamefont {Yao}},
  \bibinfo {author} {\bibfnamefont {Chih-Wei}\ \bibnamefont {Ke}}, \bibinfo
  {author} {\bibfnamefont {Yu-Cheng}\ \bibnamefont {Lin}}, \ and\ \bibinfo
  {author} {\bibfnamefont {Anders~W}\ \bibnamefont {Sandvik}},\ }\bibfield
  {title} {\enquote {\bibinfo {title} {Properties of the random-singlet phase:
  From the disordered heisenberg chain to an amorphous valence-bond solid},}\
  }\href@noop {} {\bibfield  {journal} {\bibinfo  {journal} {Physical Review
  B}\ }\textbf {\bibinfo {volume} {94}},\ \bibinfo {pages} {174442} (\bibinfo
  {year} {2016})}\BibitemShut {NoStop}%
\bibitem [{\citenamefont {Pekker}\ \emph {et~al.}(2014)\citenamefont {Pekker},
  \citenamefont {Refael}, \citenamefont {Altman}, \citenamefont {Demler},\ and\
  \citenamefont {Oganesyan}}]{pekker2014hilbert}%
  \BibitemOpen
  \bibfield  {author} {\bibinfo {author} {\bibfnamefont {David}\ \bibnamefont
  {Pekker}}, \bibinfo {author} {\bibfnamefont {Gil}\ \bibnamefont {Refael}},
  \bibinfo {author} {\bibfnamefont {Ehud}\ \bibnamefont {Altman}}, \bibinfo
  {author} {\bibfnamefont {Eugene}\ \bibnamefont {Demler}}, \ and\ \bibinfo
  {author} {\bibfnamefont {Vadim}\ \bibnamefont {Oganesyan}},\ }\bibfield
  {title} {\enquote {\bibinfo {title} {Hilbert-glass transition: New
  universality of temperature-tuned many-body dynamical quantum criticality},}\
  }\href@noop {} {\bibfield  {journal} {\bibinfo  {journal} {Physical review
  x}\ }\textbf {\bibinfo {volume} {4}},\ \bibinfo {pages} {011052} (\bibinfo
  {year} {2014})}\BibitemShut {NoStop}%
\bibitem [{\citenamefont {Vasseur}\ \emph {et~al.}(2015)\citenamefont
  {Vasseur}, \citenamefont {Potter},\ and\ \citenamefont
  {Parameswaran}}]{vasseur2015quantum}%
  \BibitemOpen
  \bibfield  {author} {\bibinfo {author} {\bibfnamefont {R.}~\bibnamefont
  {Vasseur}}, \bibinfo {author} {\bibfnamefont {A.~C.}\ \bibnamefont {Potter}},
  \ and\ \bibinfo {author} {\bibfnamefont {S.~A.}\ \bibnamefont
  {Parameswaran}},\ }\bibfield  {title} {\enquote {\bibinfo {title} {Quantum
  criticality of hot random spin chains},}\ }\href {\doibase
  10.1103/PhysRevLett.114.217201} {\bibfield  {journal} {\bibinfo  {journal}
  {Phys. Rev. Lett.}\ }\textbf {\bibinfo {volume} {114}},\ \bibinfo {pages}
  {217201} (\bibinfo {year} {2015})}\BibitemShut {NoStop}%
\bibitem [{\citenamefont {Aizenman}\ \emph {et~al.}(2001)\citenamefont
  {Aizenman}, \citenamefont {Goldstein},\ and\ \citenamefont
  {Lebowitz}}]{aizenman2001bounded}%
  \BibitemOpen
  \bibfield  {author} {\bibinfo {author} {\bibfnamefont {M}~\bibnamefont
  {Aizenman}}, \bibinfo {author} {\bibfnamefont {S}~\bibnamefont {Goldstein}},
  \ and\ \bibinfo {author} {\bibfnamefont {JL}~\bibnamefont {Lebowitz}},\
  }\bibfield  {title} {\enquote {\bibinfo {title} {Bounded fluctuations and
  translation symmetry breaking in one-dimensional particle systems},}\
  }\href@noop {} {\bibfield  {journal} {\bibinfo  {journal} {Journal of
  Statistical Physics}\ }\textbf {\bibinfo {volume} {103}},\ \bibinfo {pages}
  {601--618} (\bibinfo {year} {2001})}\BibitemShut {NoStop}%
\bibitem [{\citenamefont {Bonesteel}(1989)}]{bonesteel1989valence}%
  \BibitemOpen
  \bibfield  {author} {\bibinfo {author} {\bibfnamefont {NE}~\bibnamefont
  {Bonesteel}},\ }\bibfield  {title} {\enquote {\bibinfo {title} {Valence bonds
  and the lieb-schultz-mattis theorem},}\ }\href@noop {} {\bibfield  {journal}
  {\bibinfo  {journal} {Physical Review B}\ }\textbf {\bibinfo {volume} {40}},\
  \bibinfo {pages} {8954} (\bibinfo {year} {1989})}\BibitemShut {NoStop}%
\bibitem [{\citenamefont {Thouless}(1987)}]{thouless1987fluxoid}%
  \BibitemOpen
  \bibfield  {author} {\bibinfo {author} {\bibfnamefont {DJ}~\bibnamefont
  {Thouless}},\ }\bibfield  {title} {\enquote {\bibinfo {title} {Fluxoid
  quantization in the resonating-valence-bond model},}\ }\href@noop {}
  {\bibfield  {journal} {\bibinfo  {journal} {Physical Review B}\ }\textbf
  {\bibinfo {volume} {36}},\ \bibinfo {pages} {7187} (\bibinfo {year}
  {1987})}\BibitemShut {NoStop}%
\bibitem [{\citenamefont {Hoshen}\ and\ \citenamefont
  {Kopelman}(1976)}]{hoshen1976percolation}%
  \BibitemOpen
  \bibfield  {author} {\bibinfo {author} {\bibfnamefont {Joseph}\ \bibnamefont
  {Hoshen}}\ and\ \bibinfo {author} {\bibfnamefont {Raoul}\ \bibnamefont
  {Kopelman}},\ }\bibfield  {title} {\enquote {\bibinfo {title} {Percolation
  and cluster distribution. i. cluster multiple labeling technique and critical
  concentration algorithm},}\ }\href@noop {} {\bibfield  {journal} {\bibinfo
  {journal} {Physical Review B}\ }\textbf {\bibinfo {volume} {14}},\ \bibinfo
  {pages} {3438} (\bibinfo {year} {1976})}\BibitemShut {NoStop}%
\bibitem [{\citenamefont {Deng}\ and\ \citenamefont
  {Bl{\"o}te}(2005)}]{deng2005monte}%
  \BibitemOpen
  \bibfield  {author} {\bibinfo {author} {\bibfnamefont {Youjin}\ \bibnamefont
  {Deng}}\ and\ \bibinfo {author} {\bibfnamefont {Henk~WJ}\ \bibnamefont
  {Bl{\"o}te}},\ }\bibfield  {title} {\enquote {\bibinfo {title} {Monte carlo
  study of the site-percolation model in two and three dimensions},}\
  }\href@noop {} {\bibfield  {journal} {\bibinfo  {journal} {Physical Review
  E}\ }\textbf {\bibinfo {volume} {72}},\ \bibinfo {pages} {016126} (\bibinfo
  {year} {2005})}\BibitemShut {NoStop}%
\bibitem [{\citenamefont {Nahum}\ \emph
  {et~al.}(2013{\natexlab{a}})\citenamefont {Nahum}, \citenamefont {Serna},
  \citenamefont {Somoza},\ and\ \citenamefont {Ortuno}}]{nahum2013loop}%
  \BibitemOpen
  \bibfield  {author} {\bibinfo {author} {\bibfnamefont {Adam}\ \bibnamefont
  {Nahum}}, \bibinfo {author} {\bibfnamefont {P}~\bibnamefont {Serna}},
  \bibinfo {author} {\bibfnamefont {AM}~\bibnamefont {Somoza}}, \ and\ \bibinfo
  {author} {\bibfnamefont {M}~\bibnamefont {Ortuno}},\ }\bibfield  {title}
  {\enquote {\bibinfo {title} {Loop models with crossings},}\ }\href@noop {}
  {\bibfield  {journal} {\bibinfo  {journal} {Physical Review B}\ }\textbf
  {\bibinfo {volume} {87}},\ \bibinfo {pages} {184204} (\bibinfo {year}
  {2013}{\natexlab{a}})}\BibitemShut {NoStop}%
\bibitem [{\citenamefont {Kitaev}(2006)}]{kitaev2006anyons}%
  \BibitemOpen
  \bibfield  {author} {\bibinfo {author} {\bibfnamefont {Alexei}\ \bibnamefont
  {Kitaev}},\ }\bibfield  {title} {\enquote {\bibinfo {title} {Anyons in an
  exactly solved model and beyond},}\ }\href@noop {} {\bibfield  {journal}
  {\bibinfo  {journal} {Annals of Physics}\ }\textbf {\bibinfo {volume}
  {321}},\ \bibinfo {pages} {2--111} (\bibinfo {year} {2006})}\BibitemShut
  {NoStop}%
\bibitem [{\citenamefont {Ivanov}(2001)}]{ivanov2001non}%
  \BibitemOpen
  \bibfield  {author} {\bibinfo {author} {\bibfnamefont {Dmitri~A}\
  \bibnamefont {Ivanov}},\ }\bibfield  {title} {\enquote {\bibinfo {title}
  {Non-abelian statistics of half-quantum vortices in p-wave
  superconductors},}\ }\href@noop {} {\bibfield  {journal} {\bibinfo  {journal}
  {Physical review letters}\ }\textbf {\bibinfo {volume} {86}},\ \bibinfo
  {pages} {268} (\bibinfo {year} {2001})}\BibitemShut {NoStop}%
\bibitem [{\citenamefont {Lee}(1994)}]{lee1994renormalization}%
  \BibitemOpen
  \bibfield  {author} {\bibinfo {author} {\bibfnamefont {Benjamin~P}\
  \bibnamefont {Lee}},\ }\bibfield  {title} {\enquote {\bibinfo {title}
  {Renormalization group calculation for the reaction ka to oe},}\ }\href@noop
  {} {\bibfield  {journal} {\bibinfo  {journal} {Journal of Physics A:
  Mathematical and General}\ }\textbf {\bibinfo {volume} {27}},\ \bibinfo
  {pages} {2633} (\bibinfo {year} {1994})}\BibitemShut {NoStop}%
\bibitem [{\citenamefont {Nahum}\ \emph
  {et~al.}(2013{\natexlab{b}})\citenamefont {Nahum}, \citenamefont {Chalker},
  \citenamefont {Serna}, \citenamefont {Ortuno},\ and\ \citenamefont
  {Somoza}}]{nahum2013phase}%
  \BibitemOpen
  \bibfield  {author} {\bibinfo {author} {\bibfnamefont {Adam}\ \bibnamefont
  {Nahum}}, \bibinfo {author} {\bibfnamefont {JT}~\bibnamefont {Chalker}},
  \bibinfo {author} {\bibfnamefont {P}~\bibnamefont {Serna}}, \bibinfo {author}
  {\bibfnamefont {M}~\bibnamefont {Ortuno}}, \ and\ \bibinfo {author}
  {\bibfnamefont {AM}~\bibnamefont {Somoza}},\ }\bibfield  {title} {\enquote
  {\bibinfo {title} {Phase transitions in three-dimensional loop models and the
  c p n- 1 sigma model},}\ }\href@noop {} {\bibfield  {journal} {\bibinfo
  {journal} {Physical Review B}\ }\textbf {\bibinfo {volume} {88}},\ \bibinfo
  {pages} {134411} (\bibinfo {year} {2013}{\natexlab{b}})}\BibitemShut
  {NoStop}%
\bibitem [{\citenamefont {de~Gennes}(1979)}]{de1979scaling}%
  \BibitemOpen
  \bibfield  {author} {\bibinfo {author} {\bibfnamefont {Pierre-Gilles}\
  \bibnamefont {de~Gennes}},\ }\href@noop {} {\emph {\bibinfo {title} {Scaling
  concepts in polymer physics}}}\ (\bibinfo  {publisher} {Cornell university
  press},\ \bibinfo {year} {1979})\BibitemShut {NoStop}%
\bibitem [{\citenamefont {Wolf}(2006)}]{WolfViolation2006}%
  \BibitemOpen
  \bibfield  {author} {\bibinfo {author} {\bibfnamefont {Michael~M.}\
  \bibnamefont {Wolf}},\ }\bibfield  {title} {\enquote {\bibinfo {title}
  {Violation of the entropic area law for fermions},}\ }\href {\doibase
  10.1103/PhysRevLett.96.010404} {\bibfield  {journal} {\bibinfo  {journal}
  {Phys. Rev. Lett.}\ }\textbf {\bibinfo {volume} {96}},\ \bibinfo {pages}
  {010404} (\bibinfo {year} {2006})}\BibitemShut {NoStop}%
\bibitem [{\citenamefont {Gioev}\ and\ \citenamefont
  {Klich}(2006)}]{EntanglementGioev2006}%
  \BibitemOpen
  \bibfield  {author} {\bibinfo {author} {\bibfnamefont {Dimitri}\ \bibnamefont
  {Gioev}}\ and\ \bibinfo {author} {\bibfnamefont {Israel}\ \bibnamefont
  {Klich}},\ }\bibfield  {title} {\enquote {\bibinfo {title} {Entanglement
  entropy of fermions in any dimension and the widom conjecture},}\ }\href
  {\doibase 10.1103/PhysRevLett.96.100503} {\bibfield  {journal} {\bibinfo
  {journal} {Phys. Rev. Lett.}\ }\textbf {\bibinfo {volume} {96}},\ \bibinfo
  {pages} {100503} (\bibinfo {year} {2006})}\BibitemShut {NoStop}%
\bibitem [{\citenamefont {Swingle}(2010)}]{SwingleEntanglementFermi2010}%
  \BibitemOpen
  \bibfield  {author} {\bibinfo {author} {\bibfnamefont {Brian}\ \bibnamefont
  {Swingle}},\ }\bibfield  {title} {\enquote {\bibinfo {title} {Entanglement
  entropy and the fermi surface},}\ }\href {\doibase
  10.1103/PhysRevLett.105.050502} {\bibfield  {journal} {\bibinfo  {journal}
  {Phys. Rev. Lett.}\ }\textbf {\bibinfo {volume} {105}},\ \bibinfo {pages}
  {050502} (\bibinfo {year} {2010})}\BibitemShut {NoStop}%
\bibitem [{\citenamefont {Nahum}\ and\ \citenamefont
  {Chalker}(2012)}]{nahum2012universal}%
  \BibitemOpen
  \bibfield  {author} {\bibinfo {author} {\bibfnamefont {Adam}\ \bibnamefont
  {Nahum}}\ and\ \bibinfo {author} {\bibfnamefont {JT}~\bibnamefont
  {Chalker}},\ }\bibfield  {title} {\enquote {\bibinfo {title} {Universal
  statistics of vortex lines},}\ }\href@noop {} {\bibfield  {journal} {\bibinfo
   {journal} {Physical Review E}\ }\textbf {\bibinfo {volume} {85}},\ \bibinfo
  {pages} {031141} (\bibinfo {year} {2012})}\BibitemShut {NoStop}%
\bibitem [{\citenamefont {Jacobsen}\ \emph {et~al.}(2003)\citenamefont
  {Jacobsen}, \citenamefont {Read},\ and\ \citenamefont
  {Saleur}}]{jacobsen2003dense}%
  \BibitemOpen
  \bibfield  {author} {\bibinfo {author} {\bibfnamefont {Jesper-Lykke}\
  \bibnamefont {Jacobsen}}, \bibinfo {author} {\bibfnamefont {Nicholas}\
  \bibnamefont {Read}}, \ and\ \bibinfo {author} {\bibfnamefont {Hubert}\
  \bibnamefont {Saleur}},\ }\bibfield  {title} {\enquote {\bibinfo {title}
  {Dense loops, supersymmetry, and goldstone phases in two dimensions},}\
  }\href@noop {} {\bibfield  {journal} {\bibinfo  {journal} {Physical review
  letters}\ }\textbf {\bibinfo {volume} {90}},\ \bibinfo {pages} {090601}
  (\bibinfo {year} {2003})}\BibitemShut {NoStop}%
\bibitem [{\citenamefont {Martins}\ \emph {et~al.}(1998)\citenamefont
  {Martins}, \citenamefont {Nienhuis},\ and\ \citenamefont
  {Rietman}}]{martins1998intersecting}%
  \BibitemOpen
  \bibfield  {author} {\bibinfo {author} {\bibfnamefont {MJ}~\bibnamefont
  {Martins}}, \bibinfo {author} {\bibfnamefont {B}~\bibnamefont {Nienhuis}}, \
  and\ \bibinfo {author} {\bibfnamefont {R}~\bibnamefont {Rietman}},\
  }\bibfield  {title} {\enquote {\bibinfo {title} {Intersecting loop model as a
  solvable super spin chain},}\ }\href@noop {} {\bibfield  {journal} {\bibinfo
  {journal} {Physical review letters}\ }\textbf {\bibinfo {volume} {81}},\
  \bibinfo {pages} {504} (\bibinfo {year} {1998})}\BibitemShut {NoStop}%
\bibitem [{\citenamefont {Chalker}\ and\ \citenamefont
  {Coddington}(1988)}]{chalker1988percolation}%
  \BibitemOpen
  \bibfield  {author} {\bibinfo {author} {\bibfnamefont {JT}~\bibnamefont
  {Chalker}}\ and\ \bibinfo {author} {\bibfnamefont {PD}~\bibnamefont
  {Coddington}},\ }\bibfield  {title} {\enquote {\bibinfo {title} {Percolation,
  quantum tunnelling and the integer hall effect},}\ }\href@noop {} {\bibfield
  {journal} {\bibinfo  {journal} {Journal of Physics C: Solid State Physics}\
  }\textbf {\bibinfo {volume} {21}},\ \bibinfo {pages} {2665} (\bibinfo {year}
  {1988})}\BibitemShut {NoStop}%
\bibitem [{\citenamefont {Gruzberg}\ \emph {et~al.}(1999)\citenamefont
  {Gruzberg}, \citenamefont {Ludwig},\ and\ \citenamefont
  {Read}}]{gruzberg1999exact}%
  \BibitemOpen
  \bibfield  {author} {\bibinfo {author} {\bibfnamefont {Ilya~A}\ \bibnamefont
  {Gruzberg}}, \bibinfo {author} {\bibfnamefont {Andreas~WW}\ \bibnamefont
  {Ludwig}}, \ and\ \bibinfo {author} {\bibfnamefont {Nicholas}\ \bibnamefont
  {Read}},\ }\bibfield  {title} {\enquote {\bibinfo {title} {Exact exponents
  for the spin quantum hall transition},}\ }\href@noop {} {\bibfield  {journal}
  {\bibinfo  {journal} {Physical review letters}\ }\textbf {\bibinfo {volume}
  {82}},\ \bibinfo {pages} {4524} (\bibinfo {year} {1999})}\BibitemShut
  {NoStop}%
\bibitem [{\citenamefont {Beamond}\ \emph {et~al.}(2002)\citenamefont
  {Beamond}, \citenamefont {Cardy},\ and\ \citenamefont
  {Chalker}}]{beamond2002quantum}%
  \BibitemOpen
  \bibfield  {author} {\bibinfo {author} {\bibfnamefont {EJ}~\bibnamefont
  {Beamond}}, \bibinfo {author} {\bibfnamefont {John}\ \bibnamefont {Cardy}}, \
  and\ \bibinfo {author} {\bibfnamefont {JT}~\bibnamefont {Chalker}},\
  }\bibfield  {title} {\enquote {\bibinfo {title} {Quantum and classical
  localization, the spin quantum hall effect, and generalizations},}\
  }\href@noop {} {\bibfield  {journal} {\bibinfo  {journal} {Physical Review
  B}\ }\textbf {\bibinfo {volume} {65}},\ \bibinfo {pages} {214301} (\bibinfo
  {year} {2002})}\BibitemShut {NoStop}%
\bibitem [{\citenamefont {Bergholtz}\ \emph {et~al.}(2019)\citenamefont
  {Bergholtz}, \citenamefont {Budich},\ and\ \citenamefont
  {Kunst}}]{bergholtz2019exceptional}%
  \BibitemOpen
  \bibfield  {author} {\bibinfo {author} {\bibfnamefont {Emil~J}\ \bibnamefont
  {Bergholtz}}, \bibinfo {author} {\bibfnamefont {Jan~Carl}\ \bibnamefont
  {Budich}}, \ and\ \bibinfo {author} {\bibfnamefont {Flore~K}\ \bibnamefont
  {Kunst}},\ }\bibfield  {title} {\enquote {\bibinfo {title} {Exceptional
  topology of non-hermitian systems},}\ }\href@noop {} {\bibfield  {journal}
  {\bibinfo  {journal} {arXiv preprint arXiv:1912.10048}\ } (\bibinfo {year}
  {2019})}\BibitemShut {NoStop}%
\bibitem [{\citenamefont {Feiguin}\ \emph {et~al.}(2007)\citenamefont
  {Feiguin}, \citenamefont {Trebst}, \citenamefont {Ludwig}, \citenamefont
  {Troyer}, \citenamefont {Kitaev}, \citenamefont {Wang},\ and\ \citenamefont
  {Freedman}}]{feiguin2007interacting}%
  \BibitemOpen
  \bibfield  {author} {\bibinfo {author} {\bibfnamefont {Adrian}\ \bibnamefont
  {Feiguin}}, \bibinfo {author} {\bibfnamefont {Simon}\ \bibnamefont {Trebst}},
  \bibinfo {author} {\bibfnamefont {Andreas~WW}\ \bibnamefont {Ludwig}},
  \bibinfo {author} {\bibfnamefont {Matthias}\ \bibnamefont {Troyer}}, \bibinfo
  {author} {\bibfnamefont {Alexei}\ \bibnamefont {Kitaev}}, \bibinfo {author}
  {\bibfnamefont {Zhenghan}\ \bibnamefont {Wang}}, \ and\ \bibinfo {author}
  {\bibfnamefont {Michael~H}\ \bibnamefont {Freedman}},\ }\bibfield  {title}
  {\enquote {\bibinfo {title} {Interacting anyons in topological quantum
  liquids: The golden chain},}\ }\href@noop {} {\bibfield  {journal} {\bibinfo
  {journal} {Physical review letters}\ }\textbf {\bibinfo {volume} {98}},\
  \bibinfo {pages} {160409} (\bibinfo {year} {2007})}\BibitemShut {NoStop}%
\bibitem [{\citenamefont {Trebst}\ \emph {et~al.}(2008)\citenamefont {Trebst},
  \citenamefont {Ardonne}, \citenamefont {Feiguin}, \citenamefont {Huse},
  \citenamefont {Ludwig},\ and\ \citenamefont {Troyer}}]{trebst2008collective}%
  \BibitemOpen
  \bibfield  {author} {\bibinfo {author} {\bibfnamefont {Simon}\ \bibnamefont
  {Trebst}}, \bibinfo {author} {\bibfnamefont {Eddy}\ \bibnamefont {Ardonne}},
  \bibinfo {author} {\bibfnamefont {Adrian}\ \bibnamefont {Feiguin}}, \bibinfo
  {author} {\bibfnamefont {David~A}\ \bibnamefont {Huse}}, \bibinfo {author}
  {\bibfnamefont {Andreas~WW}\ \bibnamefont {Ludwig}}, \ and\ \bibinfo {author}
  {\bibfnamefont {Matthias}\ \bibnamefont {Troyer}},\ }\bibfield  {title}
  {\enquote {\bibinfo {title} {Collective states of interacting fibonacci
  anyons},}\ }\href@noop {} {\bibfield  {journal} {\bibinfo  {journal}
  {Physical review letters}\ }\textbf {\bibinfo {volume} {101}},\ \bibinfo
  {pages} {050401} (\bibinfo {year} {2008})}\BibitemShut {NoStop}%
\bibitem [{\citenamefont {Bonderson}\ \emph {et~al.}(2008)\citenamefont
  {Bonderson}, \citenamefont {Freedman},\ and\ \citenamefont
  {Nayak}}]{bonderson2008measurement}%
  \BibitemOpen
  \bibfield  {author} {\bibinfo {author} {\bibfnamefont {Parsa}\ \bibnamefont
  {Bonderson}}, \bibinfo {author} {\bibfnamefont {Michael}\ \bibnamefont
  {Freedman}}, \ and\ \bibinfo {author} {\bibfnamefont {Chetan}\ \bibnamefont
  {Nayak}},\ }\bibfield  {title} {\enquote {\bibinfo {title} {Measurement-only
  topological quantum computation},}\ }\href@noop {} {\bibfield  {journal}
  {\bibinfo  {journal} {Physical review letters}\ }\textbf {\bibinfo {volume}
  {101}},\ \bibinfo {pages} {010501} (\bibinfo {year} {2008})}\BibitemShut
  {NoStop}%
\bibitem [{\citenamefont {Cui}\ and\ \citenamefont
  {Wang}(2015)}]{cui2015universal}%
  \BibitemOpen
  \bibfield  {author} {\bibinfo {author} {\bibfnamefont {Shawn~X}\ \bibnamefont
  {Cui}}\ and\ \bibinfo {author} {\bibfnamefont {Zhenghan}\ \bibnamefont
  {Wang}},\ }\bibfield  {title} {\enquote {\bibinfo {title} {Universal quantum
  computation with metaplectic anyons},}\ }\href@noop {} {\bibfield  {journal}
  {\bibinfo  {journal} {Journal of Mathematical Physics}\ }\textbf {\bibinfo
  {volume} {56}},\ \bibinfo {pages} {032202} (\bibinfo {year}
  {2015})}\BibitemShut {NoStop}%
\bibitem [{\citenamefont {Levaillant}\ \emph {et~al.}(2015)\citenamefont
  {Levaillant}, \citenamefont {Bauer}, \citenamefont {Freedman}, \citenamefont
  {Wang},\ and\ \citenamefont {Bonderson}}]{levaillant2015universal}%
  \BibitemOpen
  \bibfield  {author} {\bibinfo {author} {\bibfnamefont {Claire}\ \bibnamefont
  {Levaillant}}, \bibinfo {author} {\bibfnamefont {Bela}\ \bibnamefont
  {Bauer}}, \bibinfo {author} {\bibfnamefont {Michael}\ \bibnamefont
  {Freedman}}, \bibinfo {author} {\bibfnamefont {Zhenghan}\ \bibnamefont
  {Wang}}, \ and\ \bibinfo {author} {\bibfnamefont {Parsa}\ \bibnamefont
  {Bonderson}},\ }\bibfield  {title} {\enquote {\bibinfo {title} {Universal
  gates via fusion and measurement operations on su (2) 4 anyons},}\
  }\href@noop {} {\bibfield  {journal} {\bibinfo  {journal} {Physical Review
  A}\ }\textbf {\bibinfo {volume} {92}},\ \bibinfo {pages} {012301} (\bibinfo
  {year} {2015})}\BibitemShut {NoStop}%
\bibitem [{\citenamefont {Cardy}(1996)}]{cardy1996scaling}%
  \BibitemOpen
  \bibfield  {author} {\bibinfo {author} {\bibfnamefont {John}\ \bibnamefont
  {Cardy}},\ }\href@noop {} {\emph {\bibinfo {title} {Scaling and
  renormalization in statistical physics}}},\ Vol.~\bibinfo {volume} {5}\
  (\bibinfo  {publisher} {Cambridge university press},\ \bibinfo {year}
  {1996})\BibitemShut {NoStop}%
\bibitem [{\citenamefont {Read}\ and\ \citenamefont
  {Saleur}(2001)}]{read2001exact}%
  \BibitemOpen
  \bibfield  {author} {\bibinfo {author} {\bibfnamefont {Nick}\ \bibnamefont
  {Read}}\ and\ \bibinfo {author} {\bibfnamefont {Hubert}\ \bibnamefont
  {Saleur}},\ }\bibfield  {title} {\enquote {\bibinfo {title} {Exact spectra of
  conformal supersymmetric nonlinear sigma models in two dimensions},}\
  }\href@noop {} {\bibfield  {journal} {\bibinfo  {journal} {Nuclear Physics
  B}\ }\textbf {\bibinfo {volume} {613}},\ \bibinfo {pages} {409--444}
  (\bibinfo {year} {2001})}\BibitemShut {NoStop}%
\bibitem [{\citenamefont {Nahum}(2016)}]{nahum2016universality}%
  \BibitemOpen
  \bibfield  {author} {\bibinfo {author} {\bibfnamefont {Adam}\ \bibnamefont
  {Nahum}},\ }\bibfield  {title} {\enquote {\bibinfo {title} {Universality
  class of the two-dimensional polymer collapse transition},}\ }\href@noop {}
  {\bibfield  {journal} {\bibinfo  {journal} {Physical Review E}\ }\textbf
  {\bibinfo {volume} {93}},\ \bibinfo {pages} {052502} (\bibinfo {year}
  {2016})}\BibitemShut {NoStop}%
\end{thebibliography}%

\end{document}